\documentclass[10pt, twocolumn, twoside]{IEEEtran}

\usepackage{cite}
\usepackage{enumerate}
\usepackage{graphicx}
\usepackage[cmex10]{amsmath} 
\usepackage{amssymb}
\usepackage{mathtools}
\usepackage{xspace}
\usepackage{subfigure}
\usepackage[lined,linesnumbered,ruled,commentsnumbered]{algorithm2e} 
\usepackage{colonequals}
\usepackage[mathscr]{euscript}
\usepackage{fixltx2e}    
\usepackage{amsthm}
\usepackage{mathrsfs}

\usepackage{epsfig}
\usepackage{epstopdf}

\usepackage{tikz}
\usepackage{pgfplots}

\usetikzlibrary{arrows,shapes,backgrounds,plotmarks,positioning}

\pgfplotsset{compat=newest}                         
\pgfplotsset{plot coordinates/math parser=false}
\newlength\figureheight
\newlength\figurewidth

\newtheorem{theorem}{Theorem}[section]
\newtheorem{lemma}[theorem]{Lemma}
\newtheorem{definition}[theorem]{Definition}
\newtheorem{corollary}[theorem]{Corollary}
\newtheorem{proposition}[theorem]{Proposition}

\newtheorem{example}[theorem]{Example}
\newtheorem{remark}[theorem]{Remark}
\newtheorem{experiment}[theorem]{Experiment}

\newcommand{\Chain}{\mathcal{C}}     

\newcommand{\rv}{\mathbf{r}}         
\newcommand{\R}{\mathcal{R}}         
\newcommand{\wv}{\mathbf{w}}         
\newcommand{\One}{\mathbf{1}}
\newcommand{\Zero}{\mathbf{0}}
\newcommand{\C}{\mathcal{C}}         

\newcommand{\Fu}[1]{f_{#1}}
\newcommand{\FuHash}[1]{f_{#1}^{\#}}
\newcommand{\FuHashHat}[1]{\hat{f}_{#1}^{\#}}

\newcommand{\Real}{\mathbb{R}}
\newcommand{\RealP}{\mathbb{R}_{+}}    
\newcommand{\Z}{\mathbb{Z}}            
\newcommand{\ZP}{\mathbb{Z}_+}            
\newcommand{\F}{\mathbb{F}}            
\newcommand{\Lat}{\mathcal{L}}         

\newcommand{\SFM}{\text{SFM}}

\newcommand{\DA}{\text{DA}}
\newcommand{\DC}{\text{DC}}

\newcommand{\op}{\text}
\newcommand{\RZ}[1]{\mathsf{Z}_{#1}}
\newcommand{\Rz}[1]{\mathsf{z}_{#1}}
\newcommand{\Rx}{\mathsf{x}}
\newcommand{\RW}[1]{\mathsf{W}_{#1}}
\newcommand{\RRCO}{\mathscr{R}_{\op{CO}}}
\newcommand{\RRACO}{\mathscr{R}_{\op{ACO}}^*}
\newcommand{\RRNCO}{\mathscr{R}_{\op{NCO}}^*}
\newcommand{\RACO}{R_{\op{ACO}}}
\newcommand{\RNCO}{R_{\op{NCO}}}
\newcommand{\Card}[1]{|#1|}
\newcommand{\Set}[1]{\{#1\}}

\newcommand{\Pat}{\mathcal{P}}

\newcommand{\Y}{\mathcal{Y}}
\newcommand{\U}{\mathcal{U}}

\newcommand{\X}{\mathcal{X}}
\newcommand{\EX}{\text{EX}}

\begin{document}

\title{Determining Optimal Rates for Communication for Omniscience}

\author{Ni~Ding, Chung~Chan, Qiaoqiao~Zhou,~\IEEEmembership{Member,~IEEE}, Rodney~A.~Kennedy,~\IEEEmembership{Fellow,~IEEE} and Parastoo~Sadeghi,~\IEEEmembership{Senior Member,~IEEE}

\thanks{The primary results have been published in \cite{Ding2016NetCod}. Part of the results have also been published in \cite{Ding2015ISIT,Ding2015Game}.}
\thanks{Ni Ding, Rodney A. Kennedy and Parastoo Sadeghi are with the Research School of Engineering, the Australian National University (email: $\{$ni.ding, rodney.kennedy, parastoo.sadeghi$\}$@anu.edu.au). Part of the work of Ni Ding (email: ni.ding@inc.cuhk.edu.hk) has been done when she was a junior research assistant at the Institute of Network Coding, The Chinese University of Hong Kong, from Nov 23, 2015 to Feb 6, 2016.}
\thanks{Chung Chan (email: cchan@inc.cuhk.edu.hk) and Qiaoqiao Zhou (email: zq115@ie.cuhk.edu.hk) are with the Institute of Network Coding, The Chinese University of Hong Kong. }
\thanks{Chung Chan was supported in part by the Vice-Chancellor’s One-off Discretionary Fund of The Chinese University of Hong Kong under Project VCF2014030 and Project VCF2015007 and in part by the University Grants Committee of the Hong Kong Special Administrative Region, China, under Project 14200714.}
}

\markboth{IEEE Transactions on Information Theory}%
{Ding \MakeLowercase{\textit{et al.}}: Determining Optimal Rates for Communication for Omniscience}

\maketitle

\begin{abstract}
This paper considers the communication for omniscience (CO) problem: A set of users observe a discrete memoryless multiple source and want to recover the entire multiple source via noise-free broadcast communications. We study the problem of how to determine an optimal rate vector that attains omniscience with the minimum sum-rate, the total number of communications. The results cover both asymptotic and non-asymptotic models where the transmission rates are real and integral, respectively.
We propose a modified decomposition algorithm (MDA) and a sum-rate increment algorithm (SIA) for the asymptotic and non-asymptotic models, respectively, both of which determine the value of the minimum sum-rate and a corresponding optimal rate vector in polynomial time.
For the coordinate saturation capacity (CoordSatCap) algorithm, a nesting algorithm in MDA and SIA, we propose to implement it by a fusion method and show by experimental results that this fusion method contributes to a reduction in computation complexity.
Finally, we show that the separable convex minimization problem over the optimal rate vector set in the asymptotic model can be decomposed by the fundamental partition, the optimal partition of the user set that determines the minimum sum-rate, so that the problem can be solved more efficiently.

\end{abstract}


\section{introduction}

Assume that there are a finite number of users in a system. Each of them observes a distinct component of a discrete multiple correlated source in private. The users are allowed to exchange their observations over public authenticated broadcast channels. We assume that these channels are noiseless so that all the transmissions are correctly heard, or received, by all users. The communications could be interactive and the rates of public communications are unconstrained. That is, there are no capacity upper bounds imposed on the broadcast links. The purpose is to attain \textit{omniscience}, the state that each user obtains all the components in the entire multiple source in the system. This problem is called \textit{communication for omniscience (CO)}, which was originally formulated in \cite{Csiszar2004}. The CO problem in \cite{Csiszar2004} is based on an \textit{asymptotic model} where the length of the observation sequence is allowed to approach infinity. Whereas the authors in \cite{Niti2010,Chan2011ITW,ChanSuccessiveIT} also study the \textit{non-asymptotic model} where the number of observations is finite and the communication rates are restricted to be integral. In fact, the non-asymptotic model is important in a practical problem in peer-to-peer (P2P) wireless communications as described below.

The \textit{finite linear source model} studied in \cite{Chan2011ITW} is an example of the non-asymptotic model, where the multiple random source is represented by a vector that belongs to a finite field and the users transmit linear combinations of their observations to obtain this vector. By assuming that each dimension in this vector represents a packet, the finite linear source model can describe the situation when a base station wants to disseminate a set of packets to a group of mobile clients: Each client only obtains a partial knowledge of the packet set due to the fading effects of the wireless channels. The omniscience of the packet set can be attained by letting the clients transmit linear combinations of packets, say, by some network coding scheme, e.g., \cite{Roua2010}, via the P2P channels, which could be more reliable than the retransmissions over base-to-peer (B2P) channels if the clients are geographically close to each other. The CO problem in this packet model is called \textit{coded cooperative data exchange (CCDE)} which was independently proposed in \cite{Roua2010,Court2010,Court2010M} and further studied in \cite{SprintRand2010,Ozgul2011,AbediniNonMinRank2012,Court2011,CourtIT2014}. In \cite{CourtIT2014,Taj2011}, the idea of packet-splitting was introduced to CCDE. It allows each packet to be divided into a number of chunks so that the transmissions in CCDE refer to the linear combinations of chunks and the normalized transmission rates are fractional. This can be considered as an extension of the CCDE and finite linear source model towards the asymptotic model.

An optimization problem that naturally arises is how to attain omniscience with the least cost and the cost usually refers to the overall transmission rates, or sum-rate, e.g., the total number of linear combinations of packets that are transmitted by all clients in CCDE. It is shown in \cite{Csiszar2004} that the Slepian-Wolf (SW) constraints \cite{SW1973} on all proper subsets of the user set determine the omniscience-achievability of a transmission rate vector. Hence, in \cite{Csiszar2004,Court2010,Court2011,Court2010M,CourtIT2014,MiloIT2016,ChanMMI,ChanSuccessive,ChanSuccessiveIT}, the problem of minimizing the sum-rate is formulated by linear programming (LP) and the combinatorial nature of this problem has also been revealed. Then, instead of solving the minimum sum-rate problem directly by the existing LP algorithms, the main issue is how to deal with the exponentially growing number of constraints.

In the studies on the finite linear source model in \cite{MiloFair2012,MiloIT2016} and CCDE in \cite{Court2010,Court2011,Court2010M,CourtIT2014,Ding2015ICT,Ding2015Game,Ding2015ISIT,Ding2015NetCod}, the submodularity of the minimum sum-rate problem was revealed, which is essentially due to the submodularity of the entropy function.\footnote{In \cite[Section 3]{FujishigePolyEntropy}, it is shown that the entropy function is the rank function of a polymatroid, which belongs to a subgroup of submodular functions. The entropy function reduces to the matrix rank function in the finite linear source model and the cardinality function in CCDE, both of which are submodular.}
In particular, it is shown in \cite{MiloFair2012,Ding2015ICT,Ding2015Game,Ding2015ISIT} that, in a non-asymptotic model where the entropy function takes integer values,\footnote{Finite linear source model and CCDE are examples of non-asymptotic model with integer-valued entropy function.}
all omniscience-achievable rate vectors that have the same sum-rate constitute a submodular base polyhedron. Since a rate vector in this submodular base polyhedron can be found by the Edmond greedy algorithm\footnote{The Edmond greedy algorithm in \cite{Edmonds2003Convex} is a special case of the coordinate saturation capacity algorithm \cite[Greedy Algorithm II in Section 3.2]{Fujishige2005} for normalized submodular functions \cite[Theorems 3.18 and 3.19]{Fujishige2005}. The one implemented in \cite{CourtIT2014,MiloIT2016} is modified for the crossing and intersecting submodular functions, respectively. }
and the variation range of the minimum sum-rate in a non-asymptotic model is integral and bounded,\footnote{The value of the minimum sum-rate is real in the asymptotic model and integral in the non-asymptotic model. It is nonnegative and no greater than the total amount of information in the multiple source.}
the minimum sum-rate and a corresponding optimal rate vector are determined efficiently by the sum-rate adaption algorithms proposed in \cite{CourtIT2014,MiloIT2016}.
However, it still remains unclear if all the results derived in \cite{MiloFair2012,Ding2015ICT,Ding2015Game,Ding2015ISIT} for the non-asymptotic model also hold for the asymptotic one and if there exists an algorithm that can efficiently determine an optimal rate vector that attains omniscience by the minimum sum-rate in an asymptotic model where the variation range of the minimum sum-rate is continuous. On the other hand, the study in \cite{Taj2011} shows that allowing packet-splitting in CCDE incurs less transmission costs in P2P communications. It could mean that the minimum sum-rate in the asymptotic model is no greater than the one in the non-asymptotic model in the same system, which makes it desirable to know the optimal solution for CO in the asymptotic model.

The importance of studying the CO problem is also highlighted by its dual relationship with the secret capacity, the maximum rate at which the secret key can be generated by the users in the system, in \cite{Csiszar2004}: The secret capacity equals to the total amount of information in the multiple source subtracted by the minimum sum-rate (in an asymptotic model) for them to achieve omniscience. It is also pointed out in \cite{Csiszar2004} that the secret capacity is upper bounded by a mutual dependence over the partitions of the users and this upper bound is shown to be tight in \cite{Chan2008tight}.\footnote{ The authors in \cite{Csiszar2004} derived the results on secret capacity in a general setting: A subset of the users are active while the others are the helpers that assist the active users in generating the secret key. The author in \cite{Chan2008tight} proved that the upper bound on secret capacity is tight when there is no helpers, i.e., all the users in the system are active.}
The mutual dependence is also named as shared information in \cite{Prakash2016} for the secret generation problem and multivariate mutual information (MMI) in \cite{ChanMMI,ChanSuccessive}, where the authors in \cite{ChanMMI} proposed this mutual dependence to be the generalization of Shannon’'s mutual information for multiple random variables. Then, determining secret capacity, mutual dependence, shared information or MMI relies on the solution to the CO problem in the asymptotic model and vice versa.
It is shown in \cite{ChanMMI} that the problem of obtaining the MMI reduces to the task of determining the value of the Dilworth truncation, which can be solved in strongly polynomial time due to the submodularity of the entropy function. However, for solving the CO problem, knowing the minimum sum-rate is not sufficient: We also need to know how to distribute the minimum sum-rate among the users so that omniscience is achievable. Therefore, it is required to determine an optimal rate vector that attains omniscience with the minimum sum-rate.

The work in this paper is based on the CO problem that is originally formulated in \cite{Csiszar2004}. We consider the minimum sum-rate problem: how to attain omniscience with the minimum total number of communications. The work in this paper differs form \cite{Csiszar2004,Chan2008tight,Prakash2016} in that, in addition to characterizing the minimum sum-rate or discussing how to obtain the value of it, we are particularly interested in how to determine a corresponding optimal rate vector that attains omniscience. The results cover both asymptotic and non-asymptotic models. For the non-asymptotic model, we focus on the finite linear source model and CCDE.
For solving the CO problem, we propose a modified decomposition algorithm (MDA) and a sum-rate increment algorithm (SIA) for the asymptotic and non-asymptotic models, respectively, both of which determine the value of the minimum sum-rate and a corresponding optimal rate vector in polynomial time.
For the coordinate saturation capacity (CoordSatCap) algorithm, a nesting algorithm in the MDA and SIA algorithm, we propose to implement it by a fusion method and show by experimental results that this fusion method contributes to a reduction in computation complexity as compared to the CoorSatCap algorithm that is implemented in \cite{CourtIT2014,MiloIT2016}.
We also derive results on the \textit{fundamental partition} $\Pat^*$, the finest optimal partition of the user set that determines the value of the minimum sum-rate in the asymptotic model,\footnote{It is shown in \cite{ChanMMI} that the optimal partitions that give rise to the minimum sum-rate form a Dilworth truncation lattice \cite{Lovasz1983} where the minimal/finest and maximal/coarsest minimizers exist. The name `fundamental partition' was first used in \cite{ChanMMI} to denote the finest partition in this lattice.}
which is also determined by the MDA algorithm. It is shown that, in CCDE, the omniscience of a packet set can be attained by splitting each packet into $|\Pat^*| - 1$ chunks. Finally, we reveal some decomposition properties of $\Pat^*$. We show that the separable convex minimization problem over the optimal rate vector set in the asymptotic model can be decomposed by $\Pat^*$ so that the problem can be solved more efficiently.

\subsection{Summary of Main Results}

Our main results are summarized as follows.

1) We show that all omniscience-achievable rate vectors with the sum-rate equal to a given value form a base polyhedron. By observing the nonemptiness of the base polyhedron, we prove directly based on \cite[Theorems 2.5(i) and 2.6(i)]{Fujishige2005} that the minimum sum-rate in the asymptotic model is determined by an optimization over the partitions of the user set. This result verifies the proof in \cite{Chan2008tight} on the tightness of the lower bound on the minimum sum-rate for CO that was proposed in \cite{Csiszar2004}.
For the non-asymptotic model, we show that the minimum sum-rate is the least integer that is no less than the one in the asymptotic model, which provides theoretical proof to an observation in \cite{CourtIT2014}: The difference in minimum sum-rate between the asymptotic and non-asymptotic models is no greater than one.
Since the optimal rate vectors also form a base polyhedron, we use the integrality of the extreme points in this base polyhedron to show two results for the minimum sum-rate problem in the finite linear source model, or CCDE: (a) there exists an integral optimal rate vector, which is consistent with the results in \cite{MiloIT2016,CourtIT2014}, and (b) there exists a fractional optimal rate vector that can be implemented by dividing each packet into $|\Pat^*| - 1$ chunks.

2) For determining an optimal rate vector in the asymptotic model, an MDA algorithm is proposed. It starts with a lower estimation on the minimum sum-rate and iteratively updates it by the finest minimizer of a Dilworth truncation problem until the minimum is reached and a corresponding optimal rate vector is determined.
For the CoordSatCap algorithm, which is originally proposed in \cite[Greedy Algorithm II in Section 3.2]{Fujishige2005}, for solving the Dilworth truncation problem in each iteration of the MDA algorithm, we propose a fusion method implementation (CoordSatCapFus) so that the submodular function minimization (SFM) in each iteration is solved over a merged or fused user set with the cardinality no greater than the original one. We show that the optimal solution returned by the MDA algorithm for the asymptotic model can also be utilized for solving the minimum sum-rate problem in the non-asymptotic model by no more than one additional call of the CoordSatCapFus algorithm. Independently, we also propose an SIA algorithm for determining an optimal rate vector as well as the minimum sum-rate for the non-asymptotic model.
Both the MDA and SIA algorithms complete in polynomial time based on the existing SFM techniques. We run experiments to show that the fusion method CoordSatCapFus contributes to a reduction in computation complexity as compared to the CoorSatCap algorithm, which is implemented in \cite{CourtIT2014,MiloIT2016} for the finite linear source model, and the reduction is considerable when the number of users grows.

3) For the initial estimations in the MDA and SIA algorithms, we derive a lower bound (LB) on the minimum sum-rate for both asymptotic and non-asymptotic models that can be determined in linear time. We show that this lower bound can be used as the initial estimation of the minimum sum-rate searching algorithms, e.g., the MDA and SIA algorithm proposed in this paper for the asymptotic and non-asymptotic models, respectively. The observation that a lower bound can initiate the minimum sum-rate searching algorithm is also consistent with the results in \cite{MiloIT2016,CourtIT2014} for the finite linear source model.\footnote{For the finite linear source model, or CCDE, it is suggested in \cite[Appendix F]{CourtIT2014} to adapt the sum-rate from either lower or upper bound to the minimum, while \cite[Algorithms 3]{MiloIT2016} is a binary search method starting with the initial lower and upper bounds on the minimum sum-rate.}
We run experiments to show that the proposed LB in the non-asymptotic model is much tighter than the ones in \cite{Roua2010,SprintRand2010}.

4) We also study the minimum weighted sum-rate problem in the optimal rate vector set, a problem that was originally formulated and studied for the finite linear source model in \cite{MiloIT2016,CourtIT2014}. It is shown that by choosing a proper linear ordering of the user indices the optimal rate vectors returned by the MDA and SIA algorithms also minimize a weighted sum-rate function in the optimal rate vector set for the asymptotic and non-asymptotic models, respectively.
The result that the minimum weighted sum-rate problem can be solved by a proper linear ordering for the non-asymptotic model is consistent with the results in \cite{MiloIT2016,CourtIT2014}. Our study shows that this idea can also be applied to the asymptotic model.

5) We show that the fundamental partition $\Pat^*$ is the minimal separator of a submodular function which gives rise to the decomposition property of $\Pat^*$ in the asymptotic model: The separable convex function minimization problem over the optimal rate vector set can be broken into $|\Pat^*|$ subproblems, each of which formulates the separable convex function minimization problem in one element or user subset in $\Pat^*$. These subproblems can be solved separately so that the overall complexity is reduced.

\subsection{Organization}

In Section~\ref{sec:system}, we present the system model for CO and describe the asymptotic and non-asymptotic models, the finite linear source model and CCDE. In Section~\ref{sec:MinSumRateOptRate}, we analyze the minimum sum-rate problem in both asymptotic and non-asymptotic models based on the concepts of submodularity and Dilworth truncation. In Section~\ref{sec:LB}, a LB on the minimum sum-rate is proposed for the asymptotic and non-asymptotic models, which is used in Section~\ref{sec:algo} to initiate the MDA and SIA algorithms. The complexity of both algorithms is also discussed in Section~\ref{sec:algo}. In Section~\ref{sec:FundPartMinSep}, we reveal the decomposition property of the fundamental partition in the asymptotic model.

\section{System Model}
\label{sec:system}

Let $V$ with $\Card{V}>1$ be a finite set that contains the indices of all users in the system. We call $V$ the \textit{ground set}. Let $\RZ{V}=(\RZ{i}:i\in V)$ be a vector of discrete random variables indexed by $V$. For each $i\in V$, user $i$ privately observes an $n$-sequence $\RZ{i}^n$ of the random source $\RZ{i}$ that is i.i.d.\ generated according to the joint distribution $P_{\RZ{V}}$. We allow users exchange their sources directly so as to let all users in $V$ recover the source sequence $\RZ{V}^n$. The state that each user obtains the total information in the entire multiple source is called \textit{omniscience}, and the process that users communicate with each other to attain omniscience is called \textit{communication for omniscience (CO)} \cite{Csiszar2004}.

Let $\rv_V=(r_i:i\in V)$ be a rate vector indexed by $V$. We call $\rv_V$ an \textit{achievable rate vector} if the omniscience can be attained by letting users communicate with the rates designated by $\rv_V$. Let $r$ be the \textit{sum-rate function} associated with $\rv_V$ such that
$$ r(X)=\sum_{i\in X} r_i, \quad \forall X \subseteq V $$
with the convention that $r(\emptyset)=0$. $r(V)$ is the sum-rate of $\rv_V$ over all users, or the total number of transmissions, in the system. For $X,Y \subseteq V$, let $H(\RZ{X})$ be the amount of randomness in $\RZ{X}$ measured by Shannon entropy \cite{YeungITBook}
and $H(\RZ{X}|\RZ{Y})=H(\RZ{X \cup Y})-H(\RZ{Y})$ be the conditional entropy of $\RZ{X}$ given $\RZ{Y}$. In the rest of this paper, without loss of generality, we simplify the notation $\RZ{X}$ by $X$.

It is shown in \cite{Csiszar2004} that an achievable rate vector must satisfy the Slepian-Wolf (SW) constraints \cite{SW1973}:
    \begin{equation} \label{eq:SWConstrs}
        r(X) \geq H(X|V \setminus X), \quad \forall X \subsetneq V.
    \end{equation}
The interpretation of \eqref{eq:SWConstrs} is: To attain omniscience, the total amount of information sent from user set $X$ should be at least equal to the total amount of information that is missing in $V \setminus X$. The set of all achievable rate vectors is \cite{Csiszar2004}\footnote{The achievable rate region was originally given in \cite{Csiszar2004} based on the SW constrains in a more general case: The omniscience problem in the active user set $A \subseteq V$ with the users in $V \setminus A$ serving as the helpers. The CO problem studied in this paper is the case when $A = V$.}
$$ \RRCO(V)=\Set{ \rv_V\in\Real^{|V|} \colon r(X) \geq H(X|V\setminus X),\forall X \subsetneq V }.$$
We say that $\alpha$ is an \textit{achievable sum-rate} if there exists an achievable rate vector $\rv_V \in \RRCO(V)$ such that $r(V) = \alpha$.


\subsection{Asymptotic and Non-asymptotic Models}

We consider both asymptotic and non-asymptotic models. In the asymptotic multiple random source model, we will study the CO problem by considering the asymptotic limits as the \emph{block length} $n$ goes to infinity. The communication rates in an asymptotic model could be real or fractional. The minimum sum-rate can be determined by the following linear programming (LP) \cite[Proposition 1]{Csiszar2004}
\begin{equation} \label{eq:MinSumRateACO}
    \RACO(V)=\min\Set{r(V) \colon \rv_V\in \RRCO(V)}
\end{equation}
and the set of all optimal rate vectors is
  $$ \RRACO(V)=\Set{\rv_V\in \RRCO(V) \colon r(V)=\RACO(V)}. $$

In the non-asymptotic model, the block length $n$ is finite and the communication rates are required to be integral. The minimum sum-rate can be determined by the integer linear programming (ILP) \cite{Court2011,CourtIT2014,MiloIT2016}\footnote{This ILP problem has been formulated in \cite{Court2011,CourtIT2014} in terms of the cardinality function for CCDE and in \cite{MiloIT2016} in terms of the rank function for the finite linear source model.}
\begin{equation} \label{eq:MinSumRateNCO}
    \RNCO(V)=\min\Set{r(V) \colon \rv_V\in \RRCO(V) \cap \Z^{|V|} }
\end{equation}
and the optimal rate vector set is
$$ \RRNCO(V)=\Set{\rv_V \in \RRCO(V) \cap \Z^{|V|} \colon r(V)=\RNCO(V)}. $$
The non-asymptotic model is exemplified by the finite linear source model and CCDE as explained as follows.

\subsection{Finite Linear Source Model and CCDE} \label{sec:IntroCCDE}

Let $\F_q$ be a finite field. $q$ is the order of $\F_q$ such that $q = p^N$, where $p$ is a prime number and $N$ is a positive integer. In a finite linear multiple source model, we assume that each $\RZ{i}$ can be expressed by an $l(\Rz{i})$-dimensional column vector $\Rz{i}$ in the finite field $\F_q^{l(\Rz{i})}$ such that
$$ \Rz{i} = A_i \Rx, $$
where $\Rx \in \F_q^{l(\Rx)}$ is some $l(\Rx)$-dimensional uniformly distributed random vector and $A_i \in \F_q^{l(\Rz{i}) \times l(\Rx)}$ is an $l(\Rz{i})$-by-$l(\Rx)$ matrix. For $X \subseteq V$, let $A_X= [A_i \colon i \in X]$. In the finite linear source model, the value of the entropy function at $X$ reduces to the rank of $A_X$, i.e., $H(X) = \text{rank}(A_X)$ and $H(X|Y)=\text{rank}(A_{X \cup Y}) - \text{rank}(A_Y)$. Then, $H$ is integer-valued, i.e., $H(X) \in \ZP, \forall X \subseteq V$, and we assume that $H(V) = l(\Rx)$. The users transmit linear combinations of $\Rz{i}$s in order to attain the omniscience of $\Rx$.\footnote{In a finite linear source model, it is sufficient for the user $i$ to transmit linear combinations of $\Rz{i}$s to attain omniscience \cite{Chan2011ITW}. }
Therefore, the finite linear source model is an example of the non-asymptotic model where the value of the entropy function $H$ is integral.

By realizing that each dimension in $\Rx$ can represent a packet, the finite linear source model poses a practical problem in wireless communications: the omniscience, or recovery, of a packet set in peer-to-peer (P2P) wireless network. Let all the users in $V$ be mobile clients that are geographically close to each other so that any client's broadcasts can be received losslessly by the others. Consider the problem of disseminating the packet set $\Rx$ from a base station to all mobile clients in $V$. Due to the fading effects of wireless channels, each client may just obtain a partial knowledge of $\Rx$ at the end of base-to-peer (B2P) transmissions, but the clients' knowledge could be complementary to each other. In this case, we can set free the base station and let the clients transmit linear combinations of packets, e.g., by some network coding scheme \cite{Roua2010}, so as to help each other recover $\Rx$. The omniscience problem in this packet model is how to let all users recover the packet set $\Rx$ with the least number of transmissions and this problem, which was originally formulated in \cite{Roua2010}, is called the coded cooperative data exchange (CCDE). The concept of packet-splitting was also introduced to CCDE in \cite{CourtIT2014,Taj2011}. It extends the finite linear source model from a non-asymptotic setting towards an asymptotic one, which is explained by the following example.

\begin{figure}[tpb]
	\centering
    \scalebox{1.1}{\begin{tikzpicture}

\draw (-2.7,0.3) rectangle (-1.3,-0.3);
\node at (-2,0) {client $1$};
\draw (-2,0.3)--(-2,1)--(-1.7,1)--(-2,0.8)--(-2.3,1)--(-2,1);
\node at (-2,-0.5) {\scriptsize \textcolor{blue}{$\Rz{1} = [\RW{a},\RW{b},\RW{c},\RW{d},\RW{e}]^\intercal$}};

\draw (2.7,0.3) rectangle (1.3,-0.3);
\node at (2,0) {client $2$};
\draw (2,0.3)--(2,1)--(1.7,1)--(2,0.8)--(2.3,1)--(2,1);
\node at (2,-0.5) {\scriptsize \textcolor{blue}{$\Rz{2} = [\RW{a},\RW{b},\RW{f}]^\intercal$}};

\draw (-0.7,1.6) rectangle (0.7,1);
\node at (0,1.3){client $3$};
\draw (0,1.6)--(0,2.3)--(0.3,2.3)--(0,2.1)--(-0.3,2.3)--(0,2.3);
\node at (0,0.8) {\scriptsize \textcolor{blue}{$\Rz{3} = [\RW{c},\RW{d},\RW{f}]^\intercal$}};


\end{tikzpicture}}
	\caption{The corresponding CCDE system for the CO problem in Example~\ref{ex:main}, where $\RW{j}$ denotes a packet that belongs to a field $\F_q$. There are three clients that want to obtain six packets in $\Rx = [\RW{a},\dotsc,\RW{f}]^\intercal$. User $i$ initially obtains $\Rz{i}$. The users transmit linear combinations of $\Rz{i}$s via lossless wireless broadcast channels to help the others recover all packets in $\Rx$.}
	\label{fig:CDESystem}
\end{figure}
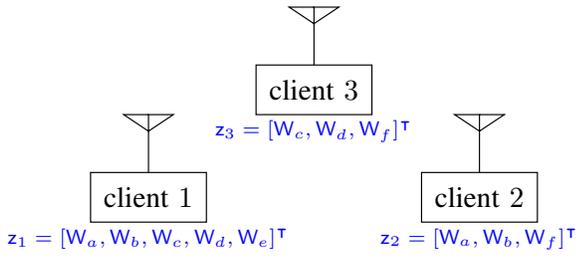

\begin{example} \label{ex:main}
There are three users $V=\Set{1,2,3}$ in the system. They observe respectively
    \begin{align}
        \RZ{1} &= (\RW{a},\RW{b},\RW{c},\RW{d},\RW{e}),   \nonumber\\
        \RZ{2} &= (\RW{a},\RW{b},\RW{f}),   \nonumber\\
        \RZ{3} &= (\RW{c},\RW{d},\RW{f}),   \nonumber
    \end{align}
where each $\RW{j}$ is an independent uniformly distributed random bit. The purpose is to let all the users attain the omniscience of $\RZ{V}$ via communications. In the corresponding CCDE system (see Fig.~\ref{fig:CDESystem}), each $\RW{j}$ represents a packet so that the column vector $\Rz{i}$ denotes all the packets received by mobile client $i$ after the B2P transmissions. All mobile clients in $V$ transmit linear combinations of $\Rz{i}$s over P2P channels in other to attain the omniscience of all packets in $\Rx=[\RW{a},\dotsc,\RW{f}]^{\intercal}$.

In this system, we have all the achievable rate vectors contained in
    \begin{align}
        \RRCO(V)=\big\{ \rv_V\in\Real^{|V|} \colon & r(\emptyset)=0,                           \nonumber \\
         &r(\Set{1}) \geq H(\Set{1}|\Set{2,3})=1 ,   \nonumber\\
         &r(\Set{2}) \geq H(\Set{2}|\Set{1,3})=0 ,   \nonumber\\
         &r(\Set{3}) \geq H(\Set{3}|\Set{1,2})=0 ,   \nonumber\\
         &r(\Set{1,2}) \geq H(\Set{1,2}|\Set{3})=3 ,   \nonumber\\
         &r(\Set{1,3}) \geq H(\Set{1,3}|\Set{2})=3 ,   \nonumber\\
         &r(\Set{2,3}) \geq H(\Set{2,3}|\Set{1})=1 \big\}.   \nonumber
    \end{align}
One can show that the minimum sum-rate is $\RACO(V)=\frac{7}{2}$ and the optimal rate vector set is $\RRACO(V)=\Set{(\frac{5}{2},\frac{1}{2},\frac{1}{2})}$ for the asymptotic model. In CCDE, the rate vector $(\frac{5}{2},\frac{1}{2},\frac{1}{2})$ can be implemented by packet-splitting. Let the users divide each packets into two chunks of equal length, e.g., $\Rz{2} = (\RW{a}^{(1)},\RW{a}^{(2)},\RW{b}^{(1)},\RW{b}^{(2)},\RW{f}^{(1)},\RW{f}^{(2)})$ where each $W_j$ is split to $W_j^{(1)}$ and $W_j^{(2)}$. Let the users transmit the rate $(5,1,1)$ with each tuple denoting the number of linear combinations of the packet chunks. We have $(\frac{5}{2},\frac{1}{2},\frac{1}{2})$ and $\frac{7}{2}$ being the normalized rate vector and sum-rate, respectively. For the non-asymptotic model, we have the minimum sum-rate $\RNCO(V)=4$ and the optimal rate vector set $\RRNCO(V)=\{(3,0,1),(2,1,1),(3,1,0)\}$.

We show an example of how to implement the rate vector $(2,1,1)$ in $\RRNCO(V)$ for the non-asymptotic model by a network coding scheme. The implementation of other optimal rate vectors in $\RRNCO(V)$ and $(\frac{5}{2},\frac{1}{2},\frac{1}{2})$ in $\RRACO(V)$ can be derived in a similar manner. By letting user $1$ transmit $\RW{b}+\RW{f}+\RW{c}$ and $\RW{e}$, user $2$ transmit $\RW{a}+\RW{f}$ and user $3$ transmit $\RW{d}+\RW{f}$, all the users are able to recover the whole packet set $\Rx$. For example, user $2$ receives $\RW{e}$, recovers $\RW{c}$ by subtracting message $\RW{b}+\RW{f}+\RW{c}$ by $\RW{b}+\RW{f}$ and recovers $\RW{d}$ by subtracting message $\RW{d}+\RW{f}$ by $\RW{f}$ so that he/she obtains all the packets in $\Rx$. The corresponding transmission rate vector in this coding scheme is $(2,1,1)$.\footnote{The coding scheme that implements an achievable rate vector $\rv_V$ is not necessarily unique.}

\end{example}

For a fractional rate vector $\rv_V$, let $k \in \ZP$ be the least common multiple (LCM) of all denominators of $r_i$s, i.e., $k$ is the minimum nonnegative integer such that $k\rv_V = (kr_i \colon i \in V)$ is integral. It means that $\rv_V$ can be implemented by $k$-packet-splitting, i.e., dividing each packet into $k$ chunks, in CCDE. For example, in Example~\ref{ex:main}, $k = 2$ is the LCM of the denominators of all dimensions in $\rv_V = (\frac{5}{2},\frac{1}{2},\frac{1}{2})$, which means $(\frac{5}{2},\frac{1}{2},\frac{1}{2})$ can be implemented by $2$-packet-splitting. Therefore, in CCDE, we are not only interested in the existence of an integral optimal rate vector in $\RRNCO(V)$ for the non-asymptotic setting, but are also concerned whether there exists a fractional optimal rate vector in $\RRACO(V)$ for the asymptotic setting and how large is the LCM $k$.

\section{Minimum sum-rate and Optimal Rate Vector} \label{sec:MinSumRateOptRate}

The fundamental problem in CO is how to obtain the value of the minimum sum-rate and an optimal rate vector: the value of $\RACO(V)$ and a rate vector in $\RRACO(V)$ for the asymptotic model and the value of $\RNCO(V)$ and a rate vector in $\RRNCO(V)$ for the non-asymptotic model. Although the minimum sum-rate problem can be formulated by LP~\eqref{eq:MinSumRateACO} and ILP~\eqref{eq:MinSumRateNCO} for asymptotic and non-asymptotic settings, respectively, it is not efficient to directly solve them since the number of the constraints grow exponentially in $|V|$.
In this section, we reveal the equivalence between the constant sum-rate achievable rate region and a base polyhedron. Directly based on the nonemptiness of this based polyhedron, we show that the minimum sum-rate in the asymptotic model is determined by an optimization problem over the partitions of the user set, which revisits the results in \cite{Csiszar2004,Chan2008tight}. We also show that the minimum sum-rate in the non-asymptotic model is the least integer that is no less than the one in the asymptotic model.

\subsection{Preliminaries}\label{sec:Prelim}

We first describe submodularity and the related concepts as follows. For a set function $f \colon 2^V \mapsto \Real$, the \textit{polyhedron and base polyhedron} of $f$ are respectively \cite[Section 2.3]{Fujishige2005} \cite[Definition 9.7.1]{Narayanan1997Book}
\begin{equation}
    \begin{aligned}
        & P(f,\leq) = \Set{\rv_V\in\Real^{|V|} \colon r(X) \leq f(X),\forall X \subseteq V},  \\
        & B(f,\leq) = \Set{\rv_V \in P(f,\leq) \colon r(V) = f(V)}.  \\
    \end{aligned}  \nonumber
\end{equation}
In the same way, we can define $P(f,\geq)$ and $B(f,\geq)$. A set function $f$ is \textit{submodular} if the submodular inequality
\begin{equation} \label{eq:SubMIneq}
    f(X) + f(Y) \geq f(X \cap Y) + f(X \cup Y)
\end{equation}
holds for all $X,Y \subseteq V$; $f$ is \textit{supermodular} if $-f$ is submodular; $f$ is \textit{modular} if it is both submodular and supermodular \cite[Section 2.3]{Fujishige2005}. $P(f,\leq)$ and $B(f,\leq)$ are submodular polyhedron and base polyhedron, respectively, if $f$ is submodular. A set function $f$ is \textit{intersecting submodular} if the submodular inequality \eqref{eq:SubMIneq} holds for all sets that are intersecting, i.e., all $X,Y \subseteq V$ such that $X \cap Y \neq \emptyset$\cite[Section 2.3]{Fujishige2005}. Note, an intersecting submodular function $f$ may or may not require \eqref{eq:SubMIneq} hold for all $X,Y \subseteq V$, which means a submodular function is also intersecting submodular, but not necessarily vice versa.

A set function $f$ is the \textit{rank function of a polymatroid} if it is (a) normalized: $f(\emptyset) = 0 $; (b) monotonic: $f(X) \geq f(Y)$ for all $X,Y \subseteq V$ such that $Y \subseteq X$; (c) submodular \cite[Section 2.2]{Fujishige2005}. If $f$ is a polymatroid rank function, the normality and monotonicity ensure the nonnegativity of $f$, i.e., $f(X) \geq 0, \forall X \subseteq V$, and $B(f,\leq) \subseteq \RealP^{|V|}$ \cite[Lemma 3.23]{Fujishige2005}. It is shown in \cite[Section 3]{FujishigePolyEntropy} that the entropy function $H$ is a polymatroid rank function. It is easy to see that $r$, the sum-rate function that is associated with a rate vector $\rv_V$, is modular and $f(X)-r(X)$ is submodular/intersecting submodular if $f$ is submodular/intersecting submodular.

\subsection{Nonemptiness of the Base Polyhedron}

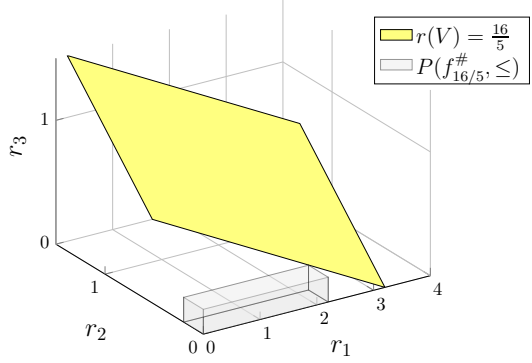
\begin{figure}[tbp]
	\centering
    \scalebox{0.7}{
%
%
%
\definecolor{mycolor1}{rgb}{1,1,0.5}%
\begin{tikzpicture}

\begin{axis}[%
width=2.8in,
height=2.5in,
area legend,
view={-33}{30},
scale only axis,
xmin=0,
xmax=4,
xlabel={\Large $r_1$},
xmajorgrids,
ymin=0,
ymax=1.5,
ylabel={\Large $r_2$},
ymajorgrids,
zmin=0,
zmax=1.5,
zlabel={\Large $r_3$},
zmajorgrids,
axis x line*=bottom,
axis y line*=left,
axis z line*=left,
legend style={at={(0.85,0.95)},anchor=north west,draw=black,fill=white,legend cell align=left}
]

\addplot3[solid,fill=mycolor1,draw=black]
table[row sep=crcr]{
x y z\\
3.2 0 0 \\
1.7 1.5 0 \\
0.2 1.5 1.5 \\
1.7 0 1.5 \\
3.2 0 0 \\
};
\addlegendentry{\large $r(V)=\frac{16}{5}$};

\addplot3[solid,fill=white!90!black,opacity=4.000000e-01,draw=black]
table[row sep=crcr]{
x y z\\
0 0 0 \\
2.2 0 0 \\
2.2 0.2 0 \\
0 0.2 0 \\
0 0 0 \\
};
\addlegendentry{\large $P(\FuHash{16/5},\leq)$};

\addplot3[solid,fill=white!90!black,opacity=4.000000e-01,draw=black,forget plot]
table[row sep=crcr]{
x y z\\
0 0 0 \\
0 0.2 0 \\
0 0.2 0.2 \\
0 0 0.2 \\
0 0 0 \\
};

\addplot3[solid,fill=white!90!black,opacity=4.000000e-01,draw=black,forget plot]
table[row sep=crcr]{
x y z\\
0 0 0 \\
2.2 0 0 \\
2.2 0 0.2 \\
0 0 0.2 \\
0 0 0 \\
};

\addplot3[solid,fill=white!90!black,opacity=4.000000e-01,draw=black,forget plot]
table[row sep=crcr]{
x y z\\
2.2 0 0 \\
2.2 0.2 0 \\
2.2 0.2 0.2 \\
2.2 0 0.2 \\
2.2 0 0 \\
};

\addplot3[solid,fill=white!90!black,opacity=4.000000e-01,draw=black,forget plot]
table[row sep=crcr]{
x y z\\
0 0.2 0 \\
0 0.2 0.2 \\
2.2 0.2 0.2 \\
2.2 0.2 0 \\
0 0.2 0 \\
};

\end{axis}
\end{tikzpicture}%
}
	\caption{For the system in Example~\ref{ex:main}, when $\alpha = \frac{16}{5}$, the polyhedron $P(\FuHash{16/5},\leq)$ does not intersect with the plane $\Set{\rv_V \in \Real^{3} \colon r(V) = \frac{16}{5}}$. Therefore, $B(\FuHash{16/5},\leq) = \Set{\rv_V \in \RRCO(V) \colon r(V) = \frac{16}{5}} = \emptyset$. It means that sum-rate $\frac{16}{5}$ is not achievable or there does not exist an achievable rate vector that has sum-rate equal to $\frac{16}{5}$.}
	\label{fig:DemoVuHash32}
\end{figure}

For $\alpha \in \RealP$, we define
$$ \Fu{\alpha}(X)=\begin{cases} H(X|V\setminus X) & X \subsetneq V \\ \alpha & X=V \end{cases}.$$
The base polyhedron of $\Fu{\alpha}$
\begin{equation}
    \begin{aligned}
        B(\Fu{\alpha},\geq) & = \Set{\rv_V \in P(\Fu{\alpha},\geq) \colon r(V)=\Fu{\alpha}(V) }  \\
                            & = \Set{\rv_V \in \RRCO(V) \colon r(V)=\alpha }
    \end{aligned}  \nonumber
\end{equation}
contains all achievable rate vectors that have sum-rate equal to $\alpha$. It is possible that $B(\Fu{\alpha},\geq) = \emptyset$, which means that the sum-rate $\alpha$ is not achievable.
Let $\FuHash{\alpha}$ be the \textit{dual set function} of $\Fu{\alpha}$ that is defined as \cite[Section 2.3]{Fujishige2005}
\begin{equation}
    \begin{aligned}
        \FuHash{\alpha}(X)  & = \Fu{\alpha}(V)-\Fu{\alpha}(V \setminus X)  \\
                            & = \alpha-\Fu{\alpha}(V \setminus X), \qquad \forall X \subseteq V.
    \end{aligned}  \nonumber
\end{equation}
Consider the constraint $r(X) \geq \Fu{\alpha}(X)$ in $B(\Fu{\alpha},\geq)$. If we restrict the rate vector $\rv_V$ to satisfy $r(X) \geq \Fu{\alpha}(X)$ for some $X\subseteq V$ and $r(V)=\alpha$, then we necessarily put constraint
$$r(V \setminus X)=r(V)-r(X) \leq \alpha-\Fu{\alpha}(V \setminus X) = \FuHash{\alpha}(V \setminus X) $$
on set $V \setminus X$. By converting the constraints in $B(\Fu{\alpha},\geq)$ in the same way for all $X \subseteq V$, we get the base polyhedron
$$ B(\FuHash{\alpha},\leq)= \Set{\rv_V \in P(\FuHash{\alpha},\leq) \colon r(V)=\FuHash{\alpha}(V)= \alpha } $$
such that $ B(\FuHash{\alpha},\leq)=B(\Fu{\alpha},\geq)$ \cite[Lemma 2.4]{Fujishige2005}.\footnote{In \cite[Lemma 2.4]{Fujishige2005}, $ B(\FuHash{\alpha},\leq)=B(\Fu{\alpha},\geq)$ holds without the submodularity or intersecting submodularity of $\FuHash{\alpha}$. } Then, the set of all achievable rate vectors with sum-rate $\alpha$ is described by $B(\FuHash{\alpha},\leq)$. It is obvious that $B(\FuHash{\alpha},\leq) \neq \emptyset$ if the polyhedron $P(\FuHash{\alpha},\leq)$ intersects with the hyperplane $\Set{\rv_V \in \Real^{|V|} \colon r(V) = \alpha }$. Also, the optimal rate sets can be described by the base polyhedra
\begin{equation}
    \begin{aligned}
        \RRACO(V) & = B(\FuHash{\RACO(V)},\leq); \\
        \RRNCO(V) & = B(\FuHash{\RNCO(V)},\leq) \cap \Z^{|V|}.
    \end{aligned} \nonumber
\end{equation}
Therefore, the minimum sum-rate, either $\RACO(V)$ or $\RNCO(V)$, can be determined by studying the condition on $\alpha$ for the nonemptiness of $B(\FuHash{\alpha},\leq)$, and a rate vector in the optimal rate set, either $\RRACO(V)$ or $\RRNCO(V)$, can be determined by any algorithm that is able to search a base point $\rv_V$ in $B(\FuHash{\RACO(V)},\leq)$ or $B(\FuHash{\RNCO(V)},\leq) \cap \Z^{|V|}$.\footnote{A base (point) is a $|V|$-dimension vector $\rv_V$ in $B(f,\leq)$, where $f$ is a set function defined on the power set $2^V$.}

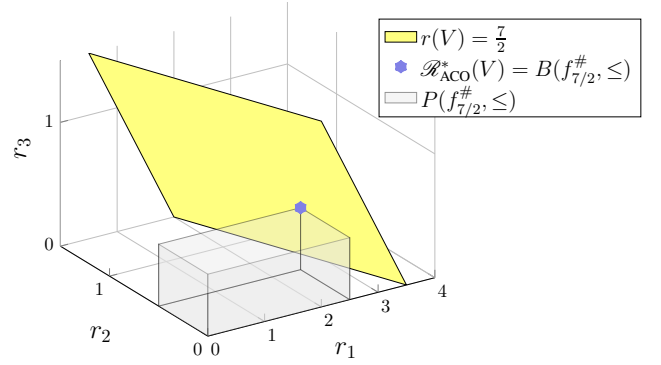
\begin{figure}[tbp]
	\centering
    \scalebox{0.7}{
%
%
%
\definecolor{mycolor1}{rgb}{1,1,0.5}%
\definecolor{mycolor2}{rgb}{0.5,0.5,0.9}%
\begin{tikzpicture}

\begin{axis}[%
width=2.8in,
height=2.5in,
view={-33}{30},
scale only axis,
xmin=0,
xmax=4,
xlabel={\Large $r_1$},
xmajorgrids,
ymin=0,
ymax=1.5,
ylabel={\Large $r_2$},
ymajorgrids,
zmin=0,
zmax=1.5,
zlabel={\Large $r_3$},
zmajorgrids,
axis x line*=bottom,
axis y line*=left,
axis z line*=left,
legend style={at={(0.85,0.95)},anchor=north west,draw=black,fill=white,legend cell align=left}
]

\addplot3[area legend,solid,fill=mycolor1,draw=black]
table[row sep=crcr]{
x y z\\
3.5 0 0 \\
2 1.5 0 \\
0.5 1.5 1.5 \\
2 0 1.5 \\
3.5 0 0 \\
};
\addlegendentry{\large $r(V)=\frac{7}{2}$};

\addplot3 [
color=mycolor2,
line width=6.0pt,
only marks,
mark=asterisk,
mark options={solid}]
table[row sep=crcr] {
2.5 0.5 0.5\\
};
\addlegendentry{\large $\RRACO(V) = B(\FuHash{7/2},\leq)$};

\addplot3[area legend,solid,fill=white!90!black,opacity=4.000000e-01,draw=black]
table[row sep=crcr]{
x y z\\
0 0 0 \\
2.5 0 0 \\
2.5 0.5 0 \\
0 0.5 0 \\
0 0 0 \\
};

\addlegendentry{\large $P(\FuHash{7/2},\leq)$};

\addplot3[solid,fill=white!90!black,opacity=4.000000e-01,draw=black,forget plot]
table[row sep=crcr]{
x y z\\
0 0 0 \\
0 0.5 0 \\
0 0.5 0.5 \\
0 0 0.5 \\
0 0 0 \\
};

\addplot3[solid,fill=white!90!black,opacity=4.000000e-01,draw=black,forget plot]
table[row sep=crcr]{
x y z\\
0 0 0 \\
2.5 0 0 \\
2.5 0 0.5 \\
0 0 0.5 \\
0 0 0 \\
};

\addplot3[solid,fill=white!90!black,opacity=4.000000e-01,draw=black,forget plot]
table[row sep=crcr]{
x y z\\
2.5 0 0 \\
2.5 0.5 0 \\
2.5 0.5 0.5 \\
2.5 0 0.5 \\
2.5 0 0 \\
};

\addplot3[solid,fill=white!90!black,opacity=4.000000e-01,draw=black,forget plot]
table[row sep=crcr]{
x y z\\
0 0.5 0 \\
0 0.5 0.5 \\
2.5 0.5 0.5 \\
2.5 0.5 0 \\
0 0.5 0 \\
};

\end{axis}
\end{tikzpicture}%
}
	\caption{For the system in Example~\ref{ex:main}, when $\alpha = \frac{7}{2}$, the the polyhedron $P(\FuHash{7/2},\leq)$ intersects with the plane $\Set{\rv_V \in \Real^{3} \colon r(V) = \frac{7}{2}}$ at $\rv_V = (\frac{5}{2},\frac{1}{2},\frac{1}{2})$, i.e., $B(\FuHash{7/2},\leq) = \Set{\rv_V \in \RRCO(V) \colon r(V) = \frac{7}{2}} = \Set{(\frac{5}{2},\frac{1}{2},\frac{1}{2})}$. In this case, sum--rate $\frac{7}{2}$ is achievable and there is only one optimal rate vector for the asymptotic model.}
	\label{fig:DemoVuHash35}
\end{figure}

\begin{figure}[tbp]
	\centering
    \scalebox{0.7}{
%
%
%
\definecolor{mycolor1}{rgb}{0.5,0.5,0.9}%
\definecolor{mycolor2}{rgb}{1,1,0.5}%
\begin{tikzpicture}

\begin{axis}[%
width=2.8in,
height=2.5in,
view={-33}{30},
scale only axis,
xmin=0,
xmax=4,
xlabel={\Large $r_1$},
xmajorgrids,
ymin=0,
ymax=1.5,
ylabel={\Large $r_2$},
ymajorgrids,
zmin=0,
zmax=1.5,
zlabel={\Large $r_3$},
zmajorgrids,
axis x line*=bottom,
axis y line*=left,
axis z line*=left,
legend style={at={(0.82,1.05)},anchor=north west,draw=black,fill=white,legend cell align=left}
]

\addplot3[area legend,solid,fill=mycolor2,draw=black]
table[row sep=crcr]{
x y z\\
4 0 0 \\
2.5 1.5 0 \\
1 1.5 1.5 \\
2.5 0 1.5 \\
4 0 0 \\
};
\addlegendentry{\large $r(V)=4$};

\addplot3[area legend,solid,fill=mycolor1,draw=black]
table[row sep=crcr]{
x y z\\
2 1 1 \\
3 0 1 \\
3 1 0 \\
2 1 1 \\
};
\addlegendentry{\large $B(\FuHash{4},\leq)$};

\addplot3[area legend,solid,fill=white!90!black,opacity=4.000000e-01,draw=black]
table[row sep=crcr]{
x y z\\
0 0 0 \\
3 0 0 \\
3 1 0 \\
0 1 0 \\
0 0 0 \\
};
\addlegendentry{\large $P(\FuHash{4},\leq)$};

\addplot3[solid,fill=white!90!black,opacity=4.000000e-01,draw=black,forget plot]
table[row sep=crcr]{
x y z\\
0 0 0 \\
0 1 0 \\
0 1 1 \\
0 0 1 \\
0 0 0 \\
};

\addplot3[solid,fill=white!90!black,opacity=4.000000e-01,draw=black,forget plot]
table[row sep=crcr]{
x y z\\
0 0 0 \\
3 0 0 \\
3 0 1 \\
0 0 1 \\
0 0 0 \\
};

\addplot3[solid,fill=white!90!black,opacity=4.000000e-01,draw=black,forget plot]
table[row sep=crcr]{
x y z\\
3 0 0 \\
3 1 0 \\
3 0 1 \\
3 0 0 \\
};

\addplot3[solid,fill=white!90!black,opacity=4.000000e-01,draw=black,forget plot]
table[row sep=crcr]{
x y z\\
0 1 0 \\
0 1 1 \\
2 1 1 \\
3 1 0 \\
0 1 0 \\
};

\addplot3[solid,fill=white!90!black,opacity=5.000000e-01,draw=black,forget plot]
table[row sep=crcr]{
x y z\\
0 0 1 \\
3 0 1 \\
2 1 1 \\
0 1 1 \\
0 0 1 \\
};

\addplot3 [
color=red,
line width=3.0pt,
only marks,
mark=triangle,
mark options={solid,,rotate=180}]
table[row sep=crcr] {
2 1 1\\
3 0 1\\
3 1 0\\
2 1 1\\
};
\addlegendentry{\large $\RRNCO(V) = B(\FuHash{4},\leq)\cap\Z^3$};

\end{axis}
\end{tikzpicture}%
}
	\caption{For the system in Example~\ref{ex:main}, when $\alpha = 4$, the polyhedron $P(\FuHash{4},\leq)$ and the plane $\Set{\rv_V \in \Real^{3} \colon r(V) = 4}$ intersect, i.e., $B(\FuHash{4},\leq)= \Set{\rv_V \in \RRCO(V) \colon r(V) = 4}\neq \emptyset$. Also, $\RRNCO(V) = B(\FuHash{4},\leq)\cap\Z^{3}=\{(2,1,1),(3,0,1),(3,1,0)\}$. In this case, sum-rate $4$ is achievable and there are three optimal rate vectors for the non-asymptotic model.}
	\label{fig:DemoVuHash4}
\end{figure}
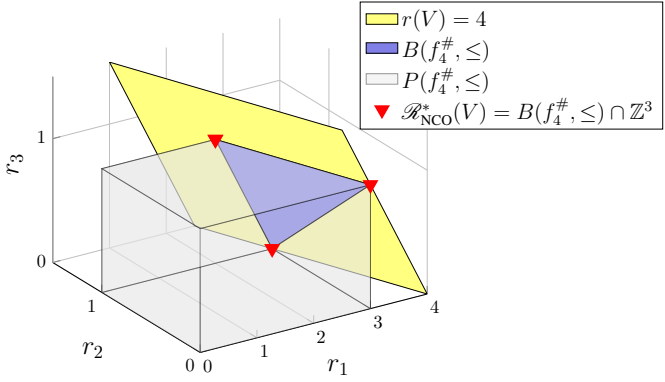

\begin{example}  \label{ex:mainBasePoly}
    For the system in Example~\ref{ex:main}, we have $B(\Fu{\alpha},\geq) = \Set{\rv \in \RRCO(V) \colon r(V) = \alpha }$. For a fixed value of $\alpha$, consider the constraint $r(\Set{1})\ge \Fu{\alpha}(\{1\})=1$ in $B(\Fu{\alpha},\geq)$. Since we restrict the sum-rate to be $r(\Set{1,2,3})=\alpha$, we have constraint $r(\Set{2,3})=\alpha-r(\Set{1})\leq \alpha-\Fu{\alpha}(\{1\})=\FuHash{\alpha}(\Set{2,3})=\alpha-1$. If we convert the constraints one by one in $B(\Fu{\alpha},\geq)$, we have the dual base polyhedron
    \begin{equation}
        \begin{aligned}
            B(\FuHash{\alpha},\leq)= \big\{ \rv_V\in\Real^{|V|} \colon & r(\emptyset)=0, r(\Set{1})\le \alpha-1,   \\
                                       &r(\Set{2})\le \alpha-3, r(\Set{3})\le \alpha-3,   \\
                                       &r(\Set{1,2})\le \alpha, r(\Set{1,3})\le \alpha,   \\
                                       &r(\Set{2,3})\le \alpha-1, \\
                                       &r(\Set{1,2,3})=\alpha   \big\}
        \end{aligned}  \nonumber
    \end{equation}
    such that $B(\FuHash{\alpha},\leq)=B(\Fu{\alpha},\geq)$. We increase the value of $\alpha$ from $0$. It can be shown that $B(\FuHash{\alpha},\leq)=\emptyset$ when $\alpha<\frac{7}{2}$, e.g., when $\alpha=\frac{16}{5}$ in Fig.~\ref{fig:DemoVuHash32}. When $\alpha=\frac{7}{2}$, we have $\RRACO(V) = B(\FuHash{7/2},\leq)=\Set{(\frac{5}{2},\frac{1}{2},\frac{1}{2})}$ as in Fig.~\ref{fig:DemoVuHash35}. We keep increasing $\alpha$ after reaching $\frac{7}{2}$. It can be shown that $B(\FuHash{\alpha},\leq) \cap \Z^{|V|} = \emptyset$ when $\alpha < 4$. When $\alpha=4$, we have $\RRNCO(V) = B(\FuHash{4},\leq) \cap \Z^{|V|} = \Set{(3,0,1),(2,1,1,),(3,1,0)}$ as shown in Fig.~\ref{fig:DemoVuHash4}.
\end{example}

\subsection{Minimum Sum-rate} \label{sec:MinSumRate}

The condition for $B(\FuHash{\alpha},\leq) \neq \emptyset$ can be easily derived based on the intersecting submodularity of $\FuHash{\alpha}$.

\begin{lemma}  \label{lemma:SubMIntSubM}
    For $\alpha \geq 0$, $\FuHash{\alpha}$ is intersecting submodular; If $\alpha \geq H(V)$, $\FuHash{\alpha}$ is submodular.
\end{lemma}
\begin{IEEEproof}
    For function $\FuHash{\alpha}$, we have
    \begin{equation}
        \begin{aligned}
            &\quad \FuHash{\alpha}(X) + \FuHash{\alpha}(Y) - \FuHash{\alpha}(X \cup Y) - \FuHash{\alpha}(X \cap Y)=  \\
            & \begin{cases}
                    H(X) + H(Y) \\
                    \quad - H(X \cup Y) - H(X \cap Y) + \alpha - H(V) & X \cap Y =\emptyset \\
                    H(X) + H(Y) - H(X \cup Y) - H(X \cap Y)                 & \text{otherwise}
              \end{cases}.
        \end{aligned} \nonumber
    \end{equation}
    Then, $\FuHash{\alpha}(X) + \FuHash{\alpha}(Y) \geq \FuHash{\alpha}(X \cup Y) + \FuHash{\alpha}(X \cap Y), \forall X,Y\subseteq V \colon X \cap Y \neq \emptyset$. Also, the inequality holds for all $X,Y \subseteq V$ when $\alpha \geq H(V)$. Lemma holds.
\end{IEEEproof}

For $X \subseteq V$, denote by $\Pi(X)$ the set that contains all partitions of $X$. A partition $\Pat$ of $X$ is the set that satisfies: (a) $C \neq \emptyset$ for all $C \in \Pat$; (b) $C \cap C' = \emptyset$ for any distinctive $C,C' \in \Pat$; (c) $\cup_{C \in \Pat} C=X$. Denote $\Pi'(X) = \Pi(X) \setminus \Set{X} = \Set{\Pat \in \Pi(X) \colon |\Pat|>1}$. For a partition $\Pat \in \Pi(X)$, let
$$ \FuHash{\alpha}[\Pat] = \sum_{C \in \Pat} \FuHash{\alpha}(C) $$
and $\FuHashHat{\alpha}$ be the \textit{Dilworth truncation} of $\FuHash{\alpha}$ that is defined as \cite{Dilworth1944}
\begin{equation} \label{eq:Dilworth}
    \FuHashHat{\alpha}(X) = \min_{\Pat \in \Pi(X)} \FuHash{\alpha}[\Pat], \quad \forall X \subseteq V.
\end{equation}
We have $\FuHashHat{\alpha}$ being a submodular function due to the intersecting submodularity of $\FuHash{\alpha}$ \cite[Theorems 2.5(i) and 2.6(i)]{Fujishige2005}. It is shown in \cite[Section 3]{Narayanan1991PLP} that, for a given value of $\alpha$, the minimal/finest and maximal/coarsest partitions that minimize $\min_{\Pat \in \Pi(X)} \FuHash{\alpha}[\Pat]$ exist.\footnote{In \cite[Section 3]{Narayanan1991PLP}, it is shown that the minimizers of $\min_{\Pat \in \Pi(X)} \FuHash{\alpha}[\Pat]$ form a partition lattice, which is called the Dilworth truncation lattice, where the minimal/finest and maximal/coarsest minimizers uniquely exist. }
We will show in the following context that a condition on the Dilworth truncation determines the nonemptiness of the base polyhedron $B(\FuHash{\alpha},\leq)$, based on which the value of the minimum sum-rate can be obtained by a maximization problem over the partition set $\Pi'(V)$. In Section~\ref{sec:MDA}, we will show that the minimum sum-rate problem can be solved in polynomial time by the efficient algorithms for solving the \textit{Dilworth truncation problem}, the minimization problem in \eqref{eq:Dilworth}.

\begin{theorem}[{\cite[Theorems 2.5(i) and 2.6(i)]{Fujishige2005}\footnote{Theorem~\ref{theo:NonEmptBase} refers to the case when $\alpha < H(V)$ in particular, since, when $\alpha \geq H(V)$, $\FuHash{\alpha}$ is submodular according to Lemma~\ref{lemma:SubMIntSubM} and $\alpha = \FuHash{\alpha}(V) = \FuHashHat{\alpha}(V) $ and $B(\FuHash{\alpha},\leq) \neq \emptyset$ for sure\cite[Theorem 2.3]{Fujishige2005}.}}] \label{theo:NonEmptBase}
    $B(\FuHash{\alpha},\leq)$ is nonempty, i.e., $\alpha$ is achievable, and $B(\FuHash{\alpha},\leq) = B(\FuHashHat{\alpha},\leq)$ if and only if
    \begin{equation} \label{eq:NonEmptBase}
        \alpha = \FuHashHat{\alpha}(V) .
    \end{equation}
\end{theorem}

In Theorem~\ref{theo:NonEmptBase}, $\FuHashHat{\alpha}(V) = \max\Set{r(V) \colon \rv_V \in P(\FuHash{\alpha},\leq)} $ determines the maximum sum-rate of all rate vectors in polyhedron $P(\FuHash{\alpha},\leq)$,\footnote{$\FuHashHat{\alpha}(V) = r(V),\forall \rv_V \in B(\FuHashHat{\alpha},\leq)$ \cite[Theorems 2.5(i) and 2.6(i)]{Fujishige2005} and, for each $\rv_V \in B(\FuHashHat{\alpha},\leq)$, we have $r(V) = \max\Set{r(V) \colon \rv_V \in P(\FuHashHat{\alpha},\leq) = P(\FuHash{\alpha},\leq)}$ \cite[Theorems 2.3 and 2.5(i)]{Fujishige2005}. A detailed explanation can also be found in Appendix~\ref{app:CoordSatCap}.}
while $\alpha$ is the sum-rate for all rate vectors in the hyperplane $\Set{\rv_V \in \Real^{|V|} \colon r(V) = \alpha}$.
There are two situations: if $\alpha > \FuHashHat{\alpha}(V)$, $P(\FuHash{\alpha},\leq)$ does not intersect with the hyperplane $\Set{\rv_V \in \Real^{|V|} \colon r(V) = \alpha}$; if $\alpha = \FuHashHat{\alpha}(V)$, $P(\FuHash{\alpha},\leq)$ intersects with the hyperplane $\Set{\rv_V \in \Real^{|V|} \colon r(V) = \alpha}$ at $B(\FuHash{\alpha},\leq)$. In the latter case, $B(\FuHash{\alpha},\leq) \neq \emptyset$. Theorem~\ref{theo:NonEmptBase} can also be interpreted by the principal sequence of partitions (PSP) in Appendix~\ref{app:PSP}.


\begin{example}
    For the system in Example~\ref{ex:main}, it can be shown that: when $\alpha < \frac{7}{2}$, we have $\alpha>\FuHashHat{\alpha}(V)$; when $\alpha \geq \frac{7}{2}$, we have $\alpha=\FuHashHat{\alpha}(V)$.\footnote{The two situations can be seen from the $\FuHashHat{\alpha}(V)$ vs. $\alpha$ plot in Fig.~\ref{fig:PSPmain}.} For example, in Fig.~\ref{fig:DemoVuHash32} when $\alpha=\frac{16}{5}$, one can show that $\max\Set{r(V) \colon \rv_V \in P(\FuHash{16/5},\leq)} = \FuHashHat{16/5}(V) = \frac{13}{5} < \alpha$. So, $P(\FuHash{16/5},\leq)$ does not intersect with hyperplane $\Set{\rv_V \in \Real^{|V|} \colon r(V) = \frac{16}{5}}$, i.e., $B(\FuHash{\alpha},\leq) = \emptyset$.

    In Fig.~\ref{fig:DemoVuHash4}, when $\alpha=4$, we have $\FuHash{4}$ being
    \begin{equation}
        \begin{aligned}
            &\FuHash{4}(\emptyset)=0,\FuHash{4}(\Set{1})=3,\FuHash{4}(\Set{2})=1,\FuHash{4}(\Set{3})=1, \\
            &\FuHash{4}(\Set{1,2})=4,\FuHash{4}(\Set{1,3})=4,\FuHash{4}(\Set{2,3})=3,   \\
            &\FuHash{4}(\Set{1,2,3})=4
        \end{aligned}  \nonumber
    \end{equation}
    and the Dilworth truncation $\FuHashHat{4}$ being
    \begin{equation}
        \begin{aligned}
            &\FuHashHat{4}(\emptyset)=0,\FuHashHat{4}(\{1\})=3,\FuHashHat{4}(\{2\})=1,\FuHashHat{4}(\{3\})=1,  \\
            &\FuHashHat{4}(\{1,2\})=4,\FuHashHat{4}(\{1,3\})=4,\FuHashHat{4}(\{2,3\})=2,   \\
            &\FuHashHat{4}(\{1,2,3\})=4.
        \end{aligned} \nonumber
    \end{equation}
    One can show that $\max\Set{r(V) \colon \rv_V \in P(\FuHash{4},\leq)} = \FuHashHat{4}(V) = 4 = \alpha$ and $B(\FuHash{4},\leq) = B(\FuHashHat{4},\leq) \neq \emptyset$.

    By comparing the values of $\FuHash{4}$ and $\FuHashHat{4}$, we can see that the Dilworth truncation tightens the constraints in the polyhedron $P(\FuHash{4},\leq)$. For example, the inequality $r(\Set{2,3}) \leq \FuHash{4}(\Set{2,3}) = 3$ in $P(\FuHash{4},\leq)$ can be tightened by $r(\Set{2}) \leq \FuHash{4}(\Set{2}) = 1$ and $r(\Set{3}) \leq \FuHash{4}(\Set{3}) = 1$ so that we have $r(\Set{2,3}) \leq \FuHashHat{4}(\Set{2,3}) = 2$ in $P(\FuHashHat{4},\leq)$. It also explains that $\FuHashHat{\alpha}(V)$ determines the maximum sum-rate over all rate vectors in the polyhedron $P(\FuHash{\alpha},\leq)$.
\end{example}


\begin{corollary} \label{coro:MinSumRate}
    For $\Pat \in \Pi'(V)$, define $$  \varphi(\Pat) = \sum_{C \in \Pat} \frac{H(V) - H(C)}{|\Pat|-1}.$$
    The minimum sum-rate in the asymptotic and non-asymptotic models are respectively
    \begin{subequations}
        \begin{align}
            \RACO(V) & = \max_{\Pat \in \Pi'(V)} \varphi(\Pat), \label{eq:MinSumRateACOPat} \\
            \RNCO(V) & = \big\lceil \max_{\Pat \in \Pi'(V)} \varphi(\Pat) \big\rceil. \label{eq:MinSumRateNCOPat}
        \end{align}
    \end{subequations}
\end{corollary}
\begin{IEEEproof}
    Equation \eqref{eq:NonEmptBase} in Theorem~\ref{theo:NonEmptBase} is equivalent to $\alpha \leq \FuHash{\alpha}[\Pat], \forall \Pat \in \Pi(V)$, which can be converted to $ \alpha \geq \varphi(\Pat), \forall \Pat \in \Pi'(V)$. It gives rise to the expressions of $\RACO(V)$ and $\RNCO(V)$ in \eqref{eq:MinSumRateACOPat} and \eqref{eq:MinSumRateNCOPat}, respectively.
\end{IEEEproof}

\begin{remark} \label{rem:MinSumRate}
    It was first shown that $\RACO(V)$ is lower bounded by $\RACO(V) \geq \max_{\Pat \in \Pi'(V)} \varphi(\Pat)$ in \cite[Example 4]{Csiszar2004} (see also Section~\ref{sec:SecretCap}). In \cite{Chan2008tight,ChanMMI}, the authors proved the tightness of this lower bound, where the same result as \eqref{eq:MinSumRateACOPat} in Corollary~\ref{coro:MinSumRate} for the asymptotic model is derived. However, the equality in \eqref{eq:MinSumRateACOPat} in \cite{Chan2008tight,ChanMMI} was proved in a different way: Instead of showing the equivalence $ B(\FuHash{\alpha},\leq) = \Set{\rv_V \in \RRCO(V) \colon r(V)=\alpha } $ and applying \cite[Theorems 2.5(i) and 2.6(i)]{Fujishige2005} to prove the nonemptiness of $B(\FuHash{\alpha},\leq)$, the authors defined a polyhedron $P'(g'_{\alpha},\leq) = \Set{ \rv_V \in \Real^{|V|} \colon r(X) \leq g'_{\alpha}(X), \forall X \subseteq V, X \neq \emptyset }$ for the submodular function $g'_{\alpha}(X) = \alpha - H(V) + H(X), \forall X \subseteq V$ and applied \cite[Theorem 48.3]{Schrijver2003} to show \eqref{eq:MinSumRateACOPat}.
    As compared to the proof in \cite[Section IV-B]{ChanMMI}, our work in this section shows that \eqref{eq:MinSumRateACOPat} straightforwardly follows from \cite[Theorems 2.5(i) and 2.6(i)]{Fujishige2005},\footnote{It is clear that $P(\FuHash{\alpha},\leq) = P'(g'_{\alpha},\leq)$ since we always have $r(\emptyset) = 0$. In this sense, the proof in \cite[Section IV-B]{ChanMMI} is essentially the same as the proof of \eqref{eq:MinSumRateACOPat} in this paper. However, the proof of Corollary~\eqref{coro:MinSumRate} is much simpler than \cite[Section IV-B]{ChanMMI}. }
    which also leads to our new derivation of $\RNCO(V)$ in \eqref{eq:MinSumRateNCOPat}. In Section~\ref{sec:MinSumRateExtend}, we show that \eqref{eq:MinSumRateNCOPat} verifies an observation in \cite[Section III-E]{CourtIT2014} that the difference between $\RACO(V)$ and $\RNCO(V)$ is bounded by one.
\end{remark}

Corollary~\ref{coro:MinSumRate} can be interpreted as follows. The minimum sum-rate can be determined by a maximization over all multi-way cuts of $V$. Any partition $\Pat \in \Pi'(V)$ can be considered as a multi-way cut of the user set $V$. For any $C \in \Pat$, the cut $\Set{C, V \setminus C}$ imposes the SW constraint $r(V \setminus C) \geq H(V \setminus C |C) = H(V) - H(C)$. By applying this to each $C \in \Pat$, we have $\sum_{C \in \Pat}r(V \setminus C) = (|\Pat| - 1) r(V) \geq \sum_{C \in \Pat} \big( H(V) - H(C) \big)$, which imposes requirement or lower bound
$$r(V) \geq \sum_{C \in \Pat} \frac{H(V) - H(C)}{|\Pat| - 1} = \varphi(\Pat)$$ on the sum-rate for attaining omniscience. Here, $|\Pat|-1$ is a normalization factor. Since the SW constraint applies to all the subsets of $V$, an achievable sum-rate must satisfy the highest requirement imposed by $\varphi(\Pat)$ over all multi-way cuts, i.e., $\varphi(\Pat)$ should be maximized over all $\Pat \in \Pi'(V)$. Therefore, we have \eqref{eq:MinSumRateACOPat} and \eqref{eq:MinSumRateNCOPat}. We call the mininal/finest maximizer of \eqref{eq:MinSumRateACOPat} the \textit{fundamental partition} and denote it by $\Pat^*$.

\begin{example}
    For the system in Example~\ref{ex:main}, by applying \eqref{eq:MinSumRateACOPat} and \eqref{eq:MinSumRateNCOPat}, we have $\RACO(V) = \frac{7}{2}$ and $\RNCO(V) = 4$, which are consistent with the results in Examples~\ref{ex:main} and \ref{ex:mainBasePoly}. In addition, we have $\Pat^* = \Set{\Set{1},\Set{2},\Set{3}}$ being the fundamental partition.
\end{example}

\subsection{Related Results} \label{sec:MinSumRateExtend}

Based on Corollary~\ref{coro:MinSumRate}, $\RNCO(V) = \lceil \RACO(V) \rceil $ so that $\RNCO(V) \geq \RACO(V)$, i.e., the minimum sum-rate in the non-asymptotic model is no less than the one in the asymptotic model, which is consistent with the results in \cite{Taj2011,CourtIT2014}.

Based on \eqref{eq:MinSumRateACOPat} and \eqref{eq:MinSumRateNCOPat} in Corollary~\ref{coro:MinSumRate}, it is straightforward that $\RACO(V) + 1 > \RNCO(V)$. There is an observation in \cite[Section III-E]{CourtIT2014} that the maximum difference between $\RACO(V)$ and $\RNCO(V)$ is one transmission. Whereas the result $\RNCO(V) - \RACO(V) < 1$ based on Corollary~\ref{coro:MinSumRate} provides theoretical proof to this observation. It states that the maximum difference between $\RACO(V)$ and $\RNCO(V)$ is strictly less than one. Besides, we also have the following results.

\begin{theorem} \label{theo:PolyMatMinSumRateSet}
    For all $\alpha \geq \RACO(V)$, $\FuHashHat{\alpha}$ is a polymatroid rank function and $B(\FuHashHat{\alpha},\leq) = B(\FuHash{\alpha},\leq) = \Set{\rv_V \in \RRCO(V) \colon r(V) = \alpha} \neq \emptyset$.
\end{theorem}
\begin{IEEEproof}
    According to the definition of function $\FuHash{\alpha}$ and the Dilworth truncation $\FuHashHat{\alpha}$, for all $\alpha \in \Real_+$, we have $\FuHashHat{\alpha}$ being normalized, i.e., $\FuHashHat{\alpha}(\emptyset) = 0$, and submodular\cite[Theorem 2.5(i)]{Fujishige2005}. The remaining task is to prove the monotonicity of $\FuHashHat{\alpha}$ when $\alpha \geq \RACO(V)$. According to \eqref{eq:MinSumRateACOPat}, we have $\alpha \geq \RACO(V) \geq \varphi(\Set{\Set{i}, V \setminus \Set{i}}) = 2H(V) - H(\Set{i} - H( V \setminus \Set{i})$ for all $i \in V$. Then, when $\alpha \geq \RACO(V)$,
    \begin{equation}
        \begin{aligned}
         \FuHashHat{\alpha}(\Set{i}) &= \FuHash{\alpha} (\Set{i}) \\
                                     &= \alpha -  H(V) + H(\Set{i}) \\
                                     &\geq H(V) - H( V \setminus \Set{i}) \geq 0,
         \end{aligned}\nonumber
    \end{equation}
    where the last inequality is due to the monotonicity of the entropy function $H$. So, for all $\alpha \geq \RACO(V)$ and $i \in V$, we have $\FuHashHat{\alpha}(\Set{i}) \geq \FuHashHat{\alpha}(\emptyset)$ and $\FuHashHat{\alpha}(X) \geq \FuHashHat{\alpha}(Y)$ for all $X,Y \subseteq V$ such that $i \in Y \subseteq X$, i.e., $\FuHashHat{\alpha}$ is monotonic. According to Theorem~\ref{theo:NonEmptBase}, when $\alpha \geq \RACO(V)$, we have $\alpha = \FuHashHat{\alpha}(V)$ and $B(\FuHashHat{\alpha},\leq) = B(\FuHash{\alpha},\leq) \neq \emptyset$.
\end{IEEEproof}


Theorem~\ref{theo:PolyMatMinSumRateSet} is important in proving the existence of a fractional and an integral rate vector in $\RRACO(V)$ and $\RRNCO(V)$, respectively, in the finite linear source model and CCDE.

\begin{corollary} \label{coro:IntNCOFracACO}
    In a finite linear source model,
    \begin{enumerate}[(a)]
        \item there exists an integral optimal rate vector in $\RRNCO(V)$.\footnote{In fact, in a finite linear source model, there exists an integral rate vector in $B(\FuHashHat{\alpha},\leq) = \Set{\rv_V \in \RRCO(V) \cap \Z^{|V|} \colon r(V) = \alpha}$ for all $\alpha \in \ZP$ such that $\alpha \geq \RNCO(V)$, which can be proved in the same way as Corollary~\ref{coro:IntNCOFracACO}(a). Corollary~\ref{coro:IntNCOFracACO}(a) is consistent with the results derived in \cite{MiloIT2016,CourtIT2014}.}
        \item there exists a fractional optimal rate vector in $\RRACO(V)$ that can be implemented by $(|\Pat^*| - 1)$-packet-splitting in CCDE.
    \end{enumerate}
\end{corollary}
\begin{IEEEproof}
    Recall that in a finite linear source model, the entropy function $H$ is integer-valued. Then, since $\FuHash{\RNCO(V)}$ is integer-valued, based on Theorem~\ref{theo:PolyMatMinSumRateSet}, $\FuHashHat{\RNCO(V)}$ is an integer-valued polymatroid rank function. Denote by $\EX(\FuHashHat{\RNCO(V)})$ the set of all extreme points, or vertices, in $B(\FuHashHat{\RNCO(V)},\leq)$. According to \cite[Theorem 3.22]{Fujishige2005}, all $\rv_V \in \EX(\FuHashHat{\RNCO(V)})$ are integral and belong to $B(\FuHashHat{\RNCO(V)},\leq) \cap \Z^{|V|} = \RRNCO(V)$.

    On the other hand, in a finite linear source model, we have $\RACO(V)=\sum_{C \in \Pat^*} \frac{H(V) - H(C)}{|\Pat^*|-1}$ according to Corollary~\ref{coro:MinSumRate}, i.e., $\RACO(V)$ is a fractional number with denominator $|\Pat^*| - 1$. So, $\RACO(V)(|\Pat^*| - 1)$ is integral. Then, $(|\Pat^*| - 1)\FuHash{\RACO(V)}$ is an integer-valued polymatroid rank function. According to \cite[Theorem 3.22]{Fujishige2005}, all $\rv_V \in \EX((|\Pat^*| - 1)\FuHashHat{\RACO(V)})$ are integral. For all $\rv_V \in \EX((|\Pat^*| - 1)\FuHashHat{\RACO(V)})$, $\frac{1}{|\Pat^*|-1}\rv_V \in \EX(\FuHashHat{\RACO(V)}) \subseteq B(\FuHashHat{\RACO(V)},\leq) = \RRACO(V)$ is a fractional optimal rate vector with $|\Pat^*| - 1$ being the LCM of the denominators of all dimensions. Corollary holds.
\end{IEEEproof}

\begin{example}  \label{ex:mainMinSumRateExpression}
    Consider the system in Example~\ref{ex:main}. We have $\RACO(V) = \frac{7}{2}$, $\RNCO(V) = 4$ and $\Pat^*=\Set{\Set{1},\Set{2},\Set{3}}$. From Fig.~\ref{fig:DemoVuHash4}, it can be seen that the set of extreme points in $B(\FuHashHat{\RNCO(V)},\leq)$ is
    $$  \EX(\FuHashHat{\RNCO(V)}) = \Big\{ (2,1,1), (3,1,0), (3,0,1) \Big\}. $$
    All $\rv_V \in \EX(\FuHashHat{\RNCO(V)})$ are the integral optimal rate vectors in $\RRNCO(V)$. From Fig.~\ref{fig:DemoVuHash35}, it can be seen that
    $$ \EX(\FuHashHat{\RACO(V)}) = \Big\{ (\frac{5}{2},\frac{1}{2},\frac{1}{2}) \Big\} = B(\FuHashHat{\RACO(V)},\leq). $$
    $(\frac{5}{2},\frac{1}{2},\frac{1}{2})$ can be implemented by $(|\Pat^*|-1)$-packet-splitting.
\end{example}

\begin{example} \label{ex:aux}
Consider a different system compared to previous examples where $V=\{1,\dotsc,5\}$ and each user observes respectively
    \begin{equation}
        \begin{aligned}
            \RZ{1} & = (\RW{b},\RW{c},\RW{d},\RW{h},\RW{i}),   \\
            \RZ{2} & = (\RW{e},\RW{f},\RW{h},\RW{i}),   \\
            \RZ{3} & = (\RW{b},\RW{c},\RW{e},\RW{j}), \\
            \RZ{4} & = (\RW{a},\RW{b},\RW{c},\RW{d},\RW{f},\RW{g},\RW{i},\RW{j}),  \\
            \RZ{5} & = (\RW{a},\RW{b},\RW{c},\RW{f},\RW{i},\RW{j}),
        \end{aligned}  \nonumber
    \end{equation}
where $\RW{j}$ is an independent uniformly distributed random bit. In this system, we have $\RACO(V)= \frac{13}{2}$, $\RNCO(V) = 7$ and $\Pat^*=\Set{\Set{1,4,5},\Set{2},\Set{3}}$. One can show that all rate vectors $\rv_V \in \EX(\FuHashHat{\RNCO(V)}) \subsetneq B(\FuHashHat{\RNCO(V)},\leq) \cap \Z^{|V|} = \RRNCO(V)$ are integral, e.g., $\rv_V = (0,1,1,5,0)$. For $\RRACO(V) = B(\FuHashHat{\RACO(V)},\leq)$, the extreme point set is
\begin{equation}
    \begin{aligned}
        \EX(\FuHashHat{\RACO(V)}) = \Big\{ & (1,\frac{1}{2},\frac{1}{2},2,\frac{5}{2}), (2,\frac{1}{2},\frac{1}{2},1,\frac{5}{2}), \\
                                       & (1,\frac{1}{2},\frac{1}{2},\frac{9}{2},0), (\frac{3}{2},\frac{1}{2},\frac{1}{2},4,0), \\
                                       & (\frac{3}{2},\frac{1}{2},\frac{1}{2},1,3) \Big\}.
    \end{aligned} \nonumber
\end{equation}
We have $|\Pat^*| - 1 = 2$ and all rate vectors in $\EX(\FuHash{\RACO(V)})$ can be implemented by $2$-packet-splitting.
\end{example}

It is shown in \cite[Section III-D]{CourtIT2014} that $(|V| - 1)$-packet-splitting is sufficient to achieve the minimum sum-rate $\RACO(V)$ in a CCDE system with high probability.\footnote{The authors in \cite{CourtIT2014} study the more general CCDE system where each user can only communicate with a subset of $V$. In the CCDE problem considered in this paper, we assume that each user can communicate losslessly with all other users, which is a special case of the general CCDE system in \cite{CourtIT2014}. The authors in \cite{CourtIT2014} call the model studied in this paper a \textit{clique} system.}
However, Corollary~\ref{coro:IntNCOFracACO}(b) states that $(|\Pat^*| - 1)$-packet-splitting with $|\Pat^*| - 1 \leq |V| - 1$ is sufficient to achieve the minimum sum-rate $\RACO(V)$ in a CCDE system for sure.

%

\begin{remark} \label{rem:IntNCOFracACO}
    The proof of Corollary~\ref{coro:IntNCOFracACO} states that determining the integral and fractional optimal rate vectors in $\RRNCO(V)$ and $\RRACO(V)$, respectively, is equivalent to searching the extreme points in $B(\FuHashHat{\RNCO(V)},\leq)$ and $B(\FuHashHat{\RACO(V)},\leq)$. In Section~\ref{sec:NCO}, we show the extreme points can be determined by the MDA and SIA algorithms for the asymptotic and non-asymptotic models, respectively.
\end{remark}

\subsection{Secrecy Capacity and Mutual Dependence} \label{sec:SecretCap}

The CO problem was first formulated in \cite{Csiszar2004} based on the study on secret capacity $\C_S(A)$, the largest rate that the secret key can be generated by the users in $A$. It is assumed that the active users form a subset $A \subseteq V$ and the others in $V \setminus A$ are the helpers that assist the active users generate the secret key. The problem studied in this paper is the case when $A = V$, for which the following results are derived in \cite{Csiszar2004}. The duality relationship between $\C_S(V)$ and $\RACO(V)$ has been revealed in \cite[Theorem 1]{Csiszar2004}:
\begin{equation} \label{eq:Dual}
    \RACO(V) = H(V) - \C_S(V).
\end{equation}
Let
\begin{equation}
    I(V)=\min_{\Pat \in \Pi'(V)} \frac{D( P_{\RZ{V}} \| \prod_{C \in \Pat} P_{\RZ{C}} )}{|\Pat|-1}, \label{eq:MMI}
\end{equation}
where $D(\cdot \| \cdot)$ is the Kullback-Leibler divergence and we have $ D( P_{\RZ{V}} \| \prod_{C \in \Pat} P_{\RZ{C}} ) = \sum_{C \in \Pat} H(C) - H(V)$.
It is shown in \cite[Example 4]{Csiszar2004} that $\C_S(V)$ is upper bounded by
\begin{equation} \label{eq:CSUB}
    \C_S(V) \leq  I(V)
\end{equation}
and $ D( P_{\RZ{V}} \| \prod_{C \in \Pat} P_{\RZ{C}} )$ is interpreted as the \textit{mutual dependence} for partition $\Pat \in \Pi'(V)$. Then, the minimum sum-rate is necessarily lower bounded by $\RACO(V) \geq  H(V) - I(V) = \max_{\Pat \in \Pi'(V)} \varphi(\Pat)$. There is also a conjecture in \cite[Example 4]{Csiszar2004} that the upper bound in \eqref{eq:CSUB} is tight.

By realizing that $|\Pat| - 1$ is a normalization factor, the author in \cite{Chan2008tight} proposed $\frac{D( P_{\RZ{V}} \| \prod_{C \in \Pat} P_{\RZ{C}} )}{|\Pat|-1}$ to be the mutual dependence measure for $\Pat \in \Pi'(V)$. The tightness of the upper bound in \eqref{eq:CSUB} is shown in \cite[Theorem 1]{Chan2008tight}. In \cite{ChanMMI}, $I(V)$ is proposed as the \textit{multivariate mutual information (MMI)} measure in $\RZ{V}$ so that the duality relationship \eqref{eq:Dual} is given in terms of $I(V)$ in \cite{ChanSuccessive,ChanSuccessiveIT,ChanMMI} as
    \begin{equation}
            \RACO(V) = H(V)-I(V), \label{eq:DualRACOMMI}
    \end{equation}
Note, $I(V)$ is also called the \textit{shared information} in \cite{Prakash2016}. The interpretation of \eqref{eq:DualRACOMMI} is: the minimum sum-rate $\RACO(V)$ must be the amount of information that is not mutual to the users in $V$. The fundamental partition defined in \cite{ChanMMI,ChanSuccessive,ChanSuccessiveIT} refers to the finest/minimal minimizer of \eqref{eq:MMI}, the same as in this paper. The dual relationship between $\RACO(V)$ and $\C_S(V)$, or $I(V)$, makes it more significant to study the minimum sum-rate problem in CO: Determining the secret capacity $\C_S(V)$ or MMI/shared information $I(V)$ relies on the efficient algorithms for solving the minimum sum-rate problem and vice versa.

\section{Lower Bound on Minimum Sum-Rate}  \label{sec:LB}

The existing algorithms for solving the minimum sum-rate problem in \cite{CourtIT2014,MiloIT2016} for the finite linear source model start with an estimation of the minimum sum-rate. In this section, we propose lower bounds (LBs) on $\RACO(V)$ and $\RNCO(V)$ that can be obtained in $O(|V|)$ time. In Section~\ref{sec:algo}, we will show that the LBs on $\RACO(V)$ and $\RNCO(V)$ can be used as an initial guess to start the MDA and SIA algorithms for searching the exact value of $\RACO(V)$ and $\RNCO(V)$ in the asymptotic and non-asymptotic models, respectively.

\begin{proposition} \label{prop:LB}
    The minimum sum-rate is lower bounded by
    \begin{equation}
        \begin{aligned}
            \RACO(V) & \geq \max_{i \in V} \big\{ \varphi(\Set{\Set{i},V \setminus \Set{i}}), \varphi(\Set{\Set{m} \colon m \in V}) \big\}, \\
            \RNCO(V) & \geq \big\lceil \max_{i \in V} \big\{ \varphi(\Set{\Set{i},V \setminus \Set{i}}), \varphi(\Set{\Set{m} \colon m \in V}) \big\} \big\rceil.
        \end{aligned}  \nonumber
    \end{equation}
\end{proposition}
\begin{IEEEproof}
The LBs on $\RACO(V)$ and $\RNCO(V)$ are obtained by \eqref{eq:MinSumRateACOPat} and \eqref{eq:MinSumRateNCOPat}, respectively, by partitions $\Pat = \Set{\Set{i}, V \setminus \Set{i}}$ for all $i \in V$ and partition $\Pat = \Set{\Set{m} \colon m \in V}$.
\end{IEEEproof}

The LBs in Proposition~\ref{prop:LB} can be obtained in $O(|V|)$ time.
\begin{remark} \label{rem:LB3user}
If $|V| = 3$, the LBs in Proposition~\ref{prop:LB} are tight for both asymptotic and non-asymptotic models: If $|V| = 3$, $\Set{\Set{i}, V \setminus \Set{i}}$ for all $i \in V$ and $\Set{\Set{m} \colon m \in V}$ constitute all partitions in $\Pi'(V)$. The tightness of the lower bound on $\RACO(V)$ when $|V| = 3$ for the asymptotic model is consistent with the result in \cite[Example 3]{Csiszar2004}.
\end{remark}

\begin{figure}[tbp]
	\centering
    \scalebox{0.7}{
%
%
\begin{tikzpicture}

\begin{axis}[%
width=4in,
height=2.3in,
view={-37.5}{30},
scale only axis,
xmin=7,
xmax=30,
xlabel={\large $H(V)$},
xmajorgrids,
ymin=3,
ymax=15,
ylabel={\large $|V|$},
ymajorgrids,
zmin=0,
zmax=6,
zlabel={\large mean error},
zmajorgrids,
axis x line*=bottom,
axis y line*=left,
axis z line*=left
]

\addplot3[%
surf,
shader=faceted,
draw=black,
colormap/jet,
mesh/rows=24]
table[row sep=crcr,header=false] {
7 3 1.01\\
7 4 0.54\\
7 5 0.31\\
7 6 0.13\\
7 7 0.14\\
7 8 0.04\\
7 9 0.07\\
7 10 0.03\\
7 11 0.01\\
7 12 0\\
7 13 0\\
7 14 0\\
7 15 0\\
8 3 1.17\\
8 4 0.55\\
8 5 0.35\\
8 6 0.25\\
8 7 0.12\\
8 8 0.07\\
8 9 0.02\\
8 10 0.05\\
8 11 0.02\\
8 12 0.02\\
8 13 0.03\\
8 14 0.02\\
8 15 0.02\\
9 3 1.23\\
9 4 0.65\\
9 5 0.36\\
9 6 0.23\\
9 7 0.13\\
9 8 0.09\\
9 9 0.04\\
9 10 0.03\\
9 11 0.03\\
9 12 0.04\\
9 13 0.02\\
9 14 0.01\\
9 15 0.02\\
10 3 1.36\\
10 4 0.6\\
10 5 0.49\\
10 6 0.26\\
10 7 0.17\\
10 8 0.09\\
10 9 0.12\\
10 10 0.02\\
10 11 0.05\\
10 12 0.03\\
10 13 0.01\\
10 14 0.02\\
10 15 0.01\\
11 3 1.63\\
11 4 0.8\\
11 5 0.37\\
11 6 0.36\\
11 7 0.14\\
11 8 0.12\\
11 9 0.14\\
11 10 0.06\\
11 11 0.05\\
11 12 0.04\\
11 13 0.09\\
11 14 0.02\\
11 15 0.02\\
12 3 1.84\\
12 4 0.89\\
12 5 0.45\\
12 6 0.3\\
12 7 0.2\\
12 8 0.18\\
12 9 0.17\\
12 10 0.05\\
12 11 0.07\\
12 12 0.02\\
12 13 0.05\\
12 14 0.03\\
12 15 0.01\\
13 3 1.9\\
13 4 1.08\\
13 5 0.52\\
13 6 0.39\\
13 7 0.22\\
13 8 0.11\\
13 9 0.11\\
13 10 0.14\\
13 11 0.05\\
13 12 0.07\\
13 13 0.03\\
13 14 0.04\\
13 15 0.07\\
14 3 2.19\\
14 4 1.11\\
14 5 0.53\\
14 6 0.34\\
14 7 0.23\\
14 8 0.19\\
14 9 0.16\\
14 10 0.11\\
14 11 0.06\\
14 12 0.1\\
14 13 0.03\\
14 14 0.01\\
14 15 0.05\\
15 3 2.3\\
15 4 1.11\\
15 5 0.67\\
15 6 0.6\\
15 7 0.31\\
15 8 0.18\\
15 9 0.15\\
15 10 0.07\\
15 11 0.11\\
15 12 0.11\\
15 13 0.06\\
15 14 0.06\\
15 15 0.06\\
16 3 2.57\\
16 4 1.27\\
16 5 0.69\\
16 6 0.48\\
16 7 0.31\\
16 8 0.17\\
16 9 0.19\\
16 10 0.11\\
16 11 0.07\\
16 12 0.08\\
16 13 0.06\\
16 14 0.05\\
16 15 0.02\\
17 3 2.57\\
17 4 1.35\\
17 5 0.67\\
17 6 0.48\\
17 7 0.39\\
17 8 0.18\\
17 9 0.16\\
17 10 0.17\\
17 11 0.13\\
17 12 0.06\\
17 13 0.06\\
17 14 0.07\\
17 15 0.03\\
18 3 2.97\\
18 4 1.42\\
18 5 0.93\\
18 6 0.44\\
18 7 0.35\\
18 8 0.24\\
18 9 0.16\\
18 10 0.17\\
18 11 0.08\\
18 12 0.09\\
18 13 0.07\\
18 14 0.07\\
18 15 0.1\\
19 3 3.01\\
19 4 1.45\\
19 5 0.91\\
19 6 0.56\\
19 7 0.41\\
19 8 0.25\\
19 9 0.22\\
19 10 0.15\\
19 11 0.13\\
19 12 0.06\\
19 13 0.1\\
19 14 0.09\\
19 15 0.03\\
20 3 3.06\\
20 4 1.53\\
20 5 1.06\\
20 6 0.68\\
20 7 0.4\\
20 8 0.25\\
20 9 0.13\\
20 10 0.14\\
20 11 0.15\\
20 12 0.09\\
20 13 0.09\\
20 14 0.09\\
20 15 0.06\\
21 3 3.69\\
21 4 1.75\\
21 5 0.95\\
21 6 0.63\\
21 7 0.33\\
21 8 0.4\\
21 9 0.29\\
21 10 0.15\\
21 11 0.15\\
21 12 0.09\\
21 13 0.11\\
21 14 0.07\\
21 15 0.06\\
22 3 3.78\\
22 4 2\\
22 5 1.06\\
22 6 0.65\\
22 7 0.53\\
22 8 0.29\\
22 9 0.17\\
22 10 0.22\\
22 11 0.09\\
22 12 0.12\\
22 13 0.09\\
22 14 0.07\\
22 15 0.08\\
23 3 3.92\\
23 4 2.16\\
23 5 1.19\\
23 6 0.72\\
23 7 0.41\\
23 8 0.3\\
23 9 0.22\\
23 10 0.25\\
23 11 0.14\\
23 12 0.12\\
23 13 0.11\\
23 14 0.06\\
23 15 0.07\\
24 3 4.25\\
24 4 1.92\\
24 5 1.24\\
24 6 0.69\\
24 7 0.73\\
24 8 0.4\\
24 9 0.29\\
24 10 0.24\\
24 11 0.17\\
24 12 0.13\\
24 13 0.09\\
24 14 0.12\\
24 15 0.12\\
25 3 4.44\\
25 4 2.26\\
25 5 1.16\\
25 6 0.73\\
25 7 0.53\\
25 8 0.34\\
25 9 0.27\\
25 10 0.23\\
25 11 0.17\\
25 12 0.12\\
25 13 0.09\\
25 14 0.11\\
25 15 0.06\\
26 3 4.44\\
26 4 2.27\\
26 5 1.42\\
26 6 0.78\\
26 7 0.52\\
26 8 0.32\\
26 9 0.24\\
26 10 0.26\\
26 11 0.2\\
26 12 0.22\\
26 13 0.06\\
26 14 0.11\\
26 15 0.09\\
27 3 4.72\\
27 4 2.11\\
27 5 1.28\\
27 6 0.81\\
27 7 0.66\\
27 8 0.36\\
27 9 0.31\\
27 10 0.21\\
27 11 0.16\\
27 12 0.08\\
27 13 0.14\\
27 14 0.09\\
27 15 0.07\\
28 3 4.63\\
28 4 2.71\\
28 5 1.56\\
28 6 0.8\\
28 7 0.58\\
28 8 0.31\\
28 9 0.36\\
28 10 0.1\\
28 11 0.18\\
28 12 0.19\\
28 13 0.11\\
28 14 0.16\\
28 15 0.12\\
29 3 5.01\\
29 4 2.97\\
29 5 1.47\\
29 6 0.91\\
29 7 0.64\\
29 8 0.45\\
29 9 0.34\\
29 10 0.3\\
29 11 0.29\\
29 12 0.23\\
29 13 0.14\\
29 14 0.12\\
29 15 0.11\\
30 3 5.44\\
30 4 2.68\\
30 5 1.72\\
30 6 1\\
30 7 0.7\\
30 8 0.44\\
30 9 0.31\\
30 10 0.22\\
30 11 0.23\\
30 12 0.18\\
30 13 0.15\\
30 14 0.11\\
30 15 0.11\\
};
\end{axis}
\end{tikzpicture}%
}
	\caption{The average error incurred by \eqref{eq:LBOld1}, the lower bound on $\RNCO(V)$ that is proposed in \cite{Roua2010}, in Experiment~\ref{exp:LB}.}
	\label{fig:AveErrLBOld1}
\end{figure}
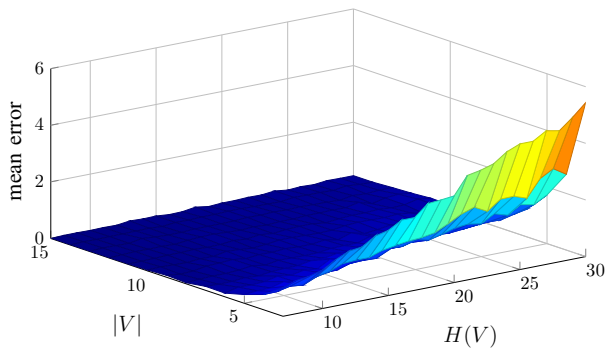

The LB on $\RNCO(V)$ has also been proposed in \cite{Roua2010,SprintRand2010} for the finite linear source model. In \cite{Roua2010}, it is shown that
\begin{equation} \label{eq:LBOld1}
    \RNCO(V) \geq H(V) - \min_{i \in V} H(\Set{i}).
\end{equation}
In addition, if $H(\Set{i})=b$ for all $i \in V$, $\RNCO(V) \geq H(V)-b+1$, which can be explained by the LB
\begin{equation} \label{eq:LBOld2}
    \begin{aligned}
        \RNCO(V) &\geq \Big\lceil \sum_{i \in V}\frac{ H(V) - H(\Set{i})}{|V|-1} \Big\rceil   \\
                 &= \lceil \varphi(\Set{\Set{i} \colon i \in V}) \rceil
    \end{aligned}
\end{equation}
proposed in \cite{SprintRand2010} in that $\lceil \varphi(\Set{\Set{i} \colon i \in V}) \rceil = \lceil \frac{|V|}{|V|-1}(H(V) - b) \rceil = H(V)-b+1$ in the case when $H(\Set{i})=b, \forall i \in V$. But,
\begin{equation}
    \begin{aligned}
        \max_{i \in V} \varphi(\Set{\Set{i},V \setminus \Set{i}}) &\geq \max_{i \in V} \big\{ H(V) - H(\Set{i}) \big\}  \\
                                                               & = H(V) -\min_{i \in V} H(\Set{i}).
    \end{aligned}\nonumber
\end{equation}
The LB on $\RNCO(V)$ in Proposition~\ref{prop:LB} is tighter than the ones in \cite{Roua2010,SprintRand2010}.

\begin{figure}[tbp]
	\centering
    \scalebox{0.7}{
%
%
\begin{tikzpicture}

\begin{axis}[%
width=4in,
height=2.3in,
view={-37.5}{30},
scale only axis,
xmin=7,
xmax=30,
xlabel={\large $H(V)$},
xmajorgrids,
ymin=3,
ymax=15,
ylabel={\large $|V|$},
ymajorgrids,
zmin=0,
zmax=4,
zlabel={\large mean error},
zmajorgrids,
axis x line*=bottom,
axis y line*=left,
axis z line*=left
]

\addplot3[%
surf,
shader=faceted,
draw=black,
colormap/jet,
mesh/rows=24]
table[row sep=crcr,header=false] {
7 3 0.22\\
7 4 0.3\\
7 5 0.5\\
7 6 0.57\\
7 7 0.69\\
7 8 0.79\\
7 9 1.01\\
7 10 0.98\\
7 11 1\\
7 12 1.07\\
7 13 1.22\\
7 14 1.24\\
7 15 1.28\\
8 3 0.23\\
8 4 0.4\\
8 5 0.49\\
8 6 0.74\\
8 7 0.7\\
8 8 0.88\\
8 9 0.87\\
8 10 1.07\\
8 11 1.19\\
8 12 1.16\\
8 13 1.21\\
8 14 1.38\\
8 15 1.36\\
9 3 0.22\\
9 4 0.37\\
9 5 0.54\\
9 6 0.75\\
9 7 0.9\\
9 8 0.89\\
9 9 1.01\\
9 10 1.13\\
9 11 1.22\\
9 12 1.27\\
9 13 1.41\\
9 14 1.43\\
9 15 1.44\\
10 3 0.24\\
10 4 0.3\\
10 5 0.62\\
10 6 0.84\\
10 7 0.79\\
10 8 0.92\\
10 9 1.19\\
10 10 1.17\\
10 11 1.43\\
10 12 1.34\\
10 13 1.4\\
10 14 1.48\\
10 15 1.49\\
11 3 0.22\\
11 4 0.51\\
11 5 0.73\\
11 6 0.81\\
11 7 0.88\\
11 8 1.03\\
11 9 1.12\\
11 10 1.29\\
11 11 1.4\\
11 12 1.46\\
11 13 1.5\\
11 14 1.55\\
11 15 1.88\\
12 3 0.27\\
12 4 0.35\\
12 5 0.67\\
12 6 0.87\\
12 7 0.91\\
12 8 1.19\\
12 9 1.3\\
12 10 1.18\\
12 11 1.44\\
12 12 1.52\\
12 13 1.61\\
12 14 1.6\\
12 15 1.69\\
13 3 0.33\\
13 4 0.46\\
13 5 0.52\\
13 6 0.62\\
13 7 0.87\\
13 8 1.22\\
13 9 1.32\\
13 10 1.33\\
13 11 1.38\\
13 12 1.52\\
13 13 1.62\\
13 14 1.76\\
13 15 1.65\\
14 3 0.33\\
14 4 0.46\\
14 5 0.73\\
14 6 0.87\\
14 7 0.94\\
14 8 1.3\\
14 9 1.28\\
14 10 1.29\\
14 11 1.66\\
14 12 1.71\\
14 13 1.77\\
14 14 1.74\\
14 15 1.97\\
15 3 0.21\\
15 4 0.5\\
15 5 0.59\\
15 6 0.91\\
15 7 1.18\\
15 8 1.17\\
15 9 1.29\\
15 10 1.51\\
15 11 1.62\\
15 12 1.66\\
15 13 1.76\\
15 14 1.9\\
15 15 1.95\\
16 3 0.25\\
16 4 0.45\\
16 5 0.67\\
16 6 1.05\\
16 7 1.04\\
16 8 1.21\\
16 9 1.4\\
16 10 1.71\\
16 11 1.68\\
16 12 1.74\\
16 13 1.75\\
16 14 1.89\\
16 15 2.04\\
17 3 0.18\\
17 4 0.55\\
17 5 0.62\\
17 6 1.07\\
17 7 1.1\\
17 8 1.32\\
17 9 1.61\\
17 10 1.73\\
17 11 1.57\\
17 12 1.75\\
17 13 1.92\\
17 14 2.04\\
17 15 2.11\\
18 3 0.24\\
18 4 0.38\\
18 5 0.69\\
18 6 0.78\\
18 7 0.99\\
18 8 1.25\\
18 9 1.52\\
18 10 1.61\\
18 11 1.65\\
18 12 1.94\\
18 13 1.87\\
18 14 2.09\\
18 15 2.07\\
19 3 0.27\\
19 4 0.51\\
19 5 0.56\\
19 6 1.11\\
19 7 0.97\\
19 8 1.34\\
19 9 1.42\\
19 10 1.59\\
19 11 1.55\\
19 12 1.8\\
19 13 1.99\\
19 14 2.15\\
19 15 2.12\\
20 3 0.12\\
20 4 0.37\\
20 5 0.81\\
20 6 0.93\\
20 7 1.24\\
20 8 1.26\\
20 9 1.62\\
20 10 1.78\\
20 11 1.69\\
20 12 2.02\\
20 13 2.03\\
20 14 2.07\\
20 15 2.24\\
21 3 0.15\\
21 4 0.49\\
21 5 0.67\\
21 6 0.9\\
21 7 1.19\\
21 8 1.32\\
21 9 1.42\\
21 10 1.54\\
21 11 1.88\\
21 12 2.08\\
21 13 2.08\\
21 14 2.14\\
21 15 2.11\\
22 3 0.15\\
22 4 0.29\\
22 5 0.54\\
22 6 0.99\\
22 7 1.09\\
22 8 1.21\\
22 9 1.42\\
22 10 1.77\\
22 11 2.05\\
22 12 2.01\\
22 13 2.05\\
22 14 2.15\\
22 15 2.24\\
23 3 0.06\\
23 4 0.42\\
23 5 0.75\\
23 6 0.89\\
23 7 1.08\\
23 8 1.43\\
23 9 1.48\\
23 10 1.82\\
23 11 1.83\\
23 12 2.04\\
23 13 2.29\\
23 14 2.2\\
23 15 2.41\\
24 3 0.21\\
24 4 0.38\\
24 5 0.7\\
24 6 1.03\\
24 7 1.27\\
24 8 1.5\\
24 9 1.69\\
24 10 1.86\\
24 11 1.83\\
24 12 2.18\\
24 13 2.11\\
24 14 2.37\\
24 15 2.37\\
25 3 0.24\\
25 4 0.4\\
25 5 0.68\\
25 6 1.11\\
25 7 1.34\\
25 8 1.32\\
25 9 1.59\\
25 10 1.86\\
25 11 1.95\\
25 12 2.18\\
25 13 2.09\\
25 14 2.35\\
25 15 2.29\\
26 3 0.09\\
26 4 0.48\\
26 5 0.53\\
26 6 1.08\\
26 7 1.08\\
26 8 1.31\\
26 9 1.63\\
26 10 1.93\\
26 11 2.04\\
26 12 2.35\\
26 13 2.45\\
26 14 2.31\\
26 15 2.59\\
27 3 0.21\\
27 4 0.34\\
27 5 0.63\\
27 6 0.91\\
27 7 1.24\\
27 8 1.54\\
27 9 1.74\\
27 10 1.91\\
27 11 2.37\\
27 12 2.01\\
27 13 2.28\\
27 14 2.21\\
27 15 2.53\\
28 3 0.23\\
28 4 0.31\\
28 5 0.65\\
28 6 0.94\\
28 7 1.37\\
28 8 1.5\\
28 9 1.86\\
28 10 1.7\\
28 11 1.99\\
28 12 2.3\\
28 13 2.31\\
28 14 2.27\\
28 15 2.73\\
29 3 0.16\\
29 4 0.33\\
29 5 0.64\\
29 6 0.79\\
29 7 1.25\\
29 8 1.55\\
29 9 1.79\\
29 10 2.02\\
29 11 1.89\\
29 12 2.26\\
29 13 2.5\\
29 14 2.61\\
29 15 2.43\\
30 3 0.14\\
30 4 0.36\\
30 5 0.54\\
30 6 0.89\\
30 7 1.09\\
30 8 1.25\\
30 9 1.7\\
30 10 1.97\\
30 11 1.91\\
30 12 2.41\\
30 13 2.31\\
30 14 2.52\\
30 15 2.68\\
};
\end{axis}
\end{tikzpicture}%
}
	\caption{The average error incurred by \eqref{eq:LBOld2}, the lower bound on $\RNCO(V)$ that is proposed in \cite{SprintRand2010}, in Experiment~\ref{exp:LB}.}
	\label{fig:AveErrLBOld2}
\end{figure}
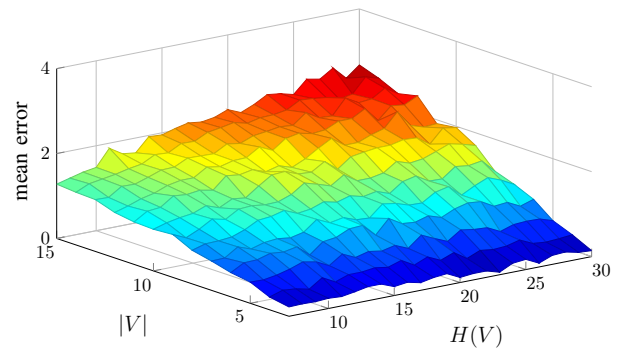

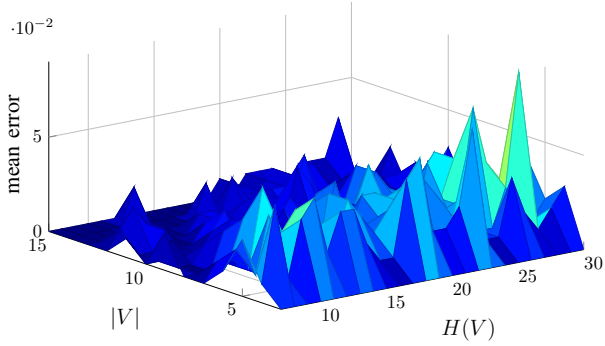
\begin{figure}[tbp]
	\centering
    \scalebox{0.7}{
%
%
\begin{tikzpicture}

\begin{axis}[%
width=4in,
height=2.3in,
view={-37.5}{30},
scale only axis,
xmin=7,
xmax=30,
xlabel={\large $H(V)$},
xmajorgrids,
ymin=3,
ymax=15,
ylabel={\large $|V|$},
ymajorgrids,
zmin=0,
zmax=0.09,
zlabel={\large mean error},
zmajorgrids,
axis x line*=bottom,
axis y line*=left,
axis z line*=left
]

\addplot3[%
surf,
shader=faceted,
draw=black,
colormap/jet,
mesh/rows=24]
table[row sep=crcr,header=false] {
7 3 0\\
7 4 0.01\\
7 5 0.03\\
7 6 0.01\\
7 7 0\\
7 8 0\\
7 9 0.02\\
7 10 0\\
7 11 0.01\\
7 12 0\\
7 13 0\\
7 14 0\\
7 15 0\\
8 3 0\\
8 4 0.05\\
8 5 0.01\\
8 6 0.02\\
8 7 0.01\\
8 8 0\\
8 9 0\\
8 10 0\\
8 11 0\\
8 12 0.01\\
8 13 0\\
8 14 0\\
8 15 0\\
9 3 0\\
9 4 0.01\\
9 5 0.04\\
9 6 0.01\\
9 7 0.01\\
9 8 0\\
9 9 0\\
9 10 0.01\\
9 11 0\\
9 12 0.02\\
9 13 0\\
9 14 0\\
9 15 0\\
10 3 0\\
10 4 0.04\\
10 5 0.03\\
10 6 0.05\\
10 7 0.03\\
10 8 0.02\\
10 9 0.01\\
10 10 0\\
10 11 0.01\\
10 12 0.01\\
10 13 0\\
10 14 0\\
10 15 0\\
11 3 0\\
11 4 0.05\\
11 5 0.04\\
11 6 0\\
11 7 0\\
11 8 0\\
11 9 0\\
11 10 0\\
11 11 0\\
11 12 0\\
11 13 0.01\\
11 14 0\\
11 15 0\\
12 3 0\\
12 4 0.02\\
12 5 0.02\\
12 6 0\\
12 7 0.02\\
12 8 0.02\\
12 9 0.02\\
12 10 0.01\\
12 11 0\\
12 12 0\\
12 13 0\\
12 14 0.02\\
12 15 0\\
13 3 0\\
13 4 0.04\\
13 5 0.03\\
13 6 0.01\\
13 7 0.01\\
13 8 0.03\\
13 9 0\\
13 10 0.02\\
13 11 0.01\\
13 12 0\\
13 13 0\\
13 14 0\\
13 15 0\\
14 3 0\\
14 4 0.04\\
14 5 0.02\\
14 6 0\\
14 7 0.02\\
14 8 0.01\\
14 9 0\\
14 10 0.01\\
14 11 0.01\\
14 12 0.01\\
14 13 0\\
14 14 0\\
14 15 0\\
15 3 0\\
15 4 0.02\\
15 5 0.04\\
15 6 0.04\\
15 7 0.03\\
15 8 0\\
15 9 0.01\\
15 10 0\\
15 11 0.01\\
15 12 0.01\\
15 13 0\\
15 14 0\\
15 15 0\\
16 3 0\\
16 4 0.01\\
16 5 0.03\\
16 6 0.02\\
16 7 0.01\\
16 8 0\\
16 9 0.02\\
16 10 0.01\\
16 11 0\\
16 12 0\\
16 13 0.02\\
16 14 0\\
16 15 0\\
17 3 0\\
17 4 0.01\\
17 5 0.02\\
17 6 0.01\\
17 7 0.04\\
17 8 0.03\\
17 9 0.01\\
17 10 0.01\\
17 11 0.02\\
17 12 0.01\\
17 13 0\\
17 14 0\\
17 15 0\\
18 3 0\\
18 4 0.05\\
18 5 0.03\\
18 6 0.02\\
18 7 0.01\\
18 8 0.02\\
18 9 0.01\\
18 10 0.02\\
18 11 0\\
18 12 0\\
18 13 0.02\\
18 14 0\\
18 15 0\\
19 3 0\\
19 4 0.06\\
19 5 0.02\\
19 6 0.02\\
19 7 0.03\\
19 8 0.01\\
19 9 0\\
19 10 0\\
19 11 0\\
19 12 0\\
19 13 0.01\\
19 14 0\\
19 15 0\\
20 3 0\\
20 4 0.02\\
20 5 0\\
20 6 0.03\\
20 7 0.03\\
20 8 0.02\\
20 9 0.02\\
20 10 0\\
20 11 0.03\\
20 12 0.01\\
20 13 0.01\\
20 14 0.01\\
20 15 0.01\\
21 3 0\\
21 4 0.02\\
21 5 0.01\\
21 6 0.04\\
21 7 0.02\\
21 8 0.04\\
21 9 0.01\\
21 10 0\\
21 11 0.02\\
21 12 0\\
21 13 0.01\\
21 14 0\\
21 15 0\\
22 3 0\\
22 4 0.03\\
22 5 0\\
22 6 0.04\\
22 7 0.03\\
22 8 0.02\\
22 9 0.01\\
22 10 0.01\\
22 11 0\\
22 12 0.03\\
22 13 0.01\\
22 14 0.01\\
22 15 0.01\\
23 3 0\\
23 4 0.07\\
23 5 0.02\\
23 6 0.03\\
23 7 0.01\\
23 8 0.02\\
23 9 0.03\\
23 10 0.03\\
23 11 0.01\\
23 12 0.01\\
23 13 0\\
23 14 0\\
23 15 0.01\\
24 3 0\\
24 4 0.01\\
24 5 0.03\\
24 6 0.02\\
24 7 0.04\\
24 8 0.02\\
24 9 0.02\\
24 10 0.01\\
24 11 0.01\\
24 12 0.01\\
24 13 0.02\\
24 14 0.01\\
24 15 0.01\\
25 3 0\\
25 4 0.01\\
25 5 0.01\\
25 6 0.02\\
25 7 0.02\\
25 8 0.03\\
25 9 0.01\\
25 10 0\\
25 11 0\\
25 12 0\\
25 13 0.01\\
25 14 0.01\\
25 15 0\\
26 3 0\\
26 4 0.04\\
26 5 0.01\\
26 6 0.07\\
26 7 0.04\\
26 8 0.04\\
26 9 0.01\\
26 10 0\\
26 11 0.01\\
26 12 0.01\\
26 13 0\\
26 14 0.01\\
26 15 0.01\\
27 3 0\\
27 4 0.02\\
27 5 0.03\\
27 6 0.04\\
27 7 0.03\\
27 8 0\\
27 9 0.03\\
27 10 0.01\\
27 11 0.01\\
27 12 0\\
27 13 0.01\\
27 14 0\\
27 15 0.01\\
28 3 0\\
28 4 0.01\\
28 5 0.09\\
28 6 0.03\\
28 7 0.02\\
28 8 0.01\\
28 9 0.02\\
28 10 0\\
28 11 0\\
28 12 0.01\\
28 13 0\\
28 14 0\\
28 15 0.01\\
29 3 0\\
29 4 0.02\\
29 5 0.01\\
29 6 0.02\\
29 7 0.01\\
29 8 0.02\\
29 9 0\\
29 10 0\\
29 11 0.02\\
29 12 0.01\\
29 13 0\\
29 14 0\\
29 15 0.03\\
30 3 0\\
30 4 0.03\\
30 5 0.02\\
30 6 0.04\\
30 7 0.03\\
30 8 0\\
30 9 0.01\\
30 10 0.03\\
30 11 0.01\\
30 12 0\\
30 13 0.02\\
30 14 0\\
30 15 0\\
};
\end{axis}
\end{tikzpicture}%
}
	\caption{The average error incurred by the lower bound on $\RNCO(V)$ in Proposition~\ref{prop:LB}, in Experiment~\ref{exp:LB}. The error is zero when $|V|=3$ according to Remark~\ref{rem:LB3user}.}
	\label{fig:AveErrLBNew}
\end{figure}

\begin{experiment} \label{exp:LB}
    We generate a number of CCDE systems as follows. The number of packets $H(V)$ varies from $6$ to $30$, while the number of users $|V|$ varies from $3$ to $15$. For each combination of $H(V)$ and $|V|$, we repeat the procedure below for $20$ times.
    \begin{itemize}
        \item randomly generate the packet sets $\Rz{i}= A_i \Rx$ for all $i \in V$ subject to the condition $l(\Rx) = H(V)$;
        \item compute the LBs on $\RNCO(V)$ based on \cite{Roua2010,SprintRand2010} and Proposition~\ref{prop:LB}.
    \end{itemize}
    We obtain the error as the absolute difference between the LB and $\RNCO(V)$. We plot the average error incurred by the LBs on $\RNCO(V)$ in \cite{Roua2010,SprintRand2010} and Proposition~\ref{prop:LB} over repetitions in Figs.~\ref{fig:AveErrLBOld1}, \ref{fig:AveErrLBOld2} and \ref{fig:AveErrLBNew}, respectively. It can be seen that the LB on $\RNCO(V)$ in Proposition~\ref{prop:LB} is much tighter than the ones in \cite{Roua2010,SprintRand2010}. In addition, the error in Fig.~\ref{fig:AveErrLBNew} is zero for $|V|=3$ according to Remark~\ref{rem:LB3user}.
\end{experiment}

\section{Algorithms for the Minimum Sum-rate Problem} \label{sec:algo}

The remaining problem is to discuss how to efficiently solve the maximization problems in~\eqref{eq:MinSumRateACOPat} and \eqref{eq:MinSumRateNCOPat} in Corollary~\ref{coro:MinSumRate} for the asymptotic and non-asymptotic settings, respectively, and determine a corresponding optimal rate vector. For this purpose, we propose the MDA and SIA algorithms in this section.

\subsection{Modified Decomposition Algorithm} \label{sec:MDA}

The MDA algorithm is given in Algorithm~\ref{algo:MDA} for solving the minimum sum-rate problem in the asymptotic model. The optimality of the MDA algorithm is summarized in Theorem~\ref{theo:MDAOpt} below. The proof is in Appendix~\ref{app:DAtoMDA}.

\begin{theorem} \label{theo:MDAOpt}
     The MDA algorithm outputs the minimum sum-rate $\RACO(V)$, the fundamental partition $\Pat^*$ and an optimal rate vector $\rv_V \in \RRACO(V)$ for the asymptotic model. The estimation sequence of $\RACO(V)$, i.e., the value of $\alpha$ in each iteration, converges monotonically upward to $\RACO(V)$. \hfill\IEEEQED
\end{theorem}

        \begin{algorithm} [t]
	       \label{algo:MDA}
	       \small
	       \SetAlgoLined
	       \SetKwInOut{Input}{input}\SetKwInOut{Output}{output}
	       \SetKwFor{For}{for}{do}{endfor}
            \SetKwRepeat{Repeat}{repeat}{until}
            \SetKwIF{If}{ElseIf}{Else}{if}{then}{else if}{else}{endif}
	       \BlankLine
           \Input{the ground set $V$, an oracle that returns the value of $H(X)$ for a given $X \subseteq V$}
	       \Output{$\alpha$ that equals to $\RACO(V)$, $\Pat^*$ which is the fundamental partition and a rate vector $\rv_V$ in the optimal rate set $\RRACO(V) = B(\FuHashHat{\RACO(V)},\leq)$}
	       \BlankLine
            initialize $\alpha$ according to Proposition~\ref{prop:LB}: $\alpha \leftarrow \max_{i \in V} \big\{ \varphi(\Set{\Set{i},V \setminus \Set{i}}), \varphi(\Set{\Set{m} \colon m \in V}) \big\}$\;
            find the minimal/finest minimizer $\Pat^*$ of $\min_{\Pat\in \Pi(V)} \FuHash{\alpha}[\Pat]$ and a rate vector $\rv_V \in B(\FuHashHat{\alpha},\leq)$\;
            $ \hat{\alpha} \leftarrow \varphi(\Pat^*)$\;
            \While{$\hat{\alpha} \neq \alpha$}{
                $\alpha \leftarrow \hat{\alpha}$\;
                find the minimal/finest minimizer $\Pat^*$ of $\min_{\Pat\in \Pi(V)} \FuHash{\alpha}[\Pat]$ and a rate vector $\rv_V \in B(\FuHashHat{\alpha},\leq)$\;
                $\hat{\alpha} \leftarrow \varphi(\Pat^*) $\;
            }
            return $\alpha$, $\Pat^*$ and $\rv_V$\;
	   \caption{Modified Decomposition Algorithm (MDA) }
	   \end{algorithm}

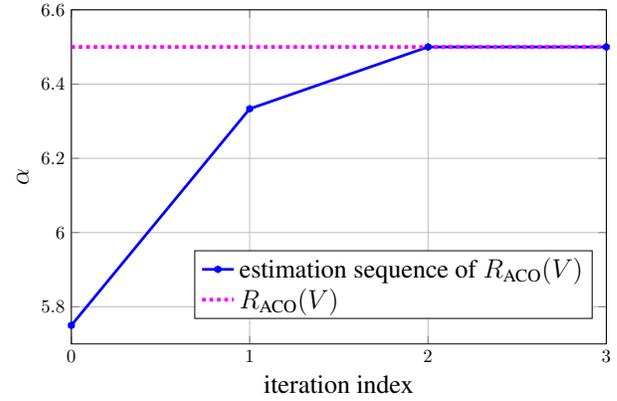
\begin{figure}[tbp]
	\centering
    \scalebox{0.7}{
%
%
%
\definecolor{mycolor1}{rgb}{1,0,1}%
\begin{tikzpicture}

\begin{axis}[%
width=4in,
height=2.5in,
scale only axis,
xmin=0,
xmax=3,
xtick={0,1,2,3},
xlabel={\Large iteration index},
xmajorgrids,
ymin=5.7,
ymax=6.6,
ymajorgrids,
ylabel={\Large $\alpha$},
legend style={at={(0.23,0.28)},anchor=north west,draw=black,fill=white,legend cell align=left}
]
\addplot [
color=blue,
solid,
line width=1.5pt,
mark=asterisk,
mark options={solid},
]
table[row sep=crcr]{
0 5.75\\
1 6.33333333333333\\
2 6.5\\
3 6.5\\
};
\addlegendentry{\Large estimation sequence of $\RACO(V)$};

\addplot [
color=mycolor1,
dotted,
line width=2pt,
]
table[row sep=crcr]{
0 6.5\\
1 6.5\\
2 6.5\\
3 6.5\\
};
\addlegendentry{\Large $\RACO(V)$};

\end{axis}
\end{tikzpicture}%
}
	\caption{The estimation sequence of $\RACO(V)$, i.e., the value of $\alpha$ in each iteration, when the MDA algorithm is applied to the system in Example~\ref{ex:aux}. According to Theorem~\ref{theo:MDAOpt}, it converges monotonically upward to $\RACO(V)$.}
	\label{fig:Converge}
\end{figure}

\begin{example}\label{ex:auxMDA}
    We apply the MDA algorithm to the system in Example~\ref{ex:aux}. We initiate $\alpha=\max_{i \in V} \big\{ \varphi(\Set{\Set{i},V \setminus \Set{i}}), \varphi(\Set{\Set{i} \colon i \in V}) \big\}=\frac{23}{4}$ as the LB on $\RACO(V)$ in Proposition~\ref{prop:LB}, and have the following results.
    \begin{itemize}
        \item When $\alpha=\frac{23}{4}$, we have $\Pat^*=\Set{\Set{4,5},\Set{1},\Set{2},\Set{3}}$ being the minimal minimizer of the Dilworth truncation problem $\min_{\Pat\in \Pi(V)} \FuHash{23/4}[\Pat]$ and $\rv_V=(\frac{3}{4},-\frac{1}{4},-\frac{1}{4},\frac{15}{4},0)$ being a vector in $B(\FuHashHat{23/4},\leq)$. We get $\hat{\alpha}=\varphi(\Pat^*)=\frac{19}{3}$. Since $\hat{\alpha} \neq \alpha$, the iteration continues;
        \item When $\alpha=\frac{19}{3}$, we have $\Pat^*=\Set{\Set{1,4,5},\Set{2},\Set{3}}$ being the minimal minimizer of $\min_{\Pat\in \Pi(V)} \FuHash{19/3}[\Pat]$ and $\rv_V=(1,\frac{1}{3},\frac{1}{3},\frac{13}{3},0) \in B(\FuHashHat{19/3},\leq)$. Since $\hat{\alpha}=\varphi(\Pat^*)=\frac{13}{2} \neq \alpha$, the iteration continues;
        \item When $\alpha=\frac{13}{2}$, we have $\Pat^*=\Set{\Set{1,4,5},\Set{2},\Set{3}}$ being the minimal minimizer of $\min_{\Pat\in \Pi(V)} \FuHash{13/2}[\Pat]$ and $\rv_V=(1,\frac{1}{2},\frac{1}{2},\frac{9}{2},0) \in B(\FuHashHat{13/2},\leq)$. Since $\hat{\alpha}=\varphi(\Pat^*)=\frac{13}{2} = \alpha $, the iteration terminates.
    \end{itemize}
    At the output, $\alpha=\frac{13}{2}$ and $\Pat^*=\Set{\Set{1,4,5},\Set{2},\Set{3}}$ coincide with the minimum sum-rate $\RACO(V)$ and the fundamental partition in Example~\ref{ex:aux}, respectively. In addition, $\rv_V=(1,\frac{1}{2},\frac{1}{2},\frac{9}{2},0) \in B(\FuHashHat{13/2},\leq)=\RRACO(V)$ is an optimal rate vector. In fact, $(1,\frac{1}{2},\frac{1}{2},\frac{9}{2},0) \in \EX(\FuHashHat{\RACO(V)})$ is one of the extreme points in the base polyhedron $B(\FuHash{\RACO(V)},\leq)$, where $\EX(\FuHash{\RACO(V)})$ is shown in Example~\ref{ex:aux}. Fig.~\ref{fig:Converge} shows that the value of $\alpha$ in each iteration of the MDA algorithm converges monotonically upwards to the minimum sum-rate $\RACO(V)$.
\end{example}

In the next subsection, we show how to solve the Dilworth truncation problem $\min_{\Pat\in \Pi(V)} \FuHash{\alpha}[\Pat]$ in steps 2 and 6 in the MDA algorithm by a fusion implementation of the coordinate-wise saturation capacity algorithm, where a rate vector $\rv_V \in B(\FuHashHat{\alpha},\leq)$ is also returned.

\subsection{Coordinate-wise Saturation Capacity Algorithm by Fusion Method} \label{sec:CoordSatCapFus}

There exist several algorithms for solving the Dilworth truncation problem $\min_{\Pat\in \Pi(V)} \FuHash{\alpha}[\Pat]$ in the literature. For example, the fusion set method is proposed in \cite{Narayanan1991PLP,Narayanan1997Book} for determining the PSP of electronic networks; The coordinate-wise saturation capacity (CoordSatCap) algorithm in \cite[Section 3.2]{Fujishige2005} has been applied to determine the PSP of a network in \cite[Algorithm A]{Baiou2011IBMRepPartition}, \cite[Section 2]{Kolmogorov2010NetPSP} and the strength of a network in \cite[Section 3]{Chen1994NetStrenght}.\footnote{One can verify that the methods in \cite[Algorithm A]{Baiou2011IBMRepPartition}, \cite[Section 2]{Kolmogorov2010NetPSP} and \cite[Section 3]{Chen1994NetStrenght} are in fact the CoordSatCap algorithm that is presented as Greedy Algorithm II in \cite[Section 3.2]{Fujishige2005}. The studies in \cite{Baiou2011IBMRepPartition,Kolmogorov2010NetPSP,Chen1994NetStrenght} are based on the cut function of a network. }

In this paper, we consider the CoordSatCap algorithm, which determines not only the minimum and the minimizer of the Dilworth truncation problem $\min_{\Pat\in \Pi(V)} \FuHash{\alpha}[\Pat]$, but also a rate vector $\rv_V$ in the base polyhedron $B(\FuHashHat{\alpha},\leq)$. 
In this section, we first describe the CoordSatCap algorithm and then show how to implement it by a fusion method.

We introduce some related definitions as follows. For $X \subseteq V$, let $\chi_X = (r_i \colon i \in V )$ be the \textit{characteristic vector} of the subset $X$ such that $r_i = 1$ if $i \in X$ and $r_i = 0$ if $i \notin X$. The notation $\chi_{\Set{i}}$ is simplified by $\chi_i$. Let $ \Phi = (\phi_1,\dotsc,\phi_{|V|})$, where $\phi_i \in V$ and $\phi_i \neq \phi_{i'}$ for all $i,i'\in\Set{1,\dotsc,|V|}$ such that $i\neq i'$. We call $\Phi$ a \textit{linear ordering} of $V$. For example, $\Phi = (2,3,1,4)$ is a linear ordering of $V = \Set{1,\dotsc,4}$.

For $U \subseteq \Pat$ where $\Pat$ is some partition in $\Pi(V)$, denote
$$\tilde{U}=\sqcup_{X \in U} X$$
where $\sqcup$ is the disjoint union, i.e., $\tilde{U} \subseteq V$ is a fusion of all the subsets in $U$. For example, for $\Pat = \Set{\Set{1,3},\Set{2,4},\Set{5},\Set{6},\Set{7}} \in \Pi({\Set{1,\dotsc,7}})$ and $U = \Set{\Set{1,3},\Set{2,4},\Set{5},\Set{6}} \subsetneq \Pat$, we have $\tilde{U} = \Set{1,\dotsc,6}$.

For a rate vector $\rv_V \in P(\FuHash{\alpha},\leq)$,
$$ \hat{\xi}_i =  \max \Set{ \xi \colon \rv_V + \xi \chi_i \in P(\FuHash{\alpha},\leq)}$$
is called the \textit{saturation capacity} in dimension $i$ \cite[Section 2.3]{Fujishige2005}. Here, $\hat{\xi}_i \in \Real_+$ denotes the maximum increment in $r_i$ such that the resulting rate vector $\rv_V + \hat{\xi}_i \chi_i$ is still in the polyhedron $P(\FuHash{\alpha},\leq)$.
The saturation capacity can be determined by solving the minimization problem \cite[Section 2.3]{Fujishige2005}
\begin{multline} \label{eq:minmax}
    \max \Set{ \xi \colon \rv_V + \xi \chi_i \in P(\FuHash{\alpha},\leq)}  = \\
    \min\Set{\FuHash{\alpha}(X) - r(X) \colon i \in X \subseteq V}.
\end{multline}
Here, $\FuHash{\alpha}$ is submodular over all $X \subseteq V$ such that $i \in X$ because of the intersecting submodularity of $\FuHash{\alpha}$, i.e., the minimization problem in \eqref{eq:minmax} is an SFM one. An SFM problem can be solved in polynomial time \cite{Khachiyan1980Ellipsoid,Grotschel2012,Grotschel1981,Iwata2007SFM,IFF2001,Fujishige2011MinNorm} (see also Section~\ref{sec:Complexity}). The minimizers of an SFM problem form a lattice, where the minimal/smallest and maximal/largest minimizer exist \cite[Lemma 2.1]{Fujishige2005}. In Step 3 of Algorithm~\ref{algo:CoordSatCap}, we obtain the minimal/smallest minimizer.

        \begin{algorithm} [t]
	       \label{algo:CoordSatCap}
	       \small
	       \SetAlgoLined
	       \SetKwInOut{Input}{input}\SetKwInOut{Output}{output}
	       \SetKwFor{For}{for}{do}{endfor}
            \SetKwRepeat{Repeat}{repeat}{until}
            \SetKwIF{If}{ElseIf}{Else}{if}{then}{else if}{else}{endif}
	       \BlankLine
           \Input{the ground set $V$, an oracle that returns the value of $H(X)$ for a given $X \subseteq V$, $\alpha$ which is an estimation of $\RACO(V)$}
	       \Output{$\rv_V$ which is a rate vector in $B(\FuHashHat{\alpha},\leq)$ and $\Pat^*$ which is the minimal/finest minimizer of $\min_{\Pat\in \Pi(V)} \FuHash{\alpha}[\Pat]$ }
	       \BlankLine
            initiate $\rv_V$ so that $\rv_V \in P(\FuHash{\alpha},\leq)$ and $\Pat^* = \Set{\Set{i} \colon i \in V}$ and choose a linear ordering $\Phi = ( \phi_1,\dotsc,\phi_{|V|})$\;
            \For{$i=1$ \emph{\KwTo} $|V|$}{
                determine the saturation capacity
                $$ \hat{\xi}_{\phi_i} \leftarrow \min\Set{\FuHash{\alpha}(X) - r(X) \colon \phi_i \in X \subseteq V} $$
                and the minimal/smallest minimizer $\hat{X}_{\phi_i}$\;
                $ \rv_V \leftarrow \rv_V + \hat{\xi}_{\phi_i} \chi_{\phi_i}$\;
                merge/fuse all subsets in $\Pat^*$ that intersect with $\hat{X}_{\phi_i}$ in to one subset $\tilde{\X} = \sqcup_{X \in \X} X$:
                \begin{equation}
                    \begin{aligned}
                        & \X \leftarrow \Set{C \in \Pat^* \colon C \cap \hat{X}_{\phi_i} \neq \emptyset};\\
                        & \Pat^* \leftarrow (\Pat^* \setminus \X) \sqcup \Set{ \tilde{\X} };
                    \end{aligned} \nonumber
                \end{equation}
            }
            return $\rv_V$ and $\Pat^*$\;
	   \caption{Coordinate-wise Saturation Capacity (CoordSatCap) Algorithm \cite{Fujishige2005}}
	   \end{algorithm}

\subsubsection{Coordinate-wise Saturation Capacity (CoordSatCap) Algorithm}\label{sec:CoordSatCap}

The main purpose of the CoordSatCap algorithm in Algorithm~\ref{algo:CoordSatCap} is to determine a rate vector, or base point, in $B(\FuHashHat{\alpha},\leq)$. The idea is to start with a point $\rv_V \in P(\FuHash{\alpha},\leq)$ and increase each dimension of $\rv_V$ in order by the saturation capacity. Finally, we have $\rv_V$ still in $P(\FuHash{\alpha},\leq)$ but reaching saturation in each dimension, i.e., $\rv_V + \epsilon \chi_i \notin P(\FuHash{\alpha},\leq)$ for all $\epsilon>0$ and $i \in V$, which means $\rv_V \in B(\FuHashHat{\alpha},\leq)$ necessarily. Based on the tight sets of this base point, the minimizers of $\min_{\Pat\in \Pi(V)} \FuHash{\alpha}[\Pat]$ can be determined, i.e., the Dilworth truncation problem is solved accordingly.\footnote{The definition of tight set and related explanations are in Appendix~\ref{app:CoordSatCap}, where we also present a brief proof/explanation that the CoordSatCap algorithm outputs a rate vector $\rv_V \in B(\FuHashHat{\alpha},\leq)$ and the minimal minimizer $\Pat^*$ of the Dilworth truncation problem $\min_{\Pat\in \Pi(V)} \FuHash{\alpha}[\Pat]$ based on the studies in \cite{Chen1994NetStrenght,Edmonds2003Convex,Fujishige2005}.}
See also Appendix~\ref{app:CoordSatCap} for the detailed explanation of the CoordSatCap algorithm.

The following lemma shows one way to initiate a rate vector $\rv_V$ such that $\rv_V \in P(\FuHash{\alpha},\leq)$. The proof is in Appendix~\ref{app:lemma:IniRate}.

\begin{lemma} \label{lemma:IniRate}
    For any $\alpha$ such that $0 \leq \alpha \leq H(V)$, $\rv_V = (\alpha - H(V)) \chi_{V} \in P(\FuHash{\alpha},\leq)$. \hfill\IEEEQED
\end{lemma}

In Algorithm~\ref{algo:CoordSatCap}, the linear ordering $\Phi$ matters when we want to minimize a weighted sum-rate objective function in the optimal rate set, which will be discussed in Section~\ref{sec:MinWeightedSum}. We remark that for the minimum (equal-weight) sum-rate problem for both asymptotic and non-asymptotic models that is considered in this section, any linear ordering $\Phi$ of the user indices can be chosen.

\begin{example} \label{ex:auxCoordSatCap1}
    Consider the Dilworth truncation problem $\min_{\Pat\in \Pi(V)} \FuHash{\alpha}[\Pat]$ for $\alpha = \frac{23}{4}$ in Example~\ref{ex:auxMDA}. We apply the CoordSatCap algorithm by initiating $\rv_V = (\alpha -H(V)) \chi_{V} = (-\frac{17}{4},\dotsc,-\frac{17}{4})$ according to Lemma~\ref{lemma:IniRate} and $\Pat^*=\Set{\Set{1},\Set{2},\Set{3},\Set{4},\Set{5}}$. The linear ordering is set to $\Phi=(4,5,3,2,1)$.
    \begin{itemize}
        \item For $\phi_1=4$, we have $\hat{\xi}_{4} = 8$ and $\hat{X}_{4}=\Set{4}$ being the minimum and minimal minimizer of $ \min\Set{\FuHash{23/4}(X) - r(X) \colon 4 \in X \subseteq V} $, respectively. By executing $\rv_V \leftarrow \rv_V + 8\chi_{4}$, we update $\rv_V$ to $(-\frac{17}{4},-\frac{17}{4},-\frac{17}{4},\frac{15}{4},-\frac{17}{4})$. Also, $\X = \Set{\Set{4}}$ and $\tilde{\X} = \Set{C \in \Pat^* \colon C \cap \Set{4} \neq \emptyset} = \Set{4}$. By executing $ \Pat^* = (\Pat^* \setminus \X) \sqcup \Set{ \tilde{\X} } $, we still have $\Pat^*=\Set{\Set{1},\Set{2},\Set{3},\Set{4},\Set{5}}$;
        \item For $\phi_2=5$, we have $\hat{\xi}_{5} = \frac{17}{4}$ and $\hat{X}_{5}=\Set{4,5}$. $\rv_V$ is updated to $(-\frac{17}{4},-\frac{17}{4},-\frac{17}{4},\frac{15}{4},0)$. Since $\X = \Set{C \in \Pat^* \colon C \cap \Set{4,5} \neq \emptyset} = \Set{\Set{4},\Set{5}}$ and $\tilde{\X} = \Set{4,5}$, by executing $ \Pat^* = (\Pat^* \setminus \X) \sqcup \Set{ \tilde{\X} } $, we update $\Pat^*$ to $\Set{\Set{1},\Set{2},\Set{3},\Set{4,5}}$;
        \item For $\phi_3=2$, we have $\hat{\xi}_{2} = 4$ and $\hat{X}_{2}=\Set{2}$. $\rv_V$ is updated to $(-\frac{17}{4},-\frac{1}{4},-\frac{17}{4},\frac{15}{4},0)$. $\Pat^* = \Set{\Set{1},\Set{2},\Set{3},\Set{4,5}}$ after executing $ \Pat^* = (\Pat^* \setminus \X) \sqcup \Set{ \tilde{\X} } $;
        \item For $\phi_4=3$, we have $\hat{\xi}_{3} = 4$ and $\hat{X}_{3}=\Set{3}$. $\rv_V$ is updated to $(-\frac{17}{4},-\frac{1}{4},-\frac{1}{4},\frac{15}{4},0)$. $\Pat^* = \Set{\Set{1},\Set{2},\Set{3},\Set{4,5}}$ after executing $ \Pat^* = (\Pat^* \setminus \X) \sqcup \Set{ \tilde{\X} } $;
        \item For $\phi_5=1$, we have $\hat{\xi}_{1} = 5$ and $\hat{X}_{1}=\Set{1}$. $\rv_V$ is updated to $(\frac{3}{4},-\frac{1}{4},-\frac{1}{4},\frac{15}{4},0)$. $\Pat^* = \Set{\Set{1},\Set{2},\Set{3},\Set{4,5}}$ after executing $ \Pat^* = (\Pat^* \setminus \X) \sqcup \Set{ \tilde{\X} } $.
    \end{itemize}
    At the output, we have $\Pat^*=\Set{\Set{1},\Set{2},\Set{3},\Set{4,5}}$ and $\rv_V = (\frac{3}{4},-\frac{1}{4},-\frac{1}{4},\frac{15}{4},0)$ being the minimal minimizer of $\min_{\Pat\in \Pi(V)} \FuHash{23/4}[\Pat]$ and a base point in $B(\FuHashHat{23/4},\leq)$, respectively.
\end{example}

\begin{figure}[tbp]
	\centering
    \scalebox{0.7}{
%
%
%
\definecolor{mycolor1}{rgb}{0.5,0.5,0.9}%
\definecolor{mycolor2}{rgb}{1,1,0.5}%
\begin{tikzpicture}

\begin{axis}[%
width=3.5in,
height=2.8in,
view={-30}{35},
scale only axis,
xmin=-4.25,
xmax=2,
xtick={-4,-3,-2,-1,0,1,2,3},
xlabel={\Large $r_1$},
xmajorgrids,
ymin=-4.25,
ymax=5,
ytick={-4,-3,-2,-1,0,1,2,3,4,5},
ylabel={\Large $r_4$},
ymajorgrids,
zmin=-4.25,
zmax=2.25,
ztick={-4,-3,-2,-1,0,1,2,3,4,5},
zlabel={\Large $r_5$},
zmajorgrids,
axis x line*=bottom,
axis y line*=left,
axis z line*=left,
legend style={at={(0.8,1.1)},anchor=north west,draw=black,fill=white,legend cell align=left}
]

\addplot3[
color=mycolor1,
line width=5.0pt]
table[row sep=crcr]{
x y z\\
0.75 3.75 0\\
0.75 2 1.75\\
};
\addlegendentry{\large $B(\FuHashHat{23/4,C},\leq)$};

\addplot3[area legend,solid,fill=white!90!black,opacity=4.000000e-01,draw=black]
table[row sep=crcr]{
x y z\\
-4.25 -4.25 1.75\\
-4.25 2 1.75\\
0.75 2 1.75\\
0.75 -4.25 1.75\\
-4.25 -4.25 1.75\\
};
\addlegendentry{\large $P(\FuHash{23/4,C},\leq)$};

\addplot3[solid,fill=white!90!black,opacity=4.000000e-01,draw=black,forget plot]
table[row sep=crcr]{
x y z\\
-4.25 2 1.75\\
0.75 2 1.75\\
0.75 3.75 0\\
-4.25 3.75 0\\
-4.25 3.75 0\\
-4.25 2 1.75\\
};

\addplot3[solid,fill=white!90!black,opacity=4.000000e-01,draw=black,forget plot]
table[row sep=crcr]{
x y z\\
0.75 -4.25 -4.25\\
0.75 -4.25 1.75\\
0.75 2 1.75\\
0.75 3.75 0\\
0.75 3.75 -4.25\\
0.75 -4.25 -4.25\\
};

\addplot3[solid,fill=white!90!black,opacity=4.000000e-01,draw=black,forget plot]
table[row sep=crcr]{
x y z\\
-4.25 3.75 -4.25\\
0.75 3.75 -4.25\\
0.75 3.75 0\\
-4.25 3.75 0\\
-4.25 3.75 -4.25\\
};

\addplot3[solid,fill=white!90!black,opacity=4.000000e-01,draw=black,forget plot]
table[row sep=crcr]{
x y z\\
-4.25 -4.25 -4.25\\
-4.25 -4.25 1.75\\
-4.25 2 1.75\\
-4.25 3.75 0\\
-4.25 3.75 -4.25\\
-4.25 -4.25 -4.25\\
};

\addplot3[solid,fill=white!90!black,opacity=4.000000e-01,draw=black,forget plot]
table[row sep=crcr]{
x y z\\
-4.25 -4.25 -4.25\\
-4.25 -4.25 1.75\\
0.75 -4.25 1.75\\
0.75 -4.25 -4.25\\
-4.25 -4.25 -4.25\\
};

\addplot3 [
->,
color=blue,
dashed,
line width=2.0pt,
mark size=3.0pt,
mark=o,
mark options={solid}]
table[row sep=crcr] {
-4.25 -4.25 -4.25\\
-4.25 3.75 -4.25\\
-4.25 3.75 0\\
0.75 3.75 0\\
};
\addlegendentry{\Large path to $(\frac{3}{4},\frac{15}{4},0)$ };

\addplot3 [
color=red,
line width=2.0pt,
only marks,
mark=triangle,
mark options={solid,,rotate=180}]
table[row sep=crcr] {
0.75 3.75 0\\
0.75 2 1.75\\
};
\addlegendentry{\large $\EX(\FuHashHat{23/4,C})$};

\end{axis}
\end{tikzpicture}%
}
	\caption{For $\alpha=\frac{23}{4}$ and $C = \Set{1,4,5}$, the figure shows the polyhedron $P(\FuHash{\alpha,C},\leq)$ and base polyhedron $B(\FuHashHat{\alpha,C},\leq)$, where $\FuHash{\alpha,C}$ is the reduction of $\FuHash{\alpha}$ on $C$. Note, in this case, the base polyhedron $B(\FuHashHat{\alpha,C},\leq)$ reduces to a line segment with two vertices, $(\frac{3}{4},\frac{15}{4},0)$ and $(\frac{3}{4},2,\frac{7}{4})$. This figure also shows the adaptation of the rate vector $\rv_C$ resulted from the CoordSatCapFus algorithm in Example~\ref{ex:auxCoordSatCap1}: the path $(-\frac{17}{4},-\frac{17}{4},-\frac{17}{4}) \rightarrow (-\frac{17}{4},\frac{15}{4},-\frac{17}{4}) \rightarrow (-\frac{17}{4},\frac{15}{4},0) \rightarrow (\frac{3}{4},\frac{15}{4},0)$. We have the rate vector $(\frac{3}{4},\frac{15}{4},0) \in B(\FuHashHat{\alpha,C},\leq)$ at the end. }
	\label{fig:CoordSatCapFusDemo1}
\end{figure}
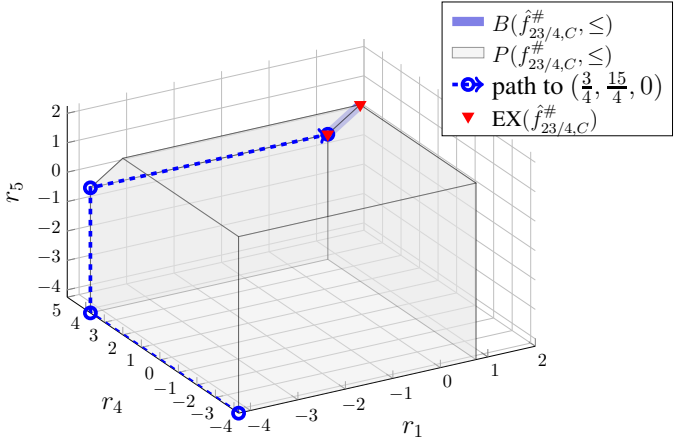

\begin{example} \label{ex:auxCoordSatCap2}
    Consider the Dilworth truncation problem $\min_{\Pat\in \Pi(V)} \FuHash{\alpha}[\Pat]$ for $\alpha = \frac{13}{2}$ in Example~\ref{ex:auxMDA}. By applying the CoordSatCap algorithm in the same way as in Example~\ref{ex:auxCoordSatCap1}, one can show that, at the output, $\Pat^* = \Set{\Set{1,4,5},\Set{2},\Set{3}}$ and $\rv_V = (1,\frac{1}{2},\frac{1}{2},\frac{9}{2},0)$, which are the minimal minimizer of $\min_{\Pat\in \Pi(V)} \FuHash{13/2}[\Pat]$ and a rate vector in $B(\FuHashHat{13/2},\leq)$, respectively.
\end{example}

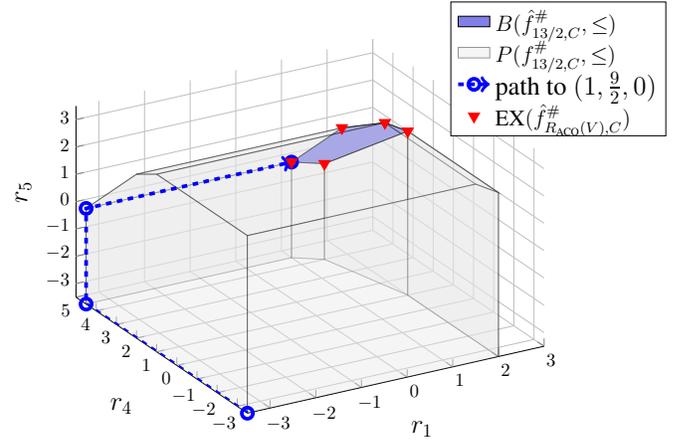
\begin{figure}[tbp]
	\centering
    \scalebox{0.7}{
%
%
%
\definecolor{mycolor1}{rgb}{0.5,0.5,0.9}%
\definecolor{mycolor2}{rgb}{1,1,0.5}%
\begin{tikzpicture}

\begin{axis}[%
width=3.5in,
height=2.8in,
view={-30}{35},
scale only axis,
xmin=-3.5,
xmax=3,
xtick={-4,-3,-2,-1,0,1,2,3},
xlabel={\Large $r_1$},
xmajorgrids,
ymin=-3.5,
ymax=5,
ytick={-4,-3,-2,-1,0,1,2,3,4,5},
ylabel={\Large $r_4$},
ymajorgrids,
zmin=-3.5,
zmax=3.5,
ztick={-4,-3,-2,-1,0,1,2,3,4,5},
zlabel={\Large $r_5$},
zmajorgrids,
axis x line*=bottom,
axis y line*=left,
axis z line*=left,
legend style={at={(0.8,1.1)},anchor=north west,draw=black,fill=white,legend cell align=left}
]

\addplot3[area legend,solid,fill=mycolor1,draw=black]
table[row sep=crcr]{
x y z\\
1 2 2.5\\
1 4.5 0\\
1.5 4 0\\
2 1 2.5\\
1.5 1 3\\
1 2 2.5\\
};
\addlegendentry{\large $B(\FuHashHat{13/2,C},\leq)$};

\addplot3[area legend,solid,fill=white!90!black,opacity=4.000000e-01,draw=black]
table[row sep=crcr]{
x y z\\
-3.5 4.5 -3.5 \\
-3.5 4.5 0 \\
1 4.5 0 \\
1 4.5 -3.5  \\
-3.5 4.5 -3.5 \\
};
\addlegendentry{\large $P(\FuHash{13/2,C},\leq)$};

\addplot3[solid,fill=white!90!black,opacity=4.000000e-01,draw=black,forget plot]
table[row sep=crcr]{
x y z\\
1 4.5 0 \\
1 4.5 -3.5  \\
1.5 4 -3.5 \\
1.5 4 0 \\
1 4.5 0 \\
};

\addplot3[solid,fill=white!90!black,opacity=4.000000e-01,draw=black,forget plot]
table[row sep=crcr]{
x y z\\
-3.5 4.5 0 \\
1 4.5 0 \\
1 2 2.5 \\
-3.5 2 2.5 \\
-3.5 4.5 0 \\
};

\addplot3[solid,fill=white!90!black,opacity=4.000000e-01,draw=black,forget plot]
table[row sep=crcr]{
x y z\\
-3.5 2 2.5 \\
1 2 2.5 \\
1.5 1 3 \\
-3.5 1 3 \\
-3.5 2 2.5\\
};

\addplot3[solid,fill=white!90!black,opacity=4.000000e-01,draw=black,forget plot]
table[row sep=crcr]{
x y z\\
-3.5 1 3\\
1.5 1 3\\
1.5 -3.5 3\\
-3.5 -3.5 3\\
-3.5 1 3\\
};

\addplot3[solid,fill=white!90!black,opacity=4.000000e-01,draw=black,forget plot]
table[row sep=crcr]{
x y z\\
1.5 -3.5 3\\
1.5 1 3\\
2 1 2.5\\
2 -3.5 2.5\\
1.5 -3.5 3\\
};

\addplot3[solid,fill=white!90!black,opacity=4.000000e-01,draw=black,forget plot]
table[row sep=crcr]{
x y z\\
2 1 -3.5\\
2 1 2.5\\
1.5 4 0\\
1.5 4 -3.5\\
2 1 -3.5\\
};

\addplot3[solid,fill=white!90!black,opacity=4.000000e-01,draw=black,forget plot]
table[row sep=crcr]{
x y z\\
2 -3.5 2.5\\
2 1 2.5\\
2 1 -3.5\\
2 -3.5 -3.5\\
2 -3.5 2.5\\
};

\addplot3[solid,fill=white!90!black,opacity=4.000000e-01,draw=black,forget plot]
table[row sep=crcr]{
x y z\\
-3.5 -3.5 3\\
-3.5 1 3\\
-3.5 2 2.5\\
-3.5 4.5 0\\
-3.5 4.5 -3.5\\
-3.5 -3.5 -3.5\\
-3.5 -3.5 3\\
};

\addplot3[solid,fill=white!90!black,opacity=4.000000e-01,draw=black,forget plot]
table[row sep=crcr]{
x y z\\
-3.5 -3.5 3\\
1.5 -3.5 3\\
2 -3.5 2.5\\
2 -3.5 -3.5\\
-3.5 -3.5 -3.5\\
-3.5 -3.5 3\\
};

\addplot3 [
->,
color=blue,
dashed,
line width=2.0pt,
mark size=3.0pt,
mark=o,
mark options={solid}]
table[row sep=crcr] {
-3.5 -3.5 -3.5\\
-3.5 4.5 -3.5\\
-3.5 4.5 0\\
1 4.5 0\\
};
\addlegendentry{\Large path to $(1,\frac{9}{2},0)$ };

\addplot3 [
color=red,
line width=2.0pt,
only marks,
mark=triangle,
mark options={solid,,rotate=180}]
table[row sep=crcr] {
1 2 2.5\\
2 1 2.5\\
1 4.5 0\\
1.5 4 0\\
1.5 1 3\\
};
\addlegendentry{\large $\EX(\FuHashHat{\RACO(V),C})$};

\end{axis}
\end{tikzpicture}%
}
	\caption{For $\alpha=\frac{13}{2}$ and $C = \Set{1,4,5}$, the figure shows the polyhedron $P(\FuHash{\alpha,C},\leq)$ and base polyhedron $B(\FuHashHat{\alpha,C},\leq)$. In this case, the base polyhedron is a 2-dimensional polygon with five vertices that constitutes the extreme points set $\EX(\FuHashHat{\RACO(V),C}) = \Set{(1,2,\frac{5}{2}),(2,1,\frac{5}{2}),(1,\frac{9}{2},0),(\frac{3}{2},4,0),(\frac{3}{2},1,3)}$. This figure also shows the adaptation of the rate vector $\rv_C$ resulted from the CoordSatCapFus algorithm in Example~\ref{ex:auxCoordSatCap2}: the path $(-\frac{7}{2},-\frac{7}{2},-\frac{7}{2}) \rightarrow (-\frac{7}{2},\frac{9}{2},-\frac{7}{2}) \rightarrow (-\frac{7}{2},\frac{9}{2},0) \rightarrow (1,\frac{9}{2},0)$. We have the rate vector $(1,\frac{9}{2},0) \in B(\FuHashHat{\alpha,C},\leq)$ at the end. }
	\label{fig:CoordSatCapFusDemo2}
\end{figure}

For $C \subseteq V$, we call $\FuHash{\alpha,C} \colon 2^C \mapsto \Real $ with $\FuHash{\alpha,C}(X) = \FuHash{\alpha}(X)$ for all $X \subseteq C$ the \textit{reduction} of $\FuHash{\alpha}$ on $C$ \cite[Section 3.1(a)]{Fujishige2005}. The polyhedron and base polyhedron of $\FuHash{\alpha,C}$ are respectively
\begin{equation}
    \begin{aligned}
        & P(\FuHash{\alpha,C},\leq) = \Set{\rv_C \in \Real^{|C|} \colon r(X) \leq \FuHash{\alpha,C}(X), X \subseteq C },  \\
        & B(\FuHash{\alpha,C},\leq) = \Set{\rv_C \in P(\FuHash{\alpha,C},\leq) \colon r(C) = \FuHash{\alpha,C}(C)}.  \\
    \end{aligned}  \nonumber
\end{equation}
In Fig.~\ref{fig:CoordSatCapFusDemo1}, we show $P(\FuHash{\alpha,C},\leq)$ and $B(\FuHashHat{\alpha,C},\leq)$ when $\alpha = \frac{23}{4}$ and $C = \Set{1,4,5}$, where we can see the the path to $\rv_C = (\frac{3}{4},\frac{15}{4},0)$ as a result of the CoordSatCap algorithm in Example~\ref{ex:auxCoordSatCap1}. In Fig.~\ref{fig:CoordSatCapFusDemo2}, we show $P(\FuHash{\alpha,C},\leq)$ and $B(\FuHashHat{\alpha,C},\leq)$ when $\alpha = \frac{13}{2}$ and $C = \Set{1,4,5}$, where we can see the the path to $\rv_C = (1,\frac{9}{2},0)$ as a result of the CoordSatCap algorithm in Example~\ref{ex:auxCoordSatCap2}.

\subsubsection{A Fusion Method Implementation}

In the CoordSatCap algorithm, the saturation capacity $\xi_{\phi_i}$ in dimension $\phi_i$ is determined by solving problem $\min\Set{\FuHash{\alpha}(X) - r(X) \colon \phi_i \in X \subseteq V}$, where each element in $V$ is the index of a user in the system. In this section, we show that this problem can be solved over a merged user set where each non-singleton element denotes a super user, i.e., the CoordSatCap algorithm can be implemented by a fusion method. The validity of this fusion method is based on Lemma~\ref{lemma:Fusion1} and Lemma~\ref{lemma:Fusion2} below with the proofs in Appendix~\ref{app:lemma:Fusion1} and Appendix~\ref{app:lemma:Fusion2}, respectively.

\begin{lemma}\label{lemma:Fusion1}
    Let the CoordSatCap algorithm start with a rate vector $\rv_V \in P(\FuHash{\alpha},\leq)$ such that $\rv_V \leq \Zero$, where $\Zero = (0,\dotsc,0) \in \Real^{|V|}$. We have
    \begin{multline} \label{eq:lemma:Fusion1}
        \min\Set{ \FuHash{\alpha}(X) - r(X) \colon \phi_i \in X \subseteq V} = \\ \min\Set{ \FuHash{\alpha}(X) - r(X) \colon \phi_i \in X \subseteq V_i},
    \end{multline}
    where $V_i = \Set{\phi_1,\dotsc,\phi_i}$ and the minimal minimizer $\hat{X}_{\phi_i} \subseteq V_i$ for all $i \in \Set{1,\dotsc,|V|}$. \hfill\IEEEQED
\end{lemma}

The equality~\eqref{eq:lemma:Fusion1} in Lemma~\ref{lemma:Fusion1} was originally derived in the proof of \cite[Theorem 3.19]{Fujishige2005}, based on which the authors in \cite{CourtIT2014,MiloIT2016} apply $\min\Set{ \FuHash{\alpha}(X) - r(X) \colon \phi_i \in X \subseteq V_i}$ in the CoordSatCap algorithm for solving the non-asymptotic minimum sum-rate problem in the finite linear source model and CCDE. 
However, the proof of \cite[Theorem 3.19]{Fujishige2005} does not show that $\hat{X}_{\phi_i} \subseteq V_i$ for all $i$.

\begin{lemma} \label{lemma:Fusion2}
    In the CoordSatCap algorithm,
    \begin{align}
        & \min\Set{\FuHash{\alpha}(X) - r(X) \colon \phi_i \in X \subseteq V} = \label{eq:lemma:Fusion21} \\
        &\qquad\qquad \min\Set{\FuHash{\alpha}(\tilde{U}) - r(\tilde{U}) \colon U \subseteq \Pat^*, \phi_i \in \tilde{U}} \label{eq:lemma:Fusion22}
    \end{align}
    for all $i \in \Set{1,\dotsc,|V|}$. Let $\hat{X}_{\phi_i}$ and $U_{\phi_i}^*$ be the minimal minimizer of the \eqref{eq:lemma:Fusion21} and \eqref{eq:lemma:Fusion22}, respectively. $\tilde{\X} = \tilde{U}_{\phi_i}^* $, where $\X = \Set{C \in \Pat^* \colon C \cap \hat{X}_{\phi_i} \neq \emptyset}$. \hfill\IEEEQED
\end{lemma}

 Both minimization problems, \eqref{eq:lemma:Fusion21} and \eqref{eq:lemma:Fusion22}, in Lemma~\ref{lemma:Fusion2} are SFM problems due to the intersecting submodularity of $\FuHash{\alpha}$. But, $\Pat^*$ is a fused user set since some of the users have been merged into a super user set, which is treated as one dimension in the SFM problem \eqref{eq:lemma:Fusion22}. So, $|\Pat^*| \leq |V|$.
 We will show in Section~\ref{sec:Complexity} that minimizing over the fused user set contributes to a reduction in computation complexity.

\begin{remark}
    Lemma~\ref{lemma:Fusion2} is based on the fact: If $\hat{X}_{\phi_i}$ is a non-singleton minimal minimizer of $\min\Set{\FuHash{\alpha}(X) - r(X) \colon \phi_i \in X \subseteq V}$ for some $\phi_i$, then, for any partition $\Pat$ that crosses $\hat{X}_{\phi_i}$, there exists a $\Pat'$ that does not cross $\hat{X}_{\phi_i}$ such that $\FuHash{\alpha}[\Pat'] < \FuHash{\alpha}[\Pat]$.\footnote{A multi-way cut or partition $\Pat \in \Pi'(V)$ does not cross a set $X \subsetneq V$ if there exist $C \in \Pat$ such that $X \subseteq C$. For example, for $X = \Set{1,3}$, $\Set{\Set{1,3,4},\Set{2}}$ and $\Set{\Set{1,3},\Set{2},\Set{4}}$ are the partitions that do not cross $X$, while $\Set{\Set{1,2},\Set{3},\Set{4}}$ and $\Set{\Set{1},\Set{2,3,4}}$ are the partitions that cross $X$.}
    It means that, to determine the minimal/finest minimizer of the Dilworth truncation problem $\min_{\Pat \in \Pi'(V)} \FuHash{\alpha}[\Pat]$, we just need to consider all the partitions in $\Pi(V)$ that do not cross $\hat{X}_{\phi_i}$.\footnote{In addition, the fundamental partition $\Pat^*$ must be a multi-way cut of $V$ that does not cross $\hat{X}_{\phi_i}$. See Appendix~\ref{app:FundPartNoncross}.}
    See Appendix~\ref{app:FundPartNoncross} for the explanation.
\end{remark}

\begin{example}
    In Exmaple~\ref{ex:auxCoordSatCap1}, we have $\hat{X}_5 = \Set{4,5}$ when $\alpha = \frac{23}{4}$. For $\Pat = \Set{\Set{1,4},\Set{2,3},\Set{5}}$ that crosses $\Set{4,5}$, we have $\Pat' = \Set{\Set{1,4,5},\Set{2,3}}$ that does not cross $\Set{4,5}$ such that $\FuHash{\alpha}[\Pat'] = \frac{15}{2} < \FuHash{\alpha}[\Pat] = \frac{37}{4}$; For $\Pat = \Set{\Set{1,4},\Set{2,3,5}}$ that crosses $\Set{4,5}$, we have $\Pat' = \Set{\Set{4,5},\Set{1},\Set{2,3}}$ that does not cross $\Set{4,5}$ such that $\FuHash{\alpha}[\Pat'] = \frac{29}{4} < \FuHash{\alpha}[\Pat] = \frac{17}{2}$. One can show that for each partition $\Pat$ that crosses $\Set{4,5}$ there always exists a partition $\Pat'$ that does not cross $\Set{4,5}$ such that $\FuHash{\alpha}[\Pat'] < \FuHash{\alpha}[\Pat]$. Therefore, we just need to consider all partitions that do not cross $\Set{4,5}$ to determine the minimal minimizer of $\min_{\Pat \in \Pi'(V)} \FuHash{\alpha}[\Pat]$. It is equivalent to considering all partitions of $\Set{\Set{4,5},\Set{1},\Set{2},\Set{3}}$, which is the value of $\Pat^*$ when $\phi_2 = 5$ in Exmaple~\ref{ex:auxCoordSatCap1}.
\end{example}

Based on Lemmas~\ref{lemma:Fusion1} and \ref{lemma:Fusion2}, the CoordSatCap algorithm in Algorithm~\ref{algo:CoordSatCap} is equivalent to the CoordSatCapFus algorithm in Algorithm~\ref{algo:CoordSatCapFus}. Steps 4, 5 and 7 in Algorithm~\ref{algo:CoordSatCapFus} are explained as follows.

According to Lemmas~\ref{lemma:Fusion1} and \ref{lemma:Fusion2}, the saturation capacity can be obtained by $ \hat{\xi}_{\phi_i} = \min\Set{ \FuHash{\alpha}(\tilde{U}) - r(\tilde{U}) \colon \Set{\phi_i} \in U \subseteq \Pat^*_i} $. Here, $\Pat^*_i = \Pat^*_{i-1} \sqcup \Set{\Set{\phi_i}} \in \Pi(V_i) $ with $\Pat^*_{i-1} \in \Pi(V_{i-1})$ being the partition obtained at the end of iteration $i-1$. For the minimal minimizer $U^*_{\phi_i}$ of this problem, we have $\hat{X}_{\phi_i} \subseteq \tilde{\X} = \tilde{U}_{\phi_i}^*$, where $\hat{X}_{\phi_i}$ is the minimal minimizer of $\min\Set{ \FuHash{\alpha}(X) - r(X) \colon \phi_i \in X \subseteq V_i}$ and $\X = \Set{C \in \Pat^*_i \colon C \cap \hat{X}_{\phi_i} \neq \emptyset}$. This explains steps 4 and 5 in Algorithm~\ref{algo:CoordSatCapFus}. The equality $\tilde{\X} = \tilde{U}_{\phi_i}^*$ also makes the update of $\Pat^*_i$ easier: We do not need to obtain $\X$ as we did in step 5 in Algorithm~\ref{algo:CoordSatCap}. Since $( \Pat^*_i \setminus U^*_{\phi_i} ) \sqcup \Set{ \tilde{U}^*_{\phi_i} } = ( \Pat^*_i \setminus \X ) \sqcup \Set{ \tilde{\X} }$, simply do $ \Pat^*_i \leftarrow ( \Pat^*_i \setminus U^*_{\phi_i} ) \sqcup \Set{ \tilde{U}^*_{\phi_i} }$, which does not require the determination of $\X$. This explains step 7 in Algorithm~\ref{algo:CoordSatCapFus}.

Since the saturation capacity in the CoordSatCapFus algorithm is always obtained by a minimization problem over a fused user set $\Pat^* \in \Pi(V_i)$ for all $i \in \Set{1,\dotsc,|V|}$, we call the CoordSatCapFus algorithm a fusion method implementation of the CoordSatCap algorithm.

        \begin{algorithm} [t]
	       \label{algo:CoordSatCapFus}
	       \small
	       \SetAlgoLined
	       \SetKwInOut{Input}{input}\SetKwInOut{Output}{output}
	       \SetKwFor{For}{for}{do}{endfor}
            \SetKwRepeat{Repeat}{repeat}{until}
            \SetKwIF{If}{ElseIf}{Else}{if}{then}{else if}{else}{endif}
	       \BlankLine
           \Input{the ground set $V$, an oracle that returns the value of $H(X)$ for a given $X \subseteq V$, $\alpha$ which is an estimation of $\RACO(V)$}
	       \Output{$\rv_V$ which is a rate vector in $B(\FuHashHat{\alpha},\leq)$ and $\Pat^*_{|V|}$ which is the minimal/finest minimizer of $\min_{\Pat\in \Pi(V)} \FuHash{\alpha}[\Pat]$ }
	       \BlankLine
            let $\rv_V \leftarrow (\alpha - H(V)) \chi_V$ so that $\rv_V \in P(\FuHash{\alpha},\leq)$ and choose a linear ordering $\Phi = ( \phi_1,\dotsc,\phi_{|V|})$\;
            initiate $ r_{\phi_1} \leftarrow \FuHash{\alpha}(\Set{\phi_1})$ and $\Pat^*_1 \leftarrow \Set{\Set{\phi_1}}$\;
            \For{$i=2$ \emph{\KwTo} $|V|$}{
                $\Pat^*_i \leftarrow \Pat^*_{i-1} \sqcup \Set{\Set{\phi_i}}$\;
                determine the saturation capacity
                $$ \hat{\xi}_{\phi_i} \leftarrow \min\Set{ \FuHash{\alpha}(\tilde{U}) - r(\tilde{U}) \colon \Set{\phi_i} \in U \subseteq \Pat^*_i} $$
                and the minimal/smallest minimizer $U^*_{\phi_i}$\;
                $ \rv_V \leftarrow \rv_V + \hat{\xi}_{\phi_i} \chi_{\phi_i}$\;
                merge/fuse all subsets in $U^*_{\phi_i}$ into one subset $\tilde{U}^*_{\phi_i} = \sqcup_{X \in U^*_{\phi_i}} X$:
                $$\Pat^*_i \leftarrow (\Pat^*_i \setminus U^*_{\phi_i}) \sqcup \Set{ \tilde{U}^*_{\phi_i} };$$
            }
            return $\rv_V$ and $\Pat^*_{|V|}$\;
	   \caption{Coordinate-wise Saturation Capacity Algorithm by Fusion Method (CoordSatCapFus)}
	   \end{algorithm}

\begin{example} \label{ex:auxCoordSatCapFus2}
     Consider the Dilworth truncation problem $\min_{\Pat\in \Pi(V)} \FuHash{\alpha}[\Pat]$ when $\alpha = \frac{23}{4}$ in Example~\ref{ex:auxMDA}. We apply the CoordSatCapFus algorithm by initiating $\rv_V = (\alpha -H(V)) \chi_{V} = (-\frac{17}{4},\dotsc,-\frac{17}{4})$. The linear ordering is set to $\Phi=(4,5,3,2,1)$.
    \begin{itemize}
        \item For $\phi_1=4$, we assign $r_4 = \FuHash{23/4}(\Set{4}) = \frac{15}{4}$ so that $\rv_V = (-\frac{17}{4},-\frac{17}{4},-\frac{17}{4},\frac{15}{4},-\frac{17}{4})$ and let $\Pat_1^*=\Set{\Set{4}}$;
        \item For $\phi_2=5$, we have $\Pat^*_2 = \Set{\Set{4},\Set{5}}$ and consider the problem $ \min\Set{ \FuHash{\alpha}(\tilde{U}) - r(\tilde{U}) \colon \Set{5} \in U \subseteq \Pat^*_2} $. Since $\Set{U \colon \Set{5} \in U \subseteq \Pat^*_2} = \Set{\Set{\Set{5}},\Set{\Set{4},\Set{5}}}$ and
            \begin{equation}
                \begin{aligned}
                    & \FuHash{23/4}(\Set{5}) - r(\Set{5}) =  6, \\
                    & \FuHash{23/4}(\Set{4,5}) - r(\Set{4,5}) = \frac{17}{4},
                \end{aligned} \nonumber
            \end{equation}
            we have $\hat{\xi}_{5} = \frac{17}{4}$ and $U^*_5 = \Set{\Set{4},\Set{5}}$. $\rv_V$ and $\Pat^*_2$ are updated to $(-\frac{17}{4},-\frac{17}{4},-\frac{17}{4},\frac{15}{4},0)$ and $\Set{\Set{4,5}}$, respectively;
        \item For $\phi_3=2$, we have $\Pat^*_3 = \Set{\Set{2},\Set{4,5}}$ and consider the problem $ \min\Set{ \FuHash{\alpha}(\tilde{U}) - r(\tilde{U}) \colon \Set{2} \in U \subseteq \Pat^*_3} $. Since $\Set{U \colon \Set{2} \in U \subseteq \Pat^*_3} = \Set{\Set{\Set{2}},\Set{\Set{2},\Set{4,5}}}$ and
            \begin{equation}
                \begin{aligned}
                    & \FuHash{23/4}(\Set{2}) - r(\Set{2}) = 4, \\
                    & \FuHash{23/4}(\Set{2,4,5}) - r(\Set{2,4,5}) = \frac{25}{4},
                \end{aligned} \nonumber
            \end{equation}
            we have $\hat{\xi}_2 = 4$ and $U^*_2 = \Set{\Set{2}}$. $\rv_V$ and $\Pat^*_3$ are updated to $(-\frac{17}{4},-\frac{1}{4},-\frac{17}{4},\frac{15}{4},0)$ and $\Set{\Set{2},\Set{4,5}}$, respectively;
        \item In the same way, we can show the followings at the end of each iteration.
                \begin{itemize}
                    \item For $\phi_4 = 3$, $\rv_V = (-\frac{1}{4},-\frac{1}{4},-\frac{17}{4},\frac{15}{4},0)$ and $\Pat^*_4 = \Set{\Set{2},\Set{3},\Set{4,5}}$;
                    \item For $\phi_5 = 1$, $\rv_V=(\frac{3}{4},-\frac{1}{4},-\frac{17}{4},\frac{15}{4},0)$ and $\Pat^*_5 = \Set{\Set{1},\Set{2},\Set{3},\Set{4,5}}$.
                \end{itemize}
    \end{itemize}
    At the end, we have $\rv = (\frac{3}{4},-\frac{1}{4},-\frac{17}{4},\frac{15}{4},0)$ and $\Pat^*_5 = \Set{\Set{1},\Set{2},\Set{3},\Set{4,5}}$, which are the same as in Example~\ref{ex:auxCoordSatCap1}.
    It can be seen that the CoordSatCapFus outputs the same results as in Example~\ref{ex:auxCoordSatCap2} for the Dilworth truncation problem $\min_{\Pat\in \Pi(V)} \FuHash{13/2}[\Pat]$.
\end{example}

\subsection{Solutions for the Finite Linear Source Model} \label{sec:NCO}

As discussed in Section~\ref{sec:IntroCCDE}, in a finite linear source model, we are particularly interested in the existence of the fractional and integral optimal rate vectors in $\RRACO(V)$ and $\RRNCO(V)$, respectively. As pointed out in Remark~\ref{rem:IntNCOFracACO}, since the extreme points in $\RRACO(V) = B(\FuHashHat{\RACO(V)},\leq)$ and $\RRNCO(V) = B(\FuHashHat{\RNCO(V)},\leq) \cap \Z^{|V|}$ are fractional and integral, respectively, in a finite linear source model, the problem reduces to determining a rate vector in $\EX(\FuHashHat{\RACO(V)})$ and $\EX(\FuHashHat{\RNCO(V)})$. For this purpose, we have the following theorem with the proof in Appendix~\ref{app:ExOutput}.

\begin{theorem} \label{theo:ExOutput}
    For all $\alpha \geq \RACO(V)$, the CoordSatCap algorithm returns $\rv_V$ that is an extreme point, or a vertex, in $B(\FuHashHat{\alpha},\leq)$, i.e., $\rv_V \in \EX(\FuHashHat{\alpha})$. \hfill\IEEEQED
\end{theorem}

According to Theorem~\ref{theo:ExOutput} and the proof of Corollary~\ref{coro:IntNCOFracACO}, we have the following results straightforwardly.

\begin{corollary} \label{coro:IntFracFinite}
    For a finite linear source model,
    \begin{enumerate}[(a)]
        \item the MDA algorithm outputs $\rv_V \in \RRACO(V)$ such that $(|\Pat^*| - 1)\rv_V$ is integral, i.e., $\rv_V$ can be implemented by $(|\Pat^*| - 1)$-packet-splitting in CCDE;
        \item with input $\alpha = \RNCO(V)$, the CoordSatCapFus algorithm returns $\rv_V \in \RRNCO(V)$ which is integral. \hfill\IEEEQED
    \end{enumerate}
\end{corollary}

If the value of $\RACO(V)$ can be determined by the MDA algorithm, we know automatically $\RNCO(V) = \lceil \RACO(V) \rceil$. Then, according to Corollary~\ref{coro:IntFracFinite}, we can determine an integral optimal rate vector in $\RRNCO(V)$ for the non-asymptotic model in a finite linear source model by no more than one additional call of the CoordSatCapFus algorithm.\footnote{If $\RACO(V) = \RNCO(V)$, $\RACO(V)$ is integral necessarily and the rate vector in $\RRACO(V)$ returned by the MDA algorithm is also an integral rate vector in $\RRNCO(V)$; If $\RACO(V) < \RNCO(V)$, an integral optimal rate vector in $\RRNCO(V)$ can be determined by an extra call $\text{CoordSatCapFus}(V,H,\lceil \RACO(V) \rceil)$. Therefore, in a finite linear source model, the non-asymptotic minimum sum-rate problem can be solved by no more than one extra call of the CoordSatCapFus algorithm after obtaining $\RACO(V)$.}

\begin{example} \label{ex:auxExOutput}
    The optimal rate vector $\rv_V = (1,\frac{1}{2},\frac{1}{2},\frac{9}{2},0) \in \RRACO(V)$ determined by the MDA algorithm in Example~\ref{ex:auxMDA} is an extreme point in $B(\FuHashHat{\RACO(V)},\leq) = \RRACO(V)$, i.e., $\rv_V = (1,\frac{1}{2},\frac{1}{2},\frac{9}{2},0) \in \EX(\FuHashHat{\RACO(V)})$, where $\EX(\FuHashHat{\RACO(V)})$ is shown in Example~\ref{ex:aux}. Recall that the fundamental partition in this system is $\Pat^*=\Set{\Set{1,4,5},\Set{2},\Set{3}}$ so that $|\Pat^*| = 3$. Therefore, $\rv_V = (1,\frac{1}{2},\frac{1}{2},\frac{9}{2},0)$ can be implemented by $2$-packet-splitting.

    Since $\RACO(V) = \frac{13}{2}$, we have $\RNCO(V) = \lceil \RACO(V) \rceil = 7$. By setting the linear ordering $\Phi = (4,5,2,3,1)$, we call $\text{CoordSatCapFus}(V,H,7)$ and have $\rv_V = (0,1,1,5,0)$ and $\Pat^*=\Set{\Set{1,2,3,4,5}}$ at the output. One can show that $\rv_V = (0,1,1,5,0) \in \EX(\FuHashHat{7}) \subsetneq B(\FuHashHat{7},\leq) \cap \Z^{|5|} = \RRNCO(V)$.\footnote{The result $\Pat^*=\Set{\Set{1,2,3,4,5}}$ is consistent with property (a) of the PSP in Theorem~\ref{theo:PSP} in Appendix~\ref{app:PSP}: Since $\RNCO(V) > \RACO(V) = \alpha_1$, $\Set{V}$ is the only minimizer of the Dilworth truncation problem $\min_{\Pat \in \Pi(V)} \FuHash{\RNCO(V)}[\Pat]$.}
\end{example}

On the other hand, we can also adopt a proper sum-rate adaptation method to solve the non-asymptotic minimum sum-rate problem in the finite linear source model. This idea was originally proposed in \cite{MiloIT2016,CourtIT2014}. The method is to iteratively update $\alpha$, the estimation of the minimum sum-rate $\RNCO(V)$, on an integer set in $\ZP$ until it reaches $\RNCO(V)$. The implementation of this method requires: (a) a method that can check if a sum-rate $\alpha$ is achievable; (b) an algorithm that can determine a rate vector $\rv_V \in B(\FuHashHat{\alpha},\leq) = \Set{\rv_V \in \RRCO(V) \colon r(V) = \alpha} $ if $\alpha$ is achievable. It is fortunate that the CoordSatCap and CoordSatCapFus algorithms can complete both tasks.

\begin{corollary} \label{coro:SumRateAchievNCO}
    For a sum-rate $\alpha$, let $\rv_V$ be the rate vector returned by the CoordSatCapFus, or CoordSatCap, algorithm: $\alpha$ is achievable if and only if $r(V) = \alpha$; If $\alpha$ is achievable, we have $\rv_V \in B(\FuHashHat{\alpha},\leq) = \Set{\rv_V \in \RRCO(V) \colon r(V) = \alpha}$ being an achievable rate vector with sum-rate $\alpha$. \hfill\IEEEQED
\end{corollary}

        \begin{algorithm} [t]
	       \label{algo:SIA}
	       \small
	       \SetAlgoLined
	       \SetKwInOut{Input}{input}\SetKwInOut{Output}{output}
	       \SetKwFor{For}{for}{do}{endfor}
            \SetKwRepeat{Repeat}{repeat}{until}
            \SetKwIF{If}{ElseIf}{Else}{if}{then}{else if}{else}{endif}
	       \BlankLine
           \Input{the ground set $V$, an oracle that returns the value of $H(X)$ for a given $X \subseteq V$}
	       \Output{a rate vector $\rv_V$ in the optimal rate set $\RRNCO(V) = B(\FuHashHat{\RNCO(V)},\leq) \cap \Z^{|V|}$ and $\alpha$ which equals to $\RNCO(V)$ }
	       \BlankLine
            initialize $\alpha$ according to Proposition~\ref{prop:LB}: $\alpha \leftarrow \big\lceil \max_{i \in V} \big\{ \varphi(\Set{\Set{i},V \setminus \Set{i}}), \varphi(\Set{\Set{m} \colon m \in V}) \big\} \big\rceil$\;
            determine a rate vector $\rv_V \in B(\FuHashHat{\alpha},\leq)$ by solving the problem $\min_{\Pat\in \Pi(V)} \FuHash{\alpha}[\Pat]$\;
            \While{$r(V)\neq \alpha$}{
                $\alpha \leftarrow \alpha + 1$\;
                determine a rate vector $\rv_V \in B(\FuHashHat{\alpha},\leq)$ by solving the problem $\min_{\Pat\in \Pi(V)} \FuHash{\alpha}[\Pat]$\;
            }
            return $\rv_V$ and $\alpha$\;
	   \caption{sum-rate increment algorithm (SIA) for solving the non-asymptotic minimum sum-rate problem in the finite linear source model}
	   \end{algorithm}

The proof of Corollary~\ref{coro:SumRateAchievNCO} is in Appendix~\ref{app:ExOutput}. According to Corollary~\ref{coro:SumRateAchievNCO}, we can start with a lower estimation $\alpha$ of $\RNCO(V)$, e.g., the LB in Proposition~\ref{prop:LB}, and increase $\alpha$ by one until it is achievable. The first achievable $\alpha$ necessarily equals to $\RNCO(V)$. Since $\RNCO(V) \in \Set{0,1,\dotsc,H(V)}$, we can adjust $\alpha$ to $\RNCO(V)$ within a finite number of iterations. This idea is implemented by the SIA algorithm in Algorithm~\ref{algo:SIA}, where the Dilworth truncation problem $\min_{\Pat\in \Pi(V)} \FuHash{\alpha}[\Pat]$ in steps 2 and 5 can be solved by the CoordSatCap or CoordSatCapFus algorithm.

\begin{example} \label{ex:auxNCO2}
    We apply the SIA algorithm to the system in Example~\ref{ex:aux}. We initiate $\alpha = \big\lceil \max_{i \in V} \big\{ \varphi(\Set{\Set{i},V \setminus \Set{i}}), \varphi(\Set{\Set{m} \colon m \in V}) \big\} \big\rceil = 6$ according to Proposition~\ref{prop:LB}. We implement the CoordSatCapFus algorithm for solving the Dilworth truncation problem $\min_{\Pat\in \Pi(V)} \FuHash{\alpha}[\Pat]$.
    \begin{itemize}
        \item For $\alpha=6$, we have $\rv_V = (1,0,0,4,0)$ and $\Pat^*=\Set{\Set{4,5},\Set{1},\Set{2},\Set{3}}$ returned by the CoordSatCapFus algorithm. Since $r(V) = 5 < \alpha$, we update $\alpha$ to $7$ and continue the iteration;
        \item For $\alpha=7$, we have $\rv_V = (0,1,1,5,0)$ and $\Pat^*=\Set{\Set{1,2,3,4,5}}$ returned by the CoordSatCapFus algorithm. Since $r(V) = 7 = \alpha$, the iteration terminates.
    \end{itemize}
    At the output, we have $\rv_V = (0,1,1,5,0)\in\RRNCO(V)$, which is consistent with the result in Example~\ref{ex:auxExOutput}. Here, the SIA algorithm solves the non-asymptotic minimum sum-rate problem in the finite linear source model without obtaining the value of $\RACO(V)$.
\end{example}

Note, in the SIA algorithm, the updates of $\alpha$ do not require the minimal/finest minimizer of the Dilworth truncation problem $\min_{\Pat \in \Pi(V)} \FuHash{\alpha}[\Pat]$.\footnote{It means that, when applied to the SIA algorithm, the initiation and updates of $\Pat^*$ in the CoordSatCap algorithm and $\Pat_i^*$ in the CoordSatCapFus algorithm are not required. In addition, the determination of the minimal minimizers, $U^*_{\phi_i}$ and $\hat{X}_{\phi_i}$ in the CoordSatCapFus and CoordSatCap algorithms, respectively, is not required. }
On the other hand, the sum-rate adaptation method is not unique. For solving the non-asymptotic minimum sum-rate problem in CCDE, the authors in \cite{MiloIT2016,CourtIT2014} proposed efficient algorithms to update $\alpha$ to $\RNCO(V)$, where the CoordSatCap algorithm based on Lemma~\ref{lemma:Fusion1} is implemented as a subroutine. 
Since the CoordSatCap and CoordSatCapFus algorithms accomplish the same tasks in Corollary~\ref{coro:SumRateAchievNCO}, we can replace the CoordSatCap algorithm by the CoordSatCapFus algorithm in the sum-rate adaptation algorithms in \cite{MiloIT2016,CourtIT2014}. In the next subsection, we will show the advantage of this replacement: the reduction in complexity.

\subsection{Complexity}\label{sec:Complexity}

Let $\delta$ be the computation complexity of evaluating the value of a submodular function $f \colon 2^V \mapsto \Real$.\footnote{We assume that the value of $f(X)$ for any $X \subseteq V$ can be obtained by an oracle call and $\delta$ refers to the upper bound on the computation time of this oracle call.}
We denote $O(\SFM(|V|))$ the complexity of solving the \textit{submodular function minimization (SFM) problem} $\min\Set{f(X) \colon X \subseteq V}$, which is strongly polynomial \cite{Khachiyan1980Ellipsoid,Grotschel2012,Grotschel1981,Iwata2007SFM,IFF2001,Fujishige2011MinNorm}. For example, the SFM algorithm proposed in \cite{Orlin2009SFM} completes in $O(|V|^5 \cdot \delta + |V|^6)$ time, which is the most efficient SFM algorithm in the literature to the best of our knowledge. Also, the minimizers of an SFM problem form a lattice, where the minimal/smallest and maximal/largest minimizers exist \cite[Lemma 2.1]{Fujishige2005}. It is shown in \cite[Section 7.1]{Fujishige2005} that the minimal and maximal minimizers can be determined by the minimum-norm point in the base polyhedron and the minimum-norm point can be determined by the SFM algorithm in \cite{Fujishige2011MinNorm}. It means that the minimal minimizer of an SFM problem, as required by step 3 in the CoorSatCap algorithm and step 5 in the CoorSatCapFus algorithm, can be determined at the same time when the SFM is solved.

Although the SFM algorithms in \cite{Khachiyan1980Ellipsoid,Grotschel2012,Grotschel1981,Iwata2007SFM,IFF2001,Fujishige2011MinNorm} vary in computation complexity, the exact completion time of an SFM algorithm depends on $|V|$. We call $|V|$ \textit{the size of the SFM problem} $\min\Set{f(X) \colon X \subseteq V}$.
In this section, we study the complexity of the MDA and SIA algorithms proposed in Sections~\ref{sec:algo} in terms of the size of the SFM problem and completion time (in seconds), respectively. It should be noted that, in this paper, we assume that the value of the entropy function $H$ at a given subset $X \subseteq V$ can be evaluated by an oracle call, which takes $X$ as an input and outputs $H(X)$, and $\delta$ refers to the complexity upper bound of this oracle call.

\subsubsection{CoordSatCapFus vs. CoordSatCap}

The main subroutine of the MDA and SIA algorithms is the CoordSatCap or CoordSatCapFus algorithm, and the core part of the CoordSatCap and CoordSatCapFus algorithms is the SFM problem that determines the saturation capacity $\hat{\xi}_{\phi_i}$. Consider the CoordSatCap algorithm where the saturation capacity is determined by
    \begin{equation}\label{eq:CoordSatCapMinVi}
        \min\Set{ \FuHash{\alpha}(X) - r(X) \colon \phi_i \in X \subseteq V_i}.
    \end{equation}
The size of this SFM problem is $|V_i|-1$. The SFM problem $\min\Set{ \FuHash{\alpha}(\tilde{U}) - r(\tilde{U}) \colon \Set{\phi_i} \in U \subseteq \Pat_i^*}$ in the CoordSatCapFus algorithm is over $\Pat_i^*$, a fused user set of $V_i$, where each non-singleton subset $X \in \Pat_i^*$ is treated as a super user that corresponds to one dimension in $\Pat_i^*$. Since $|\Pat_i^*| - 1\leq|V_i| - 1$, the computation complexity of the CoordSatCapFus algorithm is no greater than that of the CoordSatCap algorithm.

\begin{example} \label{ex:auxCompFus}
For $\phi_3 = 2$ in Example~\ref{ex:auxCoordSatCapFus2}, we have $\Pat_3^* = \Set{\Set{4,5},\Set{2}}$, where $\Set{4,5}$ forms one dimension in $\Pat_3^*$, so that the size of the SFM problem $\min\Set{ \FuHash{\alpha}(\tilde{U}) - r(\tilde{U}) \colon \Set{\phi_3} \in U \subseteq \Pat_3^*}$ is $|\Pat_3^*| - 1 = |\Set{\Set{4,5},\Set{2}}| - 1 = 1$. Suppose that we solve the problem $\min\Set{ \FuHash{\alpha}(X) - r(X) \colon \phi_3 \in X \subseteq V_3}$ instead. Then, the size of this SFM problem is $|V_3| - 1 = |\Set{2,4,5}| - 1 = 2$, which is greater than $|\Pat_3^*| - 1$.
\end{example}

However, the complexity of the CoordSatCapFus algorithm in the worst case is the same as that of the CoordSatCap algorithm, which is $O(|V| \cdot \SFM(|V|))$. The worst case is when $\Pat_i^*=\Set{\Set{\phi_1},\dotsc,\Set{\phi_i}}$ for all $i \in V$, which happens when the components in $\RZ{V}$ are mutually independent.

\subsubsection{MDA algorithm}

The MDA algorithm with the CoordSatCapFus being the subroutine completes in $O(|V|^2 \cdot \SFM(|V|))$ time. We remark that $O(|V|^2 \cdot \SFM(|V|))$ is the complexity upper bound for two reasons. On one hand, the complexity of the CoordSatCapFus algorithm is upper bounded by $O(|V| \cdot \SFM(|V|))$; On the other hand, the number of calls of the CoordSatCapFus algorithm in the MDA algorithm is upper bounded by $|V|$. Then, the complexity of solving the non-asymptotic minimum sum-rate problem in the finite linear source model by no more than one additional call of the CoordSatCapFus algorithm, as proposed in Section~\ref{sec:NCO}, is upper bounded by $O((|V|+1) \cdot |V| \cdot \SFM(|V|))$.

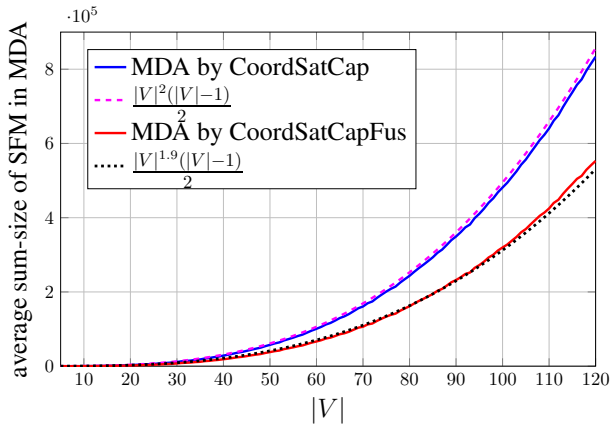
\begin{figure}[tbp]
	\centering
    \scalebox{0.7}{
%
%
\definecolor{mycolor1}{rgb}{1,0,1}%

\begin{tikzpicture}

\begin{axis}[%
width=4in,
height=2.5in,
scale only axis,
xmin=5,
xmax=120,
xmajorgrids,
xlabel={\Large $|V|$},
ymin=0,
ymax=900000,
ymajorgrids,
ylabel={\Large average sum-size of SFM in MDA},
legend style={at={(0.05,0.53)},anchor=south west,draw=black,fill=white,legend cell align=left}
]
\addplot [
line width=1.3pt,
color=blue,
solid
]
table[row sep=crcr]{
5 1.33\\
6 3.9\\
7 21\\
8 69.2\\
9 49.5\\
10 202.5\\
11 281.6\\
12 417.6\\
13 559\\
14 843.5\\
15 1011\\
16 1348.8\\
17 1669.4\\
18 1899\\
19 2299\\
20 2897\\
21 3578.4\\
22 4070\\
23 4853\\
24 5064\\
25 6037.5\\
26 7081.1\\
27 8040.6\\
28 9000.6\\
29 9816.5\\
30 11329.5\\
31 13075.8\\
32 14129.6\\
33 15180\\
34 16991.5\\
35 18179\\
36 20682\\
37 22111.2\\
38 24342.8\\
39 26091\\
40 28494\\
41 30778.7\\
42 33453\\
43 36274.8\\
44 38357\\
45 41238\\
46 44367\\
47 47197.4\\
48 49910.4\\
49 53949\\
50 57425\\
51 61179.6\\
52 63991.2\\
53 69313.4\\
54 72967.5\\
55 76901\\
56 81292.4\\
57 85021.2\\
58 90569.9\\
59 95993\\
60 100710\\
61 105767.9\\
62 112164.2\\
63 116386.2\\
64 122784\\
65 129623\\
66 134273.7\\
67 140619.6\\
68 148335.2\\
69 156078\\
70 160849.5\\
71 169533.8\\
72 174308.4\\
73 184011.1\\
74 192659\\
75 201000\\
76 208407.2\\
77 214491.2\\
78 225299.1\\
79 234551\\
80 243404\\
81 254137.5\\
82 264154.8\\
83 273385.4\\
84 283185\\
85 293216\\
86 305351.6\\
87 313983\\
88 326682.4\\
89 339713\\
90 349393.5\\
91 360924.2\\
92 373989.2\\
93 381681.3\\
94 398818.5\\
95 410495\\
96 424257.6\\
97 440224.8\\
98 452211.2\\
99 468369\\
100 481365\\
101 494081.9\\
102 509643\\
103 524991\\
104 542828\\
105 556605\\
106 574578.3\\
107 591774.2\\
108 609309\\
109 623589\\
110 640524.5\\
111 661493.4\\
112 675942.4\\
113 698317.4\\
114 718542\\
115 737817\\
116 758112.2\\
117 772562.7\\
118 795597.3\\
119 814793\\
120 833568\\
};
\addlegendentry{\Large MDA by CoordSatCap};

\addplot [
line width=1.3pt,
color=mycolor1,
dashed
]
table[row sep=crcr]{
5 50\\
6 90\\
7 147\\
8 224\\
9 324\\
10 450\\
11 605\\
12 792\\
13 1014\\
14 1274\\
15 1575\\
16 1920\\
17 2312\\
18 2754\\
19 3249\\
20 3800\\
21 4410\\
22 5082\\
23 5819\\
24 6624\\
25 7500\\
26 8450\\
27 9477\\
28 10584\\
29 11774\\
30 13050\\
31 14415\\
32 15872\\
33 17424\\
34 19074\\
35 20825\\
36 22680\\
37 24642\\
38 26714\\
39 28899\\
40 31200\\
41 33620\\
42 36162\\
43 38829\\
44 41624\\
45 44550\\
46 47610\\
47 50807\\
48 54144\\
49 57624\\
50 61250\\
51 65025\\
52 68952\\
53 73034\\
54 77274\\
55 81675\\
56 86240\\
57 90972\\
58 95874\\
59 100949\\
60 106200\\
61 111630\\
62 117242\\
63 123039\\
64 129024\\
65 135200\\
66 141570\\
67 148137\\
68 154904\\
69 161874\\
70 169050\\
71 176435\\
72 184032\\
73 191844\\
74 199874\\
75 208125\\
76 216600\\
77 225302\\
78 234234\\
79 243399\\
80 252800\\
81 262440\\
82 272322\\
83 282449\\
84 292824\\
85 303450\\
86 314330\\
87 325467\\
88 336864\\
89 348524\\
90 360450\\
91 372645\\
92 385112\\
93 397854\\
94 410874\\
95 424175\\
96 437760\\
97 451632\\
98 465794\\
99 480249\\
100 495000\\
101 510050\\
102 525402\\
103 541059\\
104 557024\\
105 573300\\
106 589890\\
107 606797\\
108 624024\\
109 641574\\
110 659450\\
111 677655\\
112 696192\\
113 715064\\
114 734274\\
115 753825\\
116 773720\\
117 793962\\
118 814554\\
119 835499\\
120 856800\\
};
\addlegendentry{\Large $\frac{|V|^{2}(|V| - 1)}{2}$};

\addplot [
line width=1.3pt,
color=red,
solid
]
table[row sep=crcr]{
5 1.2\\
6 3.6\\
7 5.1\\
8 34.2\\
9 13.6\\
10 114.7\\
11 160.1\\
12 243\\
13 332\\
14 520.5\\
15 620.6\\
16 835.6\\
17 1045.2\\
18 1169.2\\
19 1416.7\\
20 1815.8\\
21 2293.7\\
22 2587.6\\
23 3128.3\\
24 3168.8\\
25 3831.4\\
26 4558.1\\
27 5183.2\\
28 5817.8\\
29 6269\\
30 7337.5\\
31 8611.7\\
32 9222.4\\
33 9811.7\\
34 11092.5\\
35 11749.4\\
36 13582.4\\
37 14423.4\\
38 15963.4\\
39 17052.2\\
40 18695\\
41 20187.9\\
42 22024.4\\
43 23970.3\\
44 25183.6\\
45 27116\\
46 29239.4\\
47 31049.3\\
48 32705.6\\
49 35576.5\\
50 37877.6\\
51 40422\\
52 42013.6\\
53 45922.9\\
54 48239.9\\
55 50739.6\\
56 53647.4\\
57 55896.5\\
58 59791.7\\
59 63535.1\\
60 66564.6\\
61 69865.4\\
62 74397.4\\
63 76818.4\\
64 81194\\
65 85993\\
66 88627.7\\
67 92830.4\\
68 98291\\
69 103710.8\\
70 106243.7\\
71 112433.8\\
72 114861.2\\
73 121899.7\\
74 127885.2\\
75 133532.6\\
76 138180.8\\
77 141554.1\\
78 149335.1\\
79 155533.6\\
80 161359\\
81 168864.7\\
82 175626.8\\
83 181560.9\\
84 187974.4\\
85 194536.3\\
86 203053.4\\
87 208138.6\\
88 216998\\
89 226174.9\\
90 232034.5\\
91 239545\\
92 248472\\
93 252278.6\\
94 264856.9\\
95 272209.2\\
96 281498.8\\
97 292822.8\\
98 300210.8\\
99 311558.8\\
100 319762.2\\
101 327589.1\\
102 338082\\
103 348300.2\\
104 360795.2\\
105 369357.1\\
106 381741.3\\
107 393362.5\\
108 405216.4\\
109 413814.2\\
110 424876.1\\
111 439775.7\\
112 448261.4\\
113 464269.8\\
114 478133.8\\
115 490957.2\\
116 504697.6\\
117 512604.2\\
118 528785.3\\
119 541159\\
120 552950\\
};
\addlegendentry{\Large MDA by CoordSatCapFus};

\addplot [
line width = 1.5pt,
color=black,
dotted,
]
table[row sep=crcr]{
5 42.5669961260392\\
6 75.2362921870143\\
7 121.006174336977\\
8 181.944536783797\\
9 260.088266010315\\
10 357.447705625927\\
11 476.010032529047\\
12 617.741896388208\\
13 784.591537843957\\
14 978.490524579139\\
15 1201.35519910771\\
16 1455.08790384998\\
17 1741.57803066669\\
18 2062.70292964341\\
19 2420.32870333768\\
20 2816.3109066064\\
21 3252.49516770543\\
22 3730.71774307772\\
23 4252.8060157805\\
24 4820.57894561758\\
25 5435.84747758271\\
26 6100.4149140751\\
27 6816.07725543974\\
28 7584.62351265826\\
29 8407.83599542709\\
30 9287.49057838036\\
31 10225.3569478205\\
32 11223.1988309929\\
33 12282.7742096681\\
34 13405.8355195687\\
35 14594.129836982\\
36 15849.3990537394\\
37 17173.3800416017\\
38 18567.804806971\\
39 20034.4006367469\\
40 21574.8902360547\\
41 23190.9918584986\\
42 24884.4194295223\\
43 26656.8826634027\\
44 28510.0871743505\\
45 30445.7345821456\\
46 32465.5226126952\\
47 34571.1451938656\\
48 36764.2925469112\\
49 39046.6512737893\\
50 41419.9044406323\\
51 43885.7316576179\\
52 46445.8091554652\\
53 49101.8098587614\\
54 51855.4034563097\\
55 54708.2564686733\\
56 57662.0323130781\\
57 60718.3913658229\\
58 63878.9910223369\\
59 67145.4857550123\\
60 70519.5271689314\\
61 74002.7640556007\\
62 77596.8424447942\\
63 81303.405654603\\
64 85124.0943397809\\
65 89060.5465384717\\
66 93114.3977173945\\
67 97287.2808155635\\
68 101580.82628661\\
69 105996.66213977\\
70 110536.413979604\\
71 115201.705044501\\
72 119994.156244017\\
73 124915.386195116\\
74 129967.011257337\\
75 135150.645566954\\
76 140467.90107016\\
77 145920.387555312\\
78 151509.712684296\\
79 157237.482023016\\
80 163105.299071069\\
81 169114.765290631\\
82 175267.48013457\\
83 181565.04107384\\
84 188009.043624162\\
85 194601.081372033\\
86 201342.746000078\\
87 208235.627311778\\
88 215281.313255587\\
89 222481.389948471\\
90 229837.441698876\\
91 237351.051029158\\
92 245023.798697487\\
93 252857.26371924\\
94 260853.023387911\\
95 269012.653295538\\
96 277337.72735268\\
97 285829.81780795\\
98 294490.495267115\\
99 303321.328711784\\
100 312323.885517696\\
101 321499.731472615\\
102 330850.430793853\\
103 340377.546145422\\
104 350082.638654835\\
105 359967.267929564\\
106 370032.992073164\\
107 380281.36770107\\
108 390713.94995609\\
109 401332.29252358\\
110 412137.947646332\\
111 423132.466139173\\
112 434317.397403282\\
113 445694.289440239\\
114 457264.688865805\\
115 469030.140923451\\
116 480992.189497631\\
117 493152.377126819\\
118 505512.245016301\\
119 518073.333050746\\
120 530837.179806543\\
};
\addlegendentry{\Large $\frac{|V|^{1.9}(|V| - 1)}{2}$};

\end{axis}
\end{tikzpicture}%
}
	\caption{The results in Experiment~\ref{exp:CompMDA}. $H(V)$ is fixed to $50$ and the number of users $|V|$ varies from $5$ to $120$. MDA by CoordSatCapFus refers to the MDA algorithm with the CoordSatCapFus algorithm being the subroutine, while MDA by CoordSatCap refers to the MDA algorithm with the CoordSatCap algorithm where the saturation capacity is determined by problem \eqref{eq:CoordSatCapMinVi} being the subroutine. The sum-size of SFM refers to the total size of SFM algorithm in one call of the MDA algorithm. The sum-size averaged over $20$ repetitions is presented in this figure. The average sum-size of SFM in one call of the MDA algorithm is upper bounded by $\frac{|V|^2(|V|-1)}{2}$.}
	\label{fig:ComplexityMDA}
\end{figure}

\begin{experiment}  \label{exp:CompMDA}
    Let $H(V)$ be fixed to $50$ and the number of users $|V|$ vary from $5$ to $120$. For each value of $|V|$, we repeat the following procedure for 20 times:
    \begin{itemize}
        \item randomly generate a finite linear source model with the column vector $\Rz{i}= A_i \Rx$ for all $i \in V$ subject to the condition $l(\Rx) = H(V)$;
        \item solve the asymptotic minimum sum-rate problem by applying the MDA algorithm as follows:
            \begin{itemize}
                \item MDA by CoordSatCap: Algorithm~\ref{algo:MDA}, where the Dilworth truncation problem in steps 2 and 6 is solved by the CoordSatCap algorithm with the saturation capacity determined by \eqref{eq:CoordSatCapMinVi};
                \item MDA by CoordSatCapFus: Algorithm~\ref{algo:MDA}, where the Dilworth truncation problem in steps 2 and 6 is solved by the CoordSatCapFus algorithm.
            \end{itemize}
    \end{itemize}
    We sum up the sizes of the SFM algorithm in each run of the MDA algorithm. This sum-size is averaged over the 20 repetitions and shown in Fig.~\ref{fig:ComplexityMDA}. Note, the average sum-size of the MDA by CoordSatCap is upper bounded by $\frac{|V|^2(|V|-1)}{2}$.\footnote{This refers to the average size of the SFM algorithm in the MDA algorithm. The computation complexity of the MDA algorithm is still $O(|V|^2 \cdot \SFM(|V|))$.}
    It can be shown that there is a reduction from $\frac{|V|^2(|V|-1)}{2}$ to $\frac{|V|^{1.9}(|V|-1)}{2}$ in the average sum-size of the SFM problem by implementing the CoordSatCapFus algorithm. This reduction could be considerable when $|V|$ is large. For example, in Fig.~\ref{fig:ComplexityMDA}, when $|V| = 25$, the average sum-size of SFM is $6037.5$ for MDA by CoordSatCap and $3831.4$ for MDA by CoordSatCapFus.
\end{experiment}

The authors in \cite{MiloDivConq2011} proposed a divide-and-conquer (DC) algorithm for solving the asymptotic minimum sum-rate problem. This algorithm finds the fundamental partition $\Pat^*$ and recursively breaks each element in $\Pat^*$ to singletons by calling the decomposition algorithm (DA) algorithm in \cite[Section 3]{MinAveCost} \cite[Algorithm II]{Narayanan1991PLP}. The DC algorithm completes in $O(|V|^3 \cdot \SFM(|V|))$ time. The detailed description of the DC algorithm is in Appendix~\ref{app:DV}, where we also show that the recursive splitting of the subsets in $\Pat^*$ is not necessary since the asymptotic minimum sum-rate problem can be solved at the same time when the fundamental partition $\Pat^*$ is determined.

\subsubsection{SIA algorithm}

The authors in \cite[Appendix F]{CourtIT2014} \cite[Section III-C]{MiloIT2016} show that the complexity of adapting the estimation $\alpha$ on an integer set to the minimum sum-rate $\RNCO(V)$ in a finite linear source model grows logarithmically in $H(V)$. Therefore, the SIA algorithm completes in $O(\log H(V) \cdot |V| \cdot \SFM(|V|))$ time. To show the actual run-time, or completion time in seconds, of the SIA algorithm, we do the following experiment.

\begin{figure}[tbp]
	\centering
    \scalebox{0.7}{\input{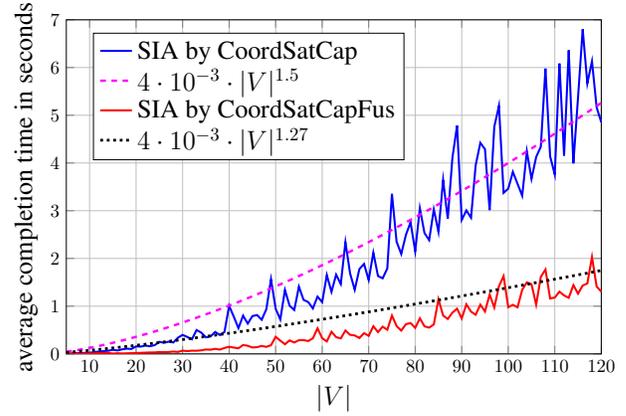}}
	\caption{The results of Experiment~\ref{exp:CompSIA}. $H(V)$ is fixed to $50$ and the number of users $|V|$ varies from $5$ to $120$. SIA by CoordSatCapFus refers to the SIA algorithm with the CoordSatCapFus algorithm being the subroutine, while SIA by CoordSatCap refers to the SIA algorithm with the CoordSatCap algorithm being the subroutine. The run-time in seconds of each call of the SIA algorithm is recorded and averaged over $20$ repetitions. We implemented the minimum norm algorithm in \cite{Fujishige2011MinNorm} as the SFM algorithm. The SIA algorithm is run in MATLAB R2013a.     }
	\label{fig:ComplexitySIA}
\end{figure}

\begin{experiment}  \label{exp:CompSIA}
    Let $H(V)$ be fixed to $50$ and the number of users $|V|$ vary from $5$ to $120$. For each value of $|V|$, we repeat the following procedure for 20 times:
    \begin{itemize}
        \item randomly generate a finite linear source model with the column vector $\Rz{i}= A_i \Rx$ for all $i \in V$ subject to the condition $l(\Rx) = H(V)$;
        \item solve the non-asymptotic minimum sum-rate problem by applying the SIA algorithm as follows:
            \begin{itemize}
                \item SIA by CoordSatCap: Algorithm~\ref{algo:SIA}, where the Dilworth truncation problem in steps 2 and 5 is solved by the CoordSatCap algorithm with the saturation capacity determined by \eqref{eq:CoordSatCapMinVi};
                \item SIA by CoordSatCapFus: Algorithm~\ref{algo:SIA}, where the Dilworth truncation problem in steps 2 and 5 is solved by the CoordSatCapFus algorithm.
            \end{itemize}
    \end{itemize}
    We implement the minimum-norm point algorithm proposed in \cite{Fujishige2011MinNorm} for solving the SFM problems in the CoordSatCap and CoordSatCapFus algorithms. The SIA algorithm is written in MATLAB and run in MATLAB R2013a. We do the experiment on a desktop computer with Intel Core i7-3770 processer, 8Gb RAM and 64-bit Windows 7 Enterprise operating system. The run-time in seconds in each call of the SIA algorithm is recorded and averaged over repetitions. The results are shown in Fig.~\ref{fig:ComplexitySIA}. The run-time of SIA by CoordSatCap is comparable to $4 \cdot 10^{-3} \cdot |V|^{1.5}$. With the fusion method, SIA by CoordSatCapFus reduces it to $4 \cdot 10^{-3} \cdot |V|^{1.27}$.
\end{experiment}

The authors in \cite[Appendix G]{CourtIT2014} show that the SIA by CoordSatCap method based on the minimum-norm algorithm \cite{Fujishige2011MinNorm} completes in $4 \cdot 10^{-3} \cdot |V|^{1.85}$ seconds on average, which is slower than the result in Experiment~\ref{exp:CompSIA}. The main reason that can cause this run-time reduction is that we do the experiment on a dataset and computer that are different from those in \cite[Appendix G]{CourtIT2014}. In addition, the LB that we used in the SIA is tighter than the one in \cite[Appendix F]{CourtIT2014} may be another reason that results in a faster run-time. On the other hand, the complexity of the minimum-norm algorithm is still unknown and may vary with different data processing softwares \cite{Fujishige2011MinNorm}. Therefore, while the average run-time just shows an example on how faster the SIA algorithm completes in practice, the complexity of the SIA algorithm is still $O(\log H(V) \cdot |V| \cdot \SFM(|V|))$, i.e., no matter how good the run-time is, it cannot be used to characterise the complexity of the SIA algorithm. However, based on Figs.~\ref{fig:ComplexityMDA} and \ref{fig:ComplexitySIA}, we can see clearly that the fusion method in CoordSatCapFus algorithm contributes to a considerable reduction in computation complexity when the number of users $|V|$ grows.

It should be noted that finite linear source model is used in Experiments~\ref{exp:CompMDA} and \ref{exp:CompSIA} since the optimality of the output rate vector can be verified by random linear network coding (RLNC) according to \cite[Theorem 6]{SprintRand2010}. The results of the MDA algorithm in Experiment~\ref{exp:CompMDA} is checked by packet-splitting and applying RLNC to the packet chunks.

\section{Minimum Weighted Sum-rate Problem}
\label{sec:MinWeightedSum}

The minimum weighted sum-rate problem, the problem of minimizing a weighted sum-rate in the optimal rate vector set, has been considered in CO in \cite{MiloIT2016,CourtIT2014} for the finite linear source mode. In this section, we show how to solve the minimum weighted sum-rate problem in the asymptotic and non-asymptotic models by choosing a proper linear ordering in the CoordSatCapFus or CoordSatCap algorithm.

Let $\wv_V = (w_i \colon i \in V) \in \RealP^{|V|}$ be a weight vector and $\wv_V^{\intercal} \rv_V = \sum_{i \in V} w_i r_i$ be the weighted sum-rate of $\rv_V$. The minimum weighted sum-rate problem in the asymptotic model and non-asymptotic models are respectively
\begin{equation}
    \begin{aligned}
        &\min\Set{ \wv_V^{\intercal} \rv_V \colon \rv_V \in \RRACO(V)}, \\
        &\min\Set{ \wv_V^{\intercal} \rv_V \colon \rv_V \in \RRNCO(V)}.
    \end{aligned} \nonumber
\end{equation}
We say that $\Phi = (\phi_1,\dotsc,\phi_{|V|})$ is a linear ordering w.r.t. $\wv_V$ if $w_{\phi_1} \leq w_{\phi_2} \leq \dotsc \leq w_{\phi_{|V|}}$. For a given weight vector $\wv_V$, a linear ordering $\Phi$ w.r.t. $\wv_V$ can be chosen, for which we have Theorem~\ref{theo:WeightedSumRate} below. The proof is in Appendix~\ref{app:theo:WeightedSumRate}.

\begin{theorem} \label{theo:WeightedSumRate}
    For a weight vector $\wv_V \in \RealP^{|V|}$, by fixing $\Phi$ to be the linear ordering w.r.t. $\wv_V$ in the CoordSatCapFus or CoordSatCap algorithm, the optimal rate $\rv_V$ returned by the MDA algorithm is the minimizer of $\min \Set{ \wv_V^{\intercal} \rv_V \colon \rv_V \in \RRACO(V) }$. In the finite linear source model and CCDE, the optimal rate $\rv_V$ returned by CoordSatCapFus or CoordSatCap algorithm with input $\alpha = \RNCO(V)$ and linear ordering w.r.t. $\wv_V$ is the minimizer of $ \min \Set{ \wv_V^{\intercal} \rv_V \colon \rv_V \in \RRNCO(V) } $. \hfill\IEEEQED
\end{theorem}

Based on Theorem~\ref{theo:WeightedSumRate}, for a finite linear source model, the minimum weighted sum-rate problem can be solved by the SIA algorithm if we choose a proper linear ordering in the CoordSatCapFus or CoordSatCap algorithm. This result is consistent with the ones in \cite{CourtIT2014,MiloIT2016}. Note, the SIA algorithm adapts the sum-rate by starting with a LB that is tighter than the ones in \cite{CourtIT2014,MiloIT2016}.\footnote{The authors in \cite{CourtIT2014} suggested LB $H(V) - \min_{i \in V} H(\Set{i})$ as the initial guess of the minimum sum-rate, which is shown to be looser than the one proposed in Proposition~\ref{sec:LB}. In \cite[Algorithms 3]{CourtIT2014}, the rate adaptation is done in the region $\Set{0,1,\dotsc,H(V)}$.}

\begin{example} \label{ex:MinWeiSumRate}
    It can be shown that the optimal rate vector $\rv_V = (1,\frac{1}{2},\frac{1}{2},\frac{9}{2},0) \in \RRACO(V)$ determined by the MDA algorithm based on the linear ordering $\Phi= (4,5,2,3,1)$ in Example~\ref{ex:auxMDA} is the minimizer of $\min \Set{ \wv_V^{\intercal} \rv_V \colon \rv_V \in \RRACO(V) }$ where $\wv_V \in \RealP^{|V|}$ could be any weight vector such that $w_{4} \leq w_{5} \leq w_{2} \leq w_{3} \leq w_{1}$, e.g., $\wv_V = (2,0.85,1.1,0.11,0.13)$. The optimal rate vector $\rv_V = (0,1,1,5,0) \in \RRNCO(V)$ based on the linear ordering $\Phi= \Set{4,5,2,3,1}$ in Examples~\ref{ex:auxExOutput} and \ref{ex:auxNCO2} is the minimizer of $\min \Set{ \wv_V^{\intercal} \rv_V \colon \rv_V \in \RRNCO(V) }$ where $\wv_V \in \RealP^{|V|}$ could be any weight vector such that $w_{4} \leq w_{5} \leq w_{2} \leq w_{3} \leq w_{1}$.
\end{example}

Note, for a weight vector $\wv_V \in \RealP^{|V|}$ such that all dimensions $w_i$ are equal, e.g., $\wv_V = \One = (1,\dotsc,1) \in \Real^{|V|}$, the minimum weighted sum-rate problem reduces to the minimum sum-rate problem. In addition, if the problem is just to determine an optimal rate vector in $\RRACO(V)$ or $\RRNCO(V)$, the linear ordering $\Phi$ in the CoordSatCap and CoordSatCapFus algorithms can be arbitrarily chosen.

\section{Fundamental Partition: Minimal Separators}
\label{sec:FundPartMinSep}

The fundamental partition $\Pat^*$ is not only the optimizer for the asymptotic minimum sum-rate problem, but also an essential solution to many problems. In network strength or optimal attack problems \cite{Cunningham1985NetStrength,Chen1994NetStrenght,Kolmogorov2010NetPSP,Baiou2011IBMRepPartition}, the fundamental partition is an optimal way for an attacker to disconnect a network, i.e., decomposing the network into the fundamental partition requires the least effort on breaking the connections/edges between nodes. The authors in \cite{MinAveCost} proposed a novel clustering criterion, which is called minimum average cost (MAC) clustering, based on the submodularity of the similarity measures that is generally used in clustering problems. The objective function in MAC is defined as the clustering cost averaged over the incremental number of clusters. 
Based on the MAC, the authors in \cite{ChanInfoCluster} proposed an information-theoretic clustering (info-clustering) framework where the MMI is used as the similarity measure and the purpose is to search a clustering solution such that the intra-cluster MMI is maximized while the inter-cluster MMI is minimized. In both MAC clustering and info-clustering, the optimal clustering is the fundamental partition $\Pat^*$.
In CO, beyond being the optimizer of the minimum sum-rate problem~\eqref{eq:MinSumRateACOPat}, the fundamental partition $\Pat^*$ has practical interpretation or usefulness in other aspects. In this section, we show that $\Pat^*$ is the set of the minimal separators of a submodular function which makes the estimation of the value of function $\FuHashHat{\RACO(V)}$ and the separable convex minimization problem over $\RRACO(V)$ decomposable.

For a normalized submodular set function $f \colon 2^V \mapsto \Real $, a nonempty proper subset $X \subsetneq V$ is a \textit{separator} of $f$ if $f(X) + f( V \setminus X ) = f(V)$ \cite[Section 3.3]{Fujishige2005}.\footnote{In \cite[Section 3.3]{Fujishige2005}, the
author defined the connectivity of a submodular system that is denoted by two tuple: the power set $2^V$ and the rank function $f \colon 2^V \mapsto \Real$ that is submodular. $(2^V,f)$ is called connected if there does not exist a nonempty subset $X \subsetneq V $ such that $f(X) + f(V \setminus X) = f(V)$. Then, $X$ is a `separator' if a submodular system is disconnected. The name `separator' is also used in \cite[Section 3]{Bilxby1985}. In this paper, without introducing the concept of the submodular system, we define the separator and separability w.r.t. a submodular set function.}
A submodular set function $f$ is called \textit{separable} if there exists a separator of $f$. For each separable submodular set function, there exists a unique set of minimal separators as defined below.

\begin{definition}[minimal separators {\cite[Theorem 3.38]{Fujishige2005}}] \label{def:MinSep}
    For a separable submodular set function $f \colon 2^V \mapsto \Real $, a partition $\Pat\in \Pi(V)$ is the set of minimal separators if, for all $X \in \Pat$, $X$ is a separator and any $X' \subsetneq X$ such that $X' \neq \emptyset$ is not a separator of $f$.
\end{definition}

\begin{theorem} \label{theo:FundPatMinSep}
    $\FuHashHat{\RACO(V)}$ is a separable submodular function and the fundamental partition $\Pat^*$ is the set of minimal separators of $\FuHashHat{\RACO(V)}$. \hfill\IEEEQED
\end{theorem}

The proof of Theorem~\ref{theo:FundPatMinSep} is in Appendix~\ref{app:FundPatMinSep}.

\subsection{Properties of Minimal/Finest Separators}

For any $X, Y \subseteq V$ such that $X \cap Y = \emptyset$, let $\rv_X \oplus \rv_Y = \rv_{X \sqcup Y}$ be the direct sum of $\rv_X$ and $\rv_Y$. For example, for $\rv_{\Set{1,3}} = (r_1,r_3) = (3,0.7)$ and $\rv_{\Set{2,5,6}} = (r_2,r_5,r_6) = (2.4,2,4)$, $\rv_{\Set{1,3}} \oplus \rv_{\Set{2,5,6}} = \rv_{\Set{1,2,3,5,6}} = (3,2.4,0.7,2,4)$.

\begin{lemma}[properties of minimal separators {\cite[Theorems 3.32 and 3.38, Lemma 3.37]{Fujishige2005}}]  \label{lemma:MinSepProp}
    For the fundamental partition $\Pat^*$ as the set of minimal separators of $\FuHashHat{\RACO(V)}$, the followings hold.
    \begin{enumerate}[(a)]
        \item $\FuHashHat{\RACO(V)}(X) = \sum_{C \in \Pat^*} \FuHashHat{\RACO(V)} (X \cap C)$ for all $X \subseteq V$;
        \item The dimension of $B(\FuHashHat{\RACO(V)},\leq)$ is $|V| - |\Pat^*|$ and
                \begin{multline}
                    B(\FuHashHat{\RACO(V)},\leq) = \\ \Set{ \oplus_{C \in \Pat^*} \rv_C \colon  \rv_C \in B(\FuHashHat{\RACO(V),C},\leq), C \in \Pat^* }, \nonumber
                \end{multline}
              where $\FuHashHat{\RACO(V),C}$ is the Dilworth truncation of $\FuHash{\RACO(V),C}$, the reduction of $\FuHash{\RACO(V)}$ on $C$;
        \item Let $\rv_V$ be any rate vector in $B(\FuHashHat{\RACO(V)},\leq)$. For any $C,C' \in \Pat^*$ such that $C \neq C'$,
              $$ \rv_V + \epsilon(\chi_i-\chi_j) \notin B(\FuHashHat{\RACO(V)},\leq) $$
              for all $\epsilon>0$, $i \in C$ and $j \in C'$. \hfill\IEEEQED
    \end{enumerate}
\end{lemma}

Based on property (a) in Lemma~\ref{lemma:MinSepProp}, by using the fundamental partition $\Pat^*$, we can break the task of evaluating the value of $\FuHashHat{\RACO(V)}$ at any subset $X \subseteq V$ into subtasks: get the values of $\FuHashHat{\RACO(V)}$ at $C \cap X$ for all $C \in \Pat^*$ and sum them up.
Here, each value of $\FuHashHat{\RACO(V)}( C \cap X )$ can be obtained by applying the CoordSatCap or CoordSatCapFus algorithm. By doing so, the complexity of evaluating $\FuHashHat{\RACO(V)}(X)$ is reduced from $O(|X| \cdot \SFM(|X|))$ to $O(\eta \cdot \SFM(\eta))$ where $\eta = \max \Set{ |C \cap X| \colon C \in \Pat^*}$. Property (b) means that a separable submodular function results in a separable base polyhedron, which gives rise to property (c) \cite[Lemma 3.41]{Fujishige2005}. Property (c) is an important result in CO in that it makes the separable convex minimization problem over the optimal rate vector set $\RRACO(V)$ decomposable. In the following context, we first show the examples of properties (a) and (b) and then discuss the decomposability of the separable convex minimization problem based on property (c).

\begin{example} \label{ex:MinSep}
    For the system in Example~\ref{ex:aux}, we know that the minimum sum-rate $\RACO(V) = \frac{13}{2}$ and the fundamental partition $\Pat^* = \Set{\Set{1,4,5},\Set{2},\Set{3}}$ for the asymptotic model by the MDA algorithm. Consider the value of the Dilworth truncation function $\FuHashHat{\RACO(V)}$ at $X = \Set{1,2,3,5}$. Based on property (a) in Lemma~\ref{lemma:MinSepProp}, we have
    \begin{equation}
        \begin{aligned}
            \FuHashHat{13/2}(\Set{1,2,3,5}) & = \sum_{C \in \Pat^*} \FuHashHat{13/2}(\Set{1,2,3,5} \cap C) \\
                                                    & = \FuHashHat{13/2}(\Set{1,5}) + \FuHashHat{13/2}(\Set{2}) + \FuHashHat{13/2}(\Set{3}) \\
                                                    & = 4 + \frac{1}{2} + \frac{1}{2}  \\
                                                    & = 5.
        \end{aligned}  \nonumber
    \end{equation}
    Here, the value of $\FuHashHat{\frac{13}{2}}(\Set{1,2,3,5})$ can be obtained in $O(4 \cdot \SFM(4))$ time, while the value of $\sum_{C \in \Pat^*} \FuHashHat{\frac{13}{2}}(\Set{1,2,3,5} \cap C)$ can be obtained in $O(2 \cdot \SFM(2))$ time. Consider the base polyhedron $B(\FuHashHat{13/2},\leq)$ for the the system in Example~\ref{ex:aux}. According to property (b) in Lemma~\ref{lemma:MinSepProp}, the dimension of $B(\FuHashHat{13/2},\leq)$ is $|V|- |\Pat^*| = 2$.

    We can visualize property (b) in Lemma~\ref{lemma:MinSepProp} via Figs.~\ref{fig:CoordSatCapFusDemo1} and \ref{fig:CoordSatCapFusDemo2}.
    In Fig.~\ref{fig:CoordSatCapFusDemo1}, when $\alpha = \frac{23}{4}$ and $C = \Set{1,4,5}$, $\FuHashHat{\alpha,C}$ is separable with the minimal separator set being $\Pat^*=\Set{\Set{1},\Set{4,5}}$ so that the dimension of $B(\FuHashHat{\alpha,C},\leq)$ is $|C| - |\Pat^*| = 1$, i.e., $B(\FuHashHat{\alpha,C},\leq)$ is a 1-dimension line segment, and $\rv_{\Set{4,5}} \oplus r_1 \in B(\FuHashHat{23/4,C},\leq)$ for $r_1 = \frac{3}{4}$ and all $\rv_{\Set{4,5}} \in B(\FuHashHat{23/4,\Set{4,5}},\leq)$.
    In Fig.~\ref{fig:CoordSatCapFusDemo2}, when $\alpha = \frac{13}{2}$ and $C = \Set{1,4,5}$, $\FuHashHat{\alpha,C}$ is nonseparable so that the dimension of $B(\FuHashHat{13/2,C},\leq)$ is $|C| - 1 = 2$, i.e., $B(\FuHashHat{\alpha,C},\leq)$ is a 2-dimension polygon on the plane $\Set{\rv_C \in \Real^{|C|}: r(C) = \FuHashHat{13/2}(C) = \frac{11}{2}}$.\footnote{A nonseparable function $f \colon 2^V \mapsto \R$ can be considered as a separable function with the minimal separator set being $\Pat^* = \Set{V}$ so that, according to property (b) in Lemma~\ref{lemma:MinSepProp}, the dimension of $B(f,\leq)$ is $|V| - 1$. Also, for a nonseparable function $f$, there does not exist $\Pat \in \Pi'(V)$ such that $f(X) = \sum_{C \in \Pat} f (X \cap C), \forall X \subseteq V$ or   $B(f,\leq) = \Set{ \oplus_{C \in \Pat} \rv_C \colon  \rv_C \in B(f,\leq), C \in \Pat }$. The latter means that if we determine $\rv_C \in B(f_C,\leq)$ for all $C \in \Pat$, the direct sum $\rv_V = \oplus_{C \in \Pat}\rv_C$ does not necessarily belong to $B(f,\leq)$.}
\end{example}

\subsection{Separable Convex Function Minimization over $\RRACO(V)$}

We call $g \colon \Real^{|V|} \mapsto \Real$ a \textit{separable convex function} if $g(\rv_V) = \sum_{i \in V} g_i(r_i)$ where $g_i \colon \Real \mapsto \Real$ is convex for all $i \in V$. For the minimization problem $\min\Set{g(\rv_V) \colon \rv_V \in \RRACO(V)}$ where $g$ is a separable convex function, the local optimality w.r.t. the elementary transform $\chi_i - \chi_j$ implies global optimality.

\begin{theorem}[{\cite[Theorem 20.3]{Fujishige2005}}]\label{theo:LocToGlobMConvex}
    For a separable convex function $g$, $\rv_V^*$ is the minimizer of $\min\Set{g(\rv_V) \colon \rv_V \in \RRACO(V)}$ if and only if, for all $i,j \in V$ and $\epsilon > 0$ such that $\rv_V^* + \epsilon (\chi_i - \chi_j) \in \RRACO(V)$,
    $$ g(\rv_V^*) \leq g(\rv_V^* + \epsilon (\chi_i - \chi_j)). \IEEEQEDhereeqn$$
\end{theorem}

\begin{corollary}\label{coro:MinSepMConvex}
    For all $C \in \Pat^*$, let $\rv_C^*$ be the minimizer of the separable convex minimization problem $\min\Set{g(\rv_C) \colon \rv_C \in B(\FuHashHat{\RACO(V),C},\leq)}$. $\rv_V^* = \oplus_{C\in\Pat^*} \rv_C^*$ is the minimizer of $\min\Set{g(\rv_V) \colon \rv_V \in \RRACO(V)}$. \hfill\IEEEQED
\end{corollary}

In Corollary~\ref{coro:MinSepMConvex}, $B(\FuHashHat{\RACO(V),C},\leq)$ is the projection of the optimal rate vector set $\RRACO(V) = B(\FuHashHat{\RACO(V)},\leq)$ on the subset $C$ and $\rv_C$ is the projection of some optimal rate vector $\rv_V \in \RRACO(V)$ on $C$. The proof of Corollary~\ref{coro:MinSepMConvex} is in Appendix~\ref{app:FundPatMinSep} based on property (c) in Lemma~\ref{lemma:MinSepProp} and Theorem~\ref{theo:LocToGlobMConvex}. Based on Corollary~\ref{coro:MinSepMConvex}, the minimization problem $\min\Set{g(\rv_V) \colon \rv_V \in \RRACO(V)}$ can be divided to $|\Pat^*|$ minimization problems
\begin{equation} \label{eq:SubProSep}
    \min\Set{g(\rv_C) \colon \rv_C \in B(\FuHashHat{\RACO(V),C},\leq)},
\end{equation}
each of which has a lower dimension than the original one. On the other hand, there exist many algorithms in the literature that efficiently solve the minimization problem \eqref{eq:SubProSep}, e.g., the algorithms in \cite{Satoko2011MConvex,Nagano2012Lex}. We show an example of Corollary~\ref{coro:MinSepMConvex} below, where $g$ is a quadratic function.

\begin{example} \label{ex:auxLexOpt}
    For the system in Example~\ref{ex:aux}, consider the quadratic programming $\min\Set{ \sum_{i \in V} \frac{r_i^2}{w_i} \colon \rv_V \in \RRACO(V)}$ where $\wv_V \in \RealP^{|V|}$ is a weight vector.\footnote{This quadratic programming problem is also called the resource allocation problem under submodular constraints in \cite{Ibaraki1988,Satoko2004}.}
    The objective function $\sum_{i \in V} \frac{r_i^2}{w_i}$ is separable convex. For this system, we have $\RACO(V) = \frac{13}{2}$ and $\Pat^*=\Set{\Set{1,4,5},\Set{2},\Set{3}}$.

    According to Corollary~\ref{coro:MinSepMConvex}, we can determine the minimizers of
    \begin{equation} \label{eq:FairC}
        \min\Set{\sum_{i \in C} \frac{r_i^2}{w_i} \colon \rv_C \in B(\FuHashHat{\RACO(V),C},\leq)}
    \end{equation}
    for all $C \in \Pat^*$ and combine the results by obtaining the direct sum of them. In fact, we just need to solve the problem \eqref{eq:FairC} for $C = \Set{1,4,5}$ since both $B(\FuHashHat{\RACO(V),\Set{2}})$ and $B(\FuHashHat{\RACO(V),\Set{3}})$ are singletons that only contain $r_2 = \frac{1}{2}$ and $r_3 = \frac{1}{2}$, respectively. For problem \eqref{eq:FairC} when $C = \Set{1,4,5}$, it can be shown that $\rv_{\Set{1,4,5}}^* = (\frac{3}{2},2,2)$ when $\wv_{\Set{1,4,5}} = (1,1,1)$ and $\rv_{\Set{1,4,5}}^* = (1,2,\frac{5}{2})$ when $\wv_{\Set{1,4,5}} = (1,3,4)$. Therefore, the minimizer of $\min\Set{ \sum_{i \in V} \frac{r_i^2}{w_i} \colon \rv_V \in \RRACO(V)}$ is $\rv_V^* = (\frac{3}{2},\frac{1}{2},\frac{1}{2},2,2)$ when $\wv_V = \One$ and $\rv_{V}^* = (1,\frac{1}{2},\frac{1}{2},2,\frac{5}{2})$ when $\wv_V = (1,2,5,3,4)$.

    It is shown in \cite{Ding2016ISIT} that the problem $\min\Set{ \sum_{i \in V} \frac{r_i^2}{w_i} \colon \rv_V \in \RRACO(V)}$ can be solved in $O(|V|^2\cdot\SFM(|V|))$ time, where $|V| = 5$ for the system in Example~\ref{ex:aux}. But, if we compute the minimizer of \eqref{eq:FairC} for each subset $C$ in the fundamental partition $\Pat^*$, the problem can be solved in $O(\eta^2\cdot\SFM(\eta))$ time, where $\eta = \max\Set{|C| \colon C \in \Pat^*} =3$. Therefore, the separate computation of the minimizer of $\min\Set{g(\rv_C) \colon \rv_C \in \RRACO(V)}$ for all $ C \in \Pat^*$ based on Corollary~\ref{coro:MinSepMConvex} reduces the computation complexity.
\end{example}

In Example~\ref{ex:auxLexOpt}, the minimzer of the quadratic programming problem $\min\Set{ \sum_{i \in V} \frac{r_i^2}{w_i} \colon \rv_V \in \RRACO(V)}$ is called the lexicographical optimizer in \cite{Fujilshige1980Lex} since it lexicographically dominates any other rate vectors in the submodular base polyhedron $B(\FuHashHat{\RACO(V)},\leq) = \RRACO(V)$.\footnote{We refer the reader to \cite{Fujilshige1980Lex,Fujishige1984} for the detailed definition and explanation of the lexicographical domination and its properties in the submodular base polyhedron.}
It is also the optimizer of many other optimization problems in $\RRACO(V)$\cite{Nagano2012Lex,Nagano2013}, e.g., $\min\Set{ \sum_{i \in V} e^{r_i+w_i} \colon \rv_V \in \RRACO(V) }$, $\max\Set{ \sum_{i \in V} w_i \ln r_i \colon \rv_V \in \RRACO(V) }$ and $\min\Set{ \sum_{i \in V} r_i \log \frac{r_i}{w_i} \colon \rv_V \in \RRACO(V) }$. The authors in \cite{MiloFair2012} proposed an integral rate incremental method for solving the fairness problem $\min\Set{ \sum_{i \in V} r_i \log r_i \colon \rv_V \in \RRNCO(V) }$, where the objective function is equivalent to $\sum_{i \in V} r_i \log \frac{r_i}{w_i}$ when $\wv = \One$. However, this method is not able to provide a solution to $\min\Set{ \sum_{i \in V} r_i \log r_i \colon \rv_V \in \RRACO(V) }$ for the asymptotic model, since the step size of each increment is uncertain when the rates are real numbers. 

\section{Conclusion}

We proposed the MDA and SIA algorithms for searching an optimal rate vector that attains omniscience with the minimum sum-rate in the asymptotic and non-asymptotic models, respectively. We also proposed a fusion method to solve the Dilworth truncation problem, a subroutine in the MDA and SIA algorithms, and ran experiment to show that this fusion method contributes to a reduction in computational complexity.
We showed that the minimum weighted sum-rate problem in both asymptotic and non-asymptotic models can be solved by choosing a proper linear ordering in the MDA and SIA algorithms, respectively.
We proved the existence of a fractional optimal rate vector in the finite linear source model which can be implemented by $(|\Pat^*| - 1)$-packet-splitting in CCDE.
In addition, we revealed the decomposition property of the fundamental partition $\Pat^*$ in the asymptotic model, where we showed that the tasks of evaluating the Dilworth truncation function and minimizing a separable parametric convex function over the optimal rate vector set can be decomposed into subtasks so that the overall complexity is reduced.

To solve a CO problem in practice, there still remains one problem: what to send in each transmission.
It is shown in \cite{MiloIT2016} that, for an optimal transmission rate vector, the coding scheme for attaining omniscience in the finite linear source model can be designed based on a simultaneous matrix completion algorithm \cite{Harvey2005}, which completes in $O(|V|^4 \cdot \gamma \cdot \log(|V| \cdot H(V)))$ time with $\gamma$ denoting the complexity of the matrix rank function. But, it is still worth discussing if there exists other less complex algorithms for the coding design.
%

\appendices

\section{Principal Sequence of Partitions (PSP)} \label{app:PSP}

We define the pairwise relationship between two partitions in $\Pi(V)$ as follows.

\begin{definition}[order $\preceq$]  \label{def:POOrder}
    For $\Pat,\Pat' \in \Pi(V)$, we denote
    \begin{itemize}
        \item $\Pat\preceq\Pat'$ if, for all $ X \in \Pat$, $\exists X' \in \Pat'$ such that $X \subseteq X'$;
        \item $\Pat = \Pat'$ if $\Pat\preceq\Pat'$  and $\Pat\succeq\Pat'$;
        \item $\Pat\prec\Pat'$ if $\Pat\preceq\Pat'$ and $\Pat\neq\Pat'$.
    \end{itemize}
\end{definition}

In other words, $\Pat\preceq\Pat'$ if $\Pat$ is finer than $\Pat'$ and $\Pat\prec\Pat'$ if $\Pat$ is strictly finer than $\Pat'$. For example, for $\Pat=\Set{\Set{1,2},\Set{3},\Set{4}}$ and $\Pat'=\Set{\Set{1,2,3},\Set{4}}$, we have $\Pat\preceq\Pat'$. In fact, $\Pat\prec\Pat'$.

\begin{theorem}[PSP {\cite[Sections 2.2 and 3]{MinAveCost}\footnote{The authors in \cite{MinAveCost} discuss the problem that is called $\beta$-minimum average clustering ($\beta$-MAC). It can be shown that the minimum sum-rate problem considered in this paper is a $\beta$-MAC problem when $\beta = 1$.}, \cite[Definition 3.8]{Narayanan1991PLP}}] \label{theo:PSP}
    $\FuHashHat{\alpha}(V)=\min_{\Pat\in\Pi(V)} \FuHash{\alpha} [\Pat]$ is a piecewise linear nondecreasing curve in $\alpha$ with $p \leq |V| - 1$ critical/turning points
    $$ H(V) = \alpha_0 > \alpha_1 > \alpha_2 > \dotsc > \alpha_{p} \geq 0$$
    that have the following properties.
    \begin{enumerate}[(a)]
        \item Denote $\Pat_j$ the finest/minimal minimizers of the Dilworth truncation problem $\min_{\Pat\in\Pi(V)} \FuHash{\alpha_j} [\Pat]$.
            All $\Pat_j$s form a partition chain/sequence $\Chain_{\Pat}$:
            $$ \Set{V} = \Pat_0 \succ \Pat_1 \succ \dotsc \succ \Pat_p = \Set{\Set{i} \colon i\in V}. $$
            If $\alpha_{j} > \alpha > \alpha_{j+1}$ for some $j \in \Set{0,\dotsc,p-1}$, the minimizer of $\min_{\Pat\in\Pi(V)} \FuHash{\alpha} [\Pat]$ is uniquely $\Pat_j$; If $\alpha < \alpha_p$, the minimizer is uniquely $\Pat_p=\Set{\Set{i} \colon i \in V}$;
        \item The gradient of $\FuHashHat{\alpha}(V)$ is decreasing in $\alpha$: The gradient is $|\Pat_p|=|V|$ initially; It changes to $|\Pat_{j-1}|$ after each critical value $\alpha_j$ and finally decreases to $1$ after $\alpha_1$. \hfill\IEEEQED
    \end{enumerate}
\end{theorem}

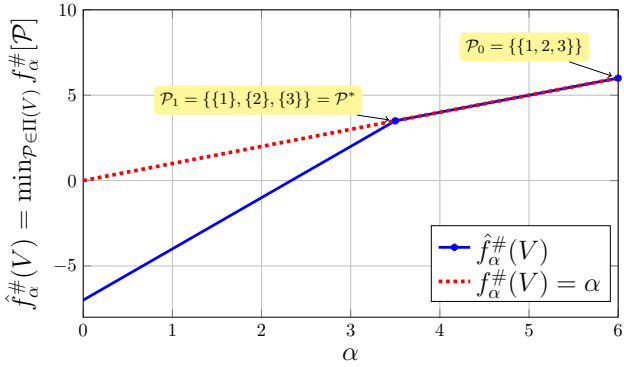
\begin{figure}[tbp]
	\centering
    \scalebox{0.7}{
%
%
\definecolor{mycolor1}{rgb}{1,0,1}%

\begin{tikzpicture}[
every pin/.style={fill=yellow!50!white,rectangle,rounded corners=3pt,font=\tiny},
every pin edge/.style={<-}]

\begin{axis}[%
width=4in,
height=2.3in,
scale only axis,
xmin=0,
xmax=6,
xlabel={\Large $\alpha$},
xmajorgrids,
ymin=-8,
ymax=10,
ylabel={\Large $\FuHashHat{\alpha}(V)=\min_{\Pat \in \Pi(V)} \FuHash{\alpha}[\Pat]$},
ymajorgrids,
legend style={at={(0.65,0.05)},anchor=south west,draw=black,fill=white,legend cell align=left}
]
\addplot [
color=blue,
solid,
line width=1.5pt,
mark=asterisk,
mark options={solid}
]
table[row sep=crcr]{
-1 -10\\
3.5 3.5\\
6 6\\
};
\addlegendentry{\Large $\FuHashHat{\alpha}(V)$};

\addplot [
color=red,
dotted,
line width=2pt,
mark options={solid}
]
table[row sep=crcr]{
0 0\\
1 1\\
2 2\\
3 3\\
4 4\\
5 5\\
6 6\\
};
\addlegendentry{\Large $\FuHash{\alpha}(V)=\alpha$};


\node at (axis cs:3.5,3.5) [pin={[pin distance = 5mm]170:\textcolor{black}{\small $\Pat_1=\Set{\Set{1},\Set{2},\Set{3}}=\Pat^*$}}] {};
\node at (axis cs:6,6) [pin={150:\textcolor{black}{\small $\Pat_0=\Set{\Set{1,2,3}}$}}] {};

\end{axis}
\end{tikzpicture}%
}
	\caption{The value of $\FuHashHat{\alpha}(V)$ as a function of $\alpha$ for the system in Example~\ref{ex:main}. $\alpha_1=\RACO(V)=\frac{7}{2}$ and $\Pat_1=\Pat^*=\Set{\Set{1},\Set{2},\Set{3}}$ according to Corollary~\ref{coro:FundPartPSP}. }
	\label{fig:PSPmain}
\end{figure}

\begin{corollary} \label{coro:FundPartPSP}
    $\alpha_1=\RACO(V)$ and $\Pat_1=\Pat^*$, i.e., the parameters in PSP that correspond to the first critical point $\alpha_1$ constitute the solutions to the asymptotic minimum sum-rate problem.
\end{corollary}
\begin{IEEEproof}
    According to Theorem~\ref{theo:NonEmptBase}, the base polyhedron $B(\FuHash{\alpha},\leq)$ is nonempty if and only if $\alpha = \FuHashHat{\alpha}(V)$. In other words, $B(\FuHash{\alpha},\leq)\neq\emptyset$ if and only if the value of $\alpha$ falls in the segment of the piecewise linear $\FuHashHat{\alpha}(V)$ vs. $\alpha$ curve where the $\FuHashHat{\alpha}(V)$ and $\FuHash{\alpha}(V) = \alpha$ overlap, which, based on property (b) in Theorem~\ref{theo:PSP}, is when $\alpha \geq \alpha_1$. Then, the minimum sum-rate $\RACO(V)$ is the smallest value of $\alpha$ such that $B(\FuHash{\alpha},\leq)\neq\emptyset$, which is $\alpha_1$. All maximizers of \eqref{eq:MinSumRateACOPat} and $\Set{V}$ constitute the set of all minimizers of $\min_{\Pat \in \Pi(V)} \FuHash{\alpha_1}[\Pat]$. So, the minimal minimizer $\Pat_1$ is the minimal maximizer of \eqref{eq:MinSumRateACOPat}, i.e., $\Pat_1$ equals to the fundamental partition $\Pat^*$.
\end{IEEEproof}

The proof of Corollary~\ref{coro:FundPartPSP} is exemplified below.

\begin{figure}[tbp]
	\centering
    \scalebox{0.7}{
%
%
\definecolor{mycolor1}{rgb}{1,0,1}%

\begin{tikzpicture}[
every pin/.style={fill=yellow!50!white,rectangle,rounded corners=3pt,font=\tiny},
every pin edge/.style={<-}]

\begin{axis}[%
width=4in,
height=2.3in,
scale only axis,
xmin=0,
xmax=10,
xlabel={\Large $\alpha$},
xmajorgrids,
ymin=-25,
ymax=17,
ylabel={\Large $\FuHashHat{\alpha}(V)=\min_{\Pat \in \Pi(V)} \FuHash{\alpha}[\Pat]$},
ymajorgrids,
legend style={at={(0.65,0.05)},anchor=south west,draw=black,fill=white,legend cell align=left}
]
\addplot [
color=blue,
solid,
line width=1.5pt,
mark=asterisk,
mark options={solid}
]
table[row sep=crcr]{
-1 -28\\
4 -3\\
6 5\\
6.5 6.5\\
10 10\\
};
\addlegendentry{\Large $\FuHashHat{\alpha}(V)$};

\addplot [
color=red,
dotted,
line width=2pt,
mark options={solid}
]
table[row sep=crcr]{
0 0\\
1 1\\
2 2\\
3 3\\
4 4\\
5 5\\
6 6\\
7 7\\
8 8\\
9 9\\
10 10\\
};
\addlegendentry{\Large $\FuHash{\alpha}(V)=\alpha$};


\node at (axis cs:4,-3) [pin={[pin distance = 1mm]-80:\textcolor{black}{\small $\Pat_3=\Set{\Set{1},\Set{2},\Set{3},\Set{4},\Set{5}}$}}] {};
\node at (axis cs:6,5) [pin={[pin distance = 1mm]-80:\textcolor{black}{\small $\Pat_2=\Set{\Set{4,5},\Set{1},\Set{2},\Set{3}}$}}] {};
\node at (axis cs:6.5,6.5) [pin={[pin distance = 5mm]170:\textcolor{black}{\small $\Pat_1=\Set{\Set{1,4,5},\Set{2},\Set{3}}=\Pat^*$}}] {};
\node at (axis cs:10,10) [pin={150:\textcolor{black}{\small $\Pat_0=\Set{\Set{1,2,3,4,5}}$}}] {};

\end{axis}
\end{tikzpicture}%
}
	\caption{The value of $\FuHashHat{\alpha}(V)$ as a function of $\alpha$ for the system in Example~\ref{ex:aux}. $\alpha_1=\RACO(V)=\frac{13}{2}$ and $\Pat_1=\Pat^*=\Set{\Set{1,4,5},\Set{2},\Set{3}}$ according to Corollary~\ref{coro:FundPartPSP}. }
	\label{fig:PSPaux}
\end{figure}
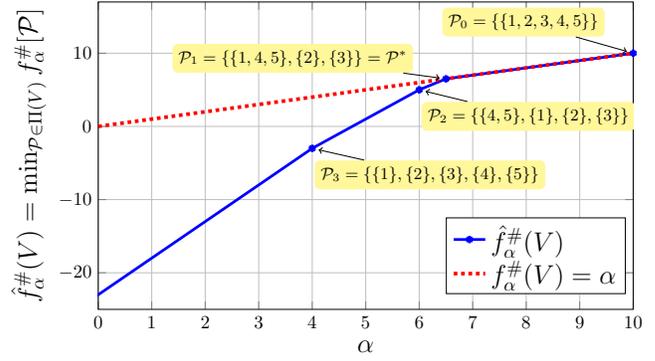

\begin{example}\label{ex:auxPSP}
    We show the plot $\FuHashHat{\alpha}(V)$ in $\alpha$ for the systems in Examples~\ref{ex:main} and \ref{ex:aux} in Figs.~\ref{fig:PSPmain} and \ref{fig:PSPaux}, respectively. It can be seen from both figures that $\FuHashHat{\alpha}(V)$ is an increasing piecewise linear function in $\alpha$. We discuss Fig.~\ref{fig:PSPaux} based on Theorem~\ref{theo:PSP} and Corollary~\ref{coro:FundPartPSP} as follows.

    In addition to $\alpha_0$ and $\Pat_0$, there are three critical points $\alpha_j$ with $\Pat_j$, the minimal minimizers of $\min_{\Pat\in\Pi(V)}\FuHash{\alpha_j}[\Pat]$, being
    \begin{equation}
        \begin{aligned}
            & \alpha_0 = 10, & \Pat_0 & = \Set{\Set{1,2,3,4,5}};\\
            & \alpha_1 = \frac{13}{2}, & \Pat_1 &= \Set{\Set{1,4,5},\Set{2},\Set{3}}; \\
            & \alpha_2 = 6, & \Pat_2 &= \Set{\Set{4,5},\Set{1},\Set{2},\Set{3}} ; \\
            & \alpha_3 = 4, & \Pat_3 &=  \Set{\Set{1},\Set{2},\Set{3},\Set{4},\Set{5}}.
        \end{aligned}\nonumber
    \end{equation}
    We have the partition sequence $\Chain_{\Pat}$: $ \Pat_0 \succ \Pat_1 \succ \Pat_2 \succ \Pat_3$.
    The gradient is: $5$ when $\alpha\in[0,\alpha_3]$; $4$ when $\alpha\in[\alpha_3,\alpha_2]$; $3$ when $\alpha\in[\alpha_2,\alpha_1]$; $1$ when $\alpha\in[\alpha_1,\alpha_0]$. In addition, one can show that the minimizer of $\min_{\Pat\in\Pi(V)}\FuHash{\alpha}[\Pat]$ is uniquely $\Pat_3$ when $\alpha\in[0,\alpha_3)$, $\Pat_2$ when $\alpha\in(\alpha_3,\alpha_2)$, $\Pat_1$ when $\alpha\in(\alpha_2,\alpha_1)$ and $\Pat_0$ when $\alpha\in(\alpha_1,\alpha_0]$. Here, $\alpha_1$ and $\Pat_1$ coincide with the minimum sum-rate $\RACO(V)=\frac{13}{2}$ and the fundamental partition $\Pat^*=\Set{\Set{4,5},\Set{1},\Set{2},\Set{3}}$ in Example~\ref{ex:aux}, respectively.

    In Fig.~\ref{fig:PSPaux}, we also plot the line $\FuHash{\alpha}(V)=\alpha$. It can be seen that $\FuHash{\alpha}(V)$ overlaps with $\FuHashHat{\alpha}(V)$, i.e., $\alpha=\FuHashHat{\alpha}(V)$, when $\alpha \in [\alpha_1,\alpha_0]$. So, $B(\FuHash{\alpha},\leq) \neq \emptyset$ when $\alpha \geq \alpha_1$. According to Theorem~\ref{theo:NonEmptBase}, $\alpha_1=\frac{13}{2}$, the minimal value of $\alpha$ in the region $[\alpha_1,\alpha_0]$, is the minimum sum-rate $\RACO(V)$.

     Consider the region when $\alpha < \alpha_1 = \frac{13}{2}$ in Fig.~\ref{fig:PSPaux}. We have $\FuHash{\alpha}(V)=\alpha > \FuHashHat{\alpha}(V)$. As discussed in Section~\ref{sec:MinSumRate}, the polyhedron $P(\FuHash{\alpha},\leq)$ does not intersect with the hyperplane $\Set{\rv_V \in \Real^{|V|} \colon r(V) = \alpha}$, i.e., $B(\FuHash{\alpha},\leq)=\emptyset$, in this region. It means that the minimum sum-rate $\alpha$ is too low for attaining the omniscience in $V$. On the contrary, when $\alpha \geq \alpha_1 = \frac{13}{2}$, $P(\FuHash{\alpha},\leq)$ intersects with the hyperplane $\Set{\rv_V \in \Real^{|V|} \colon r(V) = \alpha}$, i.e., $B(\FuHash{\alpha},\leq)=\emptyset$. It can be seen that the PSP provides another interpretation of Theorem~\ref{theo:NonEmptBase}.
\end{example}

\subsection{Proof of Theorem~\ref{theo:MDAOpt}} \label{app:DAtoMDA}

We have the following properties for $\varphi$ and the partitions $\Pat_j$s in the PSP.

\begin{lemma}[{property of $\varphi(\Pat)$ \cite[Theorem 3.14]{Narayanan1991PLP}, \cite[Section 3]{MinAveCost}}]\label{lemma:PropPSP}
    The followings hold for $\varphi(\Pat)$ and $\Pat_j \in \Chain_{\Pat}$ in the PSP for $j\in\Set{1,\dotsc,p}$.
    \begin{enumerate}[(a)]
        \item $\alpha_1 = \varphi(\Pat_1)$;
        \item For any $j$ such that $1 < j \leq p $, let $\alpha = \varphi(\Pat_{j})$. Then, $\alpha_{j}<\alpha<\alpha_{1}$.
    \end{enumerate}
\end{lemma}

Based on Lemma~\ref{lemma:PropPSP} and Theorem~\ref{theo:PSP}, consider two situations for a partition $\Pat_{j} \in \Chain_{\Pat}$ where $j\in\Set{1,\dotsc,p}$:
\begin{itemize}
    \item If $j=1$, then $\alpha_1 = \varphi(\Pat_j)$;
    \item If $j>1$, then $\alpha = \varphi(\Pat_j)$ satisfies $\alpha_j < \alpha < \alpha_1$. Let $\Pat_{j'}$ be the minimal minimizer of $\min_{\Pat \in \Pi(V)}\FuHash{\alpha}[\Pat]$. Then, $\Pat_1 \succ \Pat_{j'} \succ \Pat_{j}$.
\end{itemize}
They suggest a recursive method for determining $\Pat_1$ and $\alpha_1$. Consider the iteration
$$ \alpha^{(n+1)} = \varphi(\Pat^{(n)}), $$
where $\Pat^{(n)}$ is the minimal minimizer of $\min_{\Pat \in \Pi(V)}\FuHash{\alpha^{(n)}}[\Pat]$. Let the iteration start with $\alpha^{(0)} \leq \alpha_1$ and terminate when $\alpha^{(n+1)} = \alpha^{(n)}$. We necessarily have $\Set{\alpha^{(n)}}$ and $\Set{\Pat^{(n)}}$ converge to $\alpha_1$ and $\Pat_1$, respectively. Let the recursion terminate at $N$th iteration, where we have $\alpha^{(N+1)} = \alpha^{(N)}$. Here, $N \leq |V| - 1$ necessarily since, according to Theorem~\ref{theo:PSP}, we have at most $|V| -1$ critical values. According to Lemma~\ref{lemma:PropPSP}(b), $\alpha^{(n+1)} > \alpha^{(n)}$ and $\Pat^{(n+1)} \succ \Pat^{(n)}$ for all $n \in \Set{0,\dotsc,N}$. Recall that $\alpha_1 = \RACO(V)$ and $\Pat_1 = \Pat^*$ (Corollary~\ref{coro:FundPartPSP}). The MDA algorithm exactly implements the recursion above with $\alpha$ initiated as the LB on $\RACO(V)$ in Proposition~\ref{prop:LB}. Therefore, Theorem~\ref{theo:MDAOpt} holds. \hfill\IEEEQED

The authors in \cite[Section 3]{MinAveCost} \cite[Algorithm II]{Narayanan1991PLP} proposed a decomposition algorithm (DA) for determining all partitions $\Pat_j$s and the corresponding critical values $\alpha_j$s in the PSP. The MDA algorithm can be considered as an adapted version of the DA algorithm for determining just $\Pat_1$ and $\alpha_1$. Hence the name MDA.

\section{Tight Sets and Proof of Optimality of CoordSatCap Algorithm} \label{app:CoordSatCap}

According to Lemma~\ref{lemma:SubMIntSubM}, $\FuHash{\alpha} \colon 2^V \mapsto \Real$ is an intersecting submodular set function for all $\alpha \in \RealP$. In addition, $\FuHash{\alpha}$ is normalized, i.e., $\FuHash{\alpha}(\emptyset)=0$, and $P(\FuHash{\alpha},\leq)=P(\FuHashHat{\alpha},\leq)$ \cite[Theorems 2.5(i) and 2.6(i)]{Fujishige2005} where $\FuHashHat{\alpha}$ is submodular \cite[Theorem 12.2.4]{Narayanan1997Book}, \cite[Theorem 7]{MinAveCost}. Consider the maximum sum-rate problem
\begin{multline} \label{eq:CoordSatCapMax}
    \max\Set{r(V) \colon \rv_V \in P(\FuHash{\alpha},\leq)} \\ = \max\Set{r(V) \colon \rv_V \in P(\FuHashHat{\alpha},\leq)}.
\end{multline}
The maximizers of $\max\Set{r(V) \colon \rv_V \in P(\FuHashHat{\alpha},\leq)}$ form the base polyhedron $B(\FuHashHat{\alpha},\leq)$ \cite[Theorem 2.3]{Fujishige2005}. Then, problem \eqref{eq:CoordSatCapMax} is solved if a rate vector $\rv_V \in B(\FuHashHat{\alpha},\leq)$ is determined.
On the other hand, there is a relationship between \eqref{eq:CoordSatCapMax} and the Dilworth truncation problem \cite[Theorem 3.1]{Chen1994NetStrenght}\footnote{In \cite{Chen1994NetStrenght}, Theorem 3.1 is based on the polyhedron $P'(g'_{\alpha},\leq) = \Set{ \rv_V \in \Real^{|V|} \colon r(X) \leq g'_{\alpha}(X), \forall X \subseteq V, X \neq \emptyset }$ with the submodular function $g'_{\alpha}(X) = \alpha - H(V) + H(X), \forall X \subseteq V$, a similar approach as in \cite[Section III-E]{CourtIT2014} that is explained in Remark~\ref{rem:MinSumRate}. Note, $g'_{\alpha}$ is not normalized. Therefore, it is easy to verify that $P(\FuHash{\alpha},\leq) = P'(g'_{\alpha},\leq)$, since $r(\emptyset) = 0$ always, and the maximum and maximizers of $\max\Set{r(V) \colon \rv_V \in P(\FuHash{\alpha},\leq)}$ and $\max\Set{r(V) \colon \rv_V \in P'(g'_{\alpha},\leq)}$ coincide.}:
$$ \max\Set{r(V) \colon \rv_V \in P(\FuHash{\alpha},\leq)} =  \min_{\Pat\in\Pi(V)} \FuHash{\alpha}[\Pat].$$
Then, for a rate vector $\rv_V \in B(\FuHashHat{\alpha},\leq)$, its sum-rate $r(V)$ also determines the value of $\FuHashHat{\alpha}(V) = \min_{\Pat\in\Pi(V)} \FuHash{\alpha}[\Pat]$.

Consider the CoordSatCap algorithm, which is originally proposed as Greedy Algorithm II in \cite[Section 3.2]{Fujishige2005}. Since the algorithm starts with a rate vector $\rv_V \in P(\FuHash{\alpha},\leq)$, it can be shown by induction that, we have $\rv_V \in P(\FuHash{\alpha},\leq)$ and $\rv_V + \hat{\xi}_{\phi_i} \chi_{\phi_i} \in P(\FuHash{\alpha},\leq)$ for all $i \in \Set{1,\dotsc,|V|}$ (also refer to \cite[proof of Theorem 3.19]{Fujishige2005}). Since CoordSatCap algorithm outputs a point $\rv_V \in B(\FuHashHat{\alpha},\leq)$ \cite[Theorem 3.19]{Fujishige2005}, we finally have a rate vector $\rv_V \in P(\FuHash{\alpha},\leq)$ with saturation reached in each dimension, i.e., for all $i\in\Set{1,\dotsc,|V|}$ and $\epsilon > 0$, $\rv_V + \epsilon \chi_{\phi_i} \notin P(\FuHash{\alpha},\leq)$ \cite[Theorem 2.3]{Fujishige2005}. It means $\rv_V \in B(\FuHashHat{\alpha},\leq)$ with the sum-rate $r(V)$ being the maximum of $\max\Set{r(V) \colon \rv_V \in P(\FuHash{\alpha},\leq)}$ or the minimum of $\min_{\Pat\in\Pi(V)} \FuHash{\alpha}[\Pat]$. The minimizer of $\min_{\Pat\in\Pi(V)} \FuHash{\alpha}[\Pat]$ can be determined by the minimizers $\hat{X}_{\phi_i}$ for all $i$ as follows.

Let $\hat{X}_{\phi_i}$ be the minimizer of the saturation capacity problem $ \min\Set{\FuHash{\alpha}(X) - r(X) \colon \phi_i \in X \subseteq V} $ for $i \in \Set{1,\dotsc,|V|}$ in the CoordSatCap algorithm. For $i$, let $\rv_V$ be the rate vector after executing $\rv_V \leftarrow \rv_V + \hat{\xi}_{\phi_i} \chi_{\phi_i}$. Then, $\hat{X}_{\phi_i}$ is $\rv_V$-tight, i.e., $r(\hat{X}_{\phi_i}) = \FuHash{\alpha}(\hat{X}_{\phi_i})$ \cite[proof of Theorem 3.19]{Fujishige2005}. At the end of the CoordSatCap algorithm, we have $\rv_V \in B(\FuHashHat{\alpha},\leq)$ and it can be shown by induction that $\hat{X}_{\phi_i}$ is $\rv_V$-tight for all $i \in \Set{1,\dotsc,|V|}$. Due to the intersecting submodularity of $\FuHash{\alpha}$, $\hat{X}_{\phi_i}$s satisfy property: If $\hat{X}_{\phi_i} \cap \hat{X}_{\phi_j} \neq \emptyset$ for $i \neq j$, then $\hat{X}_{\phi_i} \cap \hat{X}_{\phi_j}$ and $\hat{X}_{\phi_i} \cup \hat{X}_{\phi_j}$ are also $\rv_V$-tight \cite[Section 3]{Chen1994NetStrenght}, \cite[Section 4.2]{MinAveCost}, \cite[Lemma 6]{CourtIT2014}. Then, consider the following process:
\begin{itemize}
    \item Let $\Pat^* \leftarrow \Set{\hat{X}_{\phi_i} \colon i \in V}$;
    \item Repeatedly merge any two elements $\hat{X}_{\phi_i}, \hat{X}_{\phi_j} \in \Pat^*$ such that $\hat{X}_{\phi_i} \cap \hat{X}_{\phi_j} \neq \emptyset$, i.e., do $\Pat^* \leftarrow ( \Pat^* \setminus \Set{\hat{X}_{\phi_i},\hat{X}_{\phi_j}} ) \sqcup \Set{ \hat{X}_{\phi_i} \cup \hat{X}_{\phi_j} } $, until there are no such elements left.
\end{itemize}
Since $\phi_i \in \hat{X}_{\phi_i}$ for all $i \in \Set{1,\dotsc,|V|}$, we finally have $\Pat^*$ being a partition of $V$, i.e. $\Pat^* \in \Pi(V)$ and each element in $\Pat^*$ is $\rv_V$-tight, i.e., $r(C) = \FuHash{\alpha}(C), \forall C \in \Pat^*$. Then, $ r(V) = \sum_{C \in \Pat^*}r(C) = r[\Pat^*] = \FuHash{\alpha}[\Pat^*]$. Recall that $\rv_V \in B(\FuHashHat{\alpha},\leq)$ at the end of the CoordSatCap algorithm, which means
$$ r[\Pat^*] = \max\Set{r(V) \colon \rv_V \in P(\FuHash{\alpha},\leq)} = \min_{\Pat\in\Pi(V)} \FuHash{\alpha}[\Pat].$$
Then, $\Pat^*$ is a minimizer of $\min_{\Pat\in\Pi(V)} \FuHash{\alpha}[\Pat]$ and $\Pat^*$ is the minimal minimizer if $\hat{X}_{\phi_i}$ is the minimal minimizer of $ \min\Set{\FuHash{\alpha}(X) - r(X) \colon \phi_i \in X \subseteq V} $ for all $i \in \Set{1,\dotsc,|V|}$. The process above is equivalent to steps 6 and 7 in the CoordSatCap algorithm. There is another explanation of the optimality of the CoordSatCap algorithm in \cite[Section 4.2]{MinAveCost}.

\subsection{Edmond Greedy Algorithm}

It is shown in \cite[Theorem 3.18]{Fujishige2005} that, for a normalized submodular set function $f$, the CoordSatCap algorithm reduces to the Edmond greedy algorithm \cite{Edmonds2003Convex}: for $i = 1$ to $|V|$, do
$$ r_{\phi_{i}} = f(V_i) - f(V_{i-1}), $$
where $V_i = \Set{\phi_1,\dotsc,\phi_i}$ and $V_{0} = \emptyset$. The resulting $\rv_V$ is a point in $B(f,\leq)$ \cite[Corollary 3.17]{Fujishige2005}. The Edmond greedy algorithm is modified as: For $i = 1$ to $|V|$, do
$$ r_{\phi_i} \leftarrow r_{\phi_i} + \min\Set{f(X) -r(X) \colon \phi_i \in X \subseteq V_i} $$
for solving the non-asymptotic minimum sum-rate problem in the finite linear source model and CCDE in \cite{MiloIT2016,CourtIT2014}, which is exactly the CoordSatCap algorithm based on Lemma~\ref{lemma:Fusion1}.

\subsection{Proof of Lemma~\ref{lemma:IniRate}} \label{app:lemma:IniRate}
    Since the entropy function $H$ is a polymatroid rank function\cite{FujishigePolyEntropy}, we have $\Zero \in P(H,\leq)$, i.e., $0 \leq H(X), \forall X \subseteq V$. Consider the rate vector $\rv_V = (\alpha - H(V)) \chi_{V}$. For $0 \leq \alpha \leq H(V)$, we have $\alpha - H(V) \leq 0$. Then, for $X=\emptyset$, we have $r(\emptyset)=0=\FuHash{\alpha}(\emptyset)$; for all $X \subseteq V$ such that $X \neq \emptyset$, we have
    \begin{equation}
        \begin{aligned}
            r(X) &= (\alpha-H(V))|X| \\
                 &\leq \alpha-H(V) \\
                 &\leq \alpha-H(V)+H(X) = \FuHash{\alpha}(X)
        \end{aligned} \nonumber
    \end{equation}
    So, $\rv_V = (\alpha - H(V)) \chi_{V} \in P(\FuHash{\alpha},\leq)$. \hfill\IEEEQED

\subsection{Proof of Lemma~\ref{lemma:Fusion1}} \label{app:lemma:Fusion1}
    Recall that the entropy function $H$ is monotonic\cite{FujishigePolyEntropy}, i.e., $H(X)\leq H(Y),\forall X,Y \subseteq V \colon X \subseteq Y$. Then, $\FuHash{\alpha}(X) \leq \FuHash{\alpha}(Y),\forall X,Y \subseteq V \colon \emptyset \neq X \subseteq Y$. If $\rv_V \in P(\FuHash{\alpha},\leq)$ such that $r_i \leq 0, \forall i \in V$ initially, we have
    \begin{multline}
        \FuHash{\alpha}(X) - r(X) - ( \FuHash{\alpha}( X \cap V_i ) - r( X \cap V_i ) )=  \\
         \FuHash{\alpha}(X) - \FuHash{\alpha}( X \cap V_i ) - r( X \setminus (X \cap V_i)) \geq 0   \nonumber
    \end{multline}
    holds for all $X \subseteq V$ such that $\phi_i \in X$. Therefore, equality \eqref{eq:lemma:Fusion1} holds for all $i \in \Set{1,\dotsc,|V|}$. Let $\hat{X}$ be a minimizer of $\min\Set{ \FuHash{\alpha}(X) - r(X) \colon \phi_i \in X \subseteq V_i}$. Due to the intersecting submodularity of $\FuHash{\alpha}$, whenever there is a minimizer $\hat{Y}$ of $\min\Set{ \FuHash{\alpha}(X) - r(X) \colon \phi_i \in X \subseteq V}$ such that $\hat{Y} \nsubseteq V_i$, we have $\hat{Y} \cap \hat{X} \subseteq V_i$ and $\phi_i \in \hat{Y} \cap \hat{X} \neq \emptyset$. So, $\hat{Y} \cap \hat{X}$ is also the minimizer of $\min\Set{ \FuHash{\alpha}(X) - r(X) \colon \phi_i \in X \subseteq V}$.\footnote{ Let $\rv_V$ be any rate vector in $\in P(\FuHash{\alpha},\leq)$. The following property can be shown by the intersecting submodularity of $\FuHash{\alpha}$\cite[Section 3]{Chen1994NetStrenght}, \cite[Section 4.2]{MinAveCost}, \cite[Lemma 6]{CourtIT2014}: If $\hat{X}$ and $\hat{Y}$ are two distinctive minimizers of $\min\Set{ \FuHash{\alpha}(X) - r(X) \colon \phi_i \in X \subseteq V}$, then $\hat{X} \cap \hat{Y}$ and $\hat{X} \cup \hat{Y}$ are also the minimizers of $\min\Set{ \FuHash{\alpha}(X) - r(X) \colon \phi_i \in X \subseteq V}$. Note, this property is the same as the tight set argument in Appendix~\ref{app:CoordSatCap} for the proof of the optimality of the CoordSatCap algorithm.}
    Therefore, if $\hat{X}_{\phi_i}$ is the minimal minimizer of $\min\Set{ \FuHash{\alpha}(X) - r(X) \colon \phi_i \in X \subseteq V_i}$, it is also the minimal minimizer of $\min\Set{ \FuHash{\alpha}(X) - r(X) \colon \phi_i \in X \subseteq V}$. \hfill\IEEEQED

\subsection{Proof of Lemma~\ref{lemma:Fusion2}} \label{app:lemma:Fusion2}

    For all $i \in \Set{1,\dotsc,|V|}$, we have $r(C) = \FuHash{\alpha}(C), \forall C \in \Pat^* \colon |C| > 1$ in the CoordSatCap algorithm (due to the previous updates of $\rv_V$ and $\rv_V$-tightness as stated in Appendix~\ref{app:CoordSatCap}) and, since $\rv_V \in P(\FuHash{\alpha},\leq)$, $r(C') + \FuHash{\alpha}( C \setminus C') \geq r(C') + r( C \setminus C') = \FuHash{\alpha}(C)$, i.e., $r(C') \geq \FuHash{\alpha}(C) -\FuHash{\alpha}( C \setminus C'),\forall C' \subsetneq C \colon C' \neq \emptyset, |C| > 1$.

    For any $X \subseteq V$, let $\Y = \Set{C \in \Pat^* \colon C \cap X \neq \emptyset}$. We have
    \begin{equation}
        \begin{aligned}
            & \quad \FuHash{\alpha}(X) - r(X) - \FuHash{\alpha}(\tilde{\Y}) + r(\tilde{\Y})  \\
            & = \FuHash{\alpha}(X) - \FuHash{\alpha}(\tilde{\Y}) + r( \tilde{\Y} \setminus X )  \\
            & = \FuHash{\alpha}(X) - \FuHash{\alpha}(\tilde{\Y}) + \sum_{C \in \Y} r(C \setminus (C \cap X))  \\
            & \geq \FuHash{\alpha}(X) - \FuHash{\alpha}(\tilde{\Y}) + \sum_{C \in \Y}  \big( \FuHash{\alpha}(C) - \FuHash{\alpha}(C \cap X)  \big)  \geq 0,
        \end{aligned} \nonumber
    \end{equation}
    where the last inequality is due to the intersecting submodularity of $\FuHash{\alpha}$. The minimality of $\tilde{U}_i^*$ over all $X \subseteq V$ such that $\phi_i \in X$ and $\tilde{\X}=\tilde{U}_i^*$ can be proved by the induction below.

    Consider the set $\X = \Set{C \in \Pat^* \colon C \cap \hat{X}_{\phi_i} \neq \emptyset}$. For all $C \in \X \colon |C| = 1$, we have $C \subseteq \hat{X}_{\phi_i}$ and $r(C \cup \hat{X}_{\phi_i}) + \hat{\xi}_{\phi_i} = r(\hat{X}_{\phi_i}) + \hat{\xi}_{\phi_i} = \FuHash{\alpha}(\hat{X}_{\phi_i})$; For all $C \in \X \colon |C| > 1$, we have $r(C) = \FuHash{\alpha}(C)$ and $r(\hat{X}_{\phi_i}) + \hat{\xi}_{\phi_i} = \FuHash{\alpha}(\hat{X}_{\phi_i})$ so that $r( C \cup \hat{X}_{\phi_i}) + \hat{\xi}_{\phi_i} = \FuHash{\alpha}( C \cup \hat{X}_{\phi_i})$ (This is also due to the $\rv_V$-tightness as stated in Appendix~\ref{app:CoordSatCap}). By induction, we have $r(\tilde{\X}) + \hat{\xi}_{\phi_i} = \FuHash{\alpha}(\tilde{\X})$ and $\tilde{\X} = \tilde{U}^*_{\phi_i}$, where $\tilde{U}^*_{\phi_i}$ is the minimal minimizer of $\min\Set{\FuHash{\alpha}(\tilde{U}) - r(\tilde{U}) \colon U \subseteq \Pat^*, \phi_i \in \tilde{U}}$. \hfill\IEEEQED

\subsection{Fundamental Partition $\Pat^*$ noncrossing $\hat{X}_{\phi_i}$} \label{app:FundPartNoncross}

For $\alpha \in [0,\ \RACO(V)]$,\footnote{In the MDA algorithm, we always have $\alpha \in [0,\ \RACO(V)]$.}
let $\hat{X}_{\phi_i}$ be the minimal minimizer of $\min\Set{\FuHash{\alpha}(X) - r(X) \colon \phi_i \in X \subseteq V}$ that is non-singleton. According to Theorem~\ref{theo:PSP}(a), the minimal minimizer of $\min_{\Pat \in \Pi(V)} \FuHash{\alpha}[\Pat]$ must be a $\Pat_j$ for some $j \in \Set{1,\dotsc,p}$ in the PSP and, according to Appendix~\ref{app:CoordSatCap}, $\Pat_j$ must be a multi-way cut that does not cross $\hat{X}_{\phi_i}$. We also have $\alpha \geq \FuHash{\alpha}[\Pat_j]$ which is equivalent to $\alpha \leq \varphi(\Pat_j)$. For a $\Pat$ that is crossing $\hat{X}_{\phi_i}$, we have $\FuHash{\alpha}[\Pat_j] < \FuHash{\alpha}[\Pat]$, which is equivalent to
$$ \varphi(\Pat) < (1 - \theta) \alpha + \theta \varphi(\Pat_j) \leq  \varphi(\Pat_j), $$
where $\theta = \frac{|\Pat_j| - 1}{|\Pat| - 1}$. It means that for any partition $\Pat$ that is crossing $\hat{X}_{\phi_i}$ there always exist a partition $\Pat_j$ for some $j \in \Set{1,\dotsc,p}$ in the PSP that is not crossing $\hat{X}_{\phi_i}$ such that $\varphi(\Pat) < \varphi(\Pat_j)$. Since $\Pat_j \preceq \Pat_1 = \Pat^*$ for all $j \in \Set{1,\dotsc,p}$, the fundamental partition does not cross $\hat{X}_{\phi_i}$, necessarily.


\section{Proofs of Theorem~\ref{theo:ExOutput} and Corollary~\ref{coro:SumRateAchievNCO}} \label{app:ExOutput}

According to Theorem~\ref{theo:PolyMatMinSumRateSet}, $\FuHashHat{\alpha}$ is a polymatroid rank function for all $\alpha \geq \RACO(V)$. In this case, we have $B(\FuHashHat{\alpha},\leq) \subseteq \RealP^{|V|}$ so that $\Zero \leq \rv'_V, \forall \rv'_V \in \EX(\FuHashHat{\alpha})$ \cite[Lemma 3.23]{Fujishige2005}. Based on Lemma~\ref{lemma:IniRate}, for all $\alpha \leq H(V)$, if we start with a rate vector $\rv_V \leq \Zero \leq \rv'_V, \forall \rv'_V \in \EX(\FuHashHat{\alpha})$, we necessarily have $\rv_V \in \EX(\FuHashHat{\alpha})$ at the output of the CoordSatCap algorithm. This proves Theorem~\ref{theo:ExOutput}. A similar proof can be found in \cite[Section 4.2]{MinAveCost}. \hfill\IEEEQED

As explained in Appendix~\ref{app:CoordSatCap}, for a given value of $\alpha \geq 0$, the CoordSatCap algorithm outputs $\rv_V \in B(\FuHashHat{\alpha},\leq)$, where the sum-rate $r(V)=\FuHashHat{\alpha}(V)=\min_{\Pat \in \Pi(V)} \FuHash{\alpha}[\Pat]$. According Theorem~\ref{theo:NonEmptBase}, we have $r(V) = \alpha$ being the necessary and sufficient condition for a sum-rate $\alpha$ to be achievable. Also, if $\alpha$ is achievable, we have the output rate vector $\rv_V \in B(\FuHash{\alpha},\leq) = B(\FuHashHat{\alpha},\leq) = \Set{\rv_V \in \RRCO(V) \colon r(V) = \alpha}$. Since the CoordSatCap and CoordSatCapFus algorithms return the same results, corollary also holds for the CoordSatCapFus algorithm. This proves Corollary~\ref{coro:SumRateAchievNCO}. \hfill\IEEEQED

\section{Submodular Function Minimization Algorithms} \label{app:SFM}

The SFM problem under consideration is $\min \Set{f(X) \colon X \subseteq V}$ where $f \colon 2^V \mapsto \Real$ is a submodular set function. The minimizers of this problem form a set lattice\cite{Fujishige2005}, i.e, the set of minimizers is closed w.r.t. the operations $\cap$ and $\cup$: For a lattice $\Lat$, we have $X \cap Y \in \Lat$ and $X \cup Y \in \Lat$ for all $X,Y \in \Lat$. The minimal/smallest and maximal/largest elements of $\Lat$ are $\cap_{X \in \Lat} X$ and $\cup_{X \in \Lat} X$, respectively. For example, $\Lat_1 = \Set{\Set{1},\Set{1,2},\Set{1,3},\Set{1,2,3}}$ is a lattice, while $\Lat_2 = \Set{\Set{1,2},\Set{1,3},\Set{1,2,3}}$ is not. $\Set{1}$ and $\Set{1,2,3}$ are the minimal/smallest and maximal/largest elements of $\Lat_1$, respectively.

There exist polynomial time algorithms for solving the SFM problem\cite{Fujishige2005}. Let $\delta$ be the upper bound on the complexity of evaluating the value of $f(X)$ for $X \subseteq V$. We briefly review the SFM algorithms in the existing literature as follows. The authors in \cite{Grotschel1981} proposed the idea of solving the SFM problem by utilizing the ellipsoid method proposed in \cite{Khachiyan1980Ellipsoid} for parametric optimization, the corresponding algorithm of which is presented in \cite{Grotschel2012}. Although this algorithm completes in polynomial time, it is not very efficient in practice as pointed out in \cite{Grotschel1981,Iwata2007SFM}. The authors in \cite{IFF2001} proposed a polynomial time algorithm for solving the SFM problems base on a combinatorial approach in \cite{Cunningham1985SFM}. The complexity of which is upper bounded by $O(|V|^8 \cdot \delta)$. In \cite{Schrijver2000SFM}, the Iwata-Fleisher-Fujishige (IFF) algorithm improves the complexity to $O(|V|^7 \cdot \log |V| \cdot \delta)$. The most recent polynomial time SFM algorithm is the one in \cite{Orlin2009SFM}, which completes in $O(|V|^5 \cdot \delta + |V|^6)$ time. On the other hand, the authors in \cite{Fujishige2011MinNorm} proposed an SFM algorithm based on the approach in \cite{Wolfe1976} for determining the minimum-norm point in the convex hull. This algorithm is called the minimum-norm point algorithm in the SFM toolbox \cite{Krause2010SFO}. The experimental results in \cite{Fujishige2011MinNorm} show that this algorithm is strongly polynomial and runs faster than those algorithms in \cite{IFF2001,Schrijver2000SFM,Orlin2009SFM}. However, the complexity of the minimum-norm algorithm in terms of $\delta$ is still not determined, which is also stated as an open problem in \cite{Fujishige2011MinNorm}. While it is widely known that SFM problems can be solved in polynomial time, the exact complexity depends on which SFM algorithm is implemented. Therefore, in this paper, we just denote the complexity of solving an SFM problem by $\SFM(|V|)$, where $|V|$ is the cardinality of $V$.

There is another concern on the SFM algorithm: Can the minimal minimizer of an SFM problem be determined efficiently? It should be noted that while the minimum of an SFM problem is unique, the minimizers form a lattice which is not necessarily a singleton. Therefore, an arbitrary minimizer returned by an SFM is not necessarily the minimal in the lattice. However, it is shown in \cite{Nagano2007LineSearch} that this problem is not difficult to solve since any SFM algorithm can be modified so that the maximal/minimal minimizer can be determined. For example, the minimal minimizer can be searched by $|V|$ runs of the IFF algorithm in \cite{Schrijver2000SFM}. In addition, it is shown in \cite{Fujishige2011MinNorm} that both maximal and minimal minimizers of an SFM problem can be directly determined by the minimum-norm point in the base polyhedron, which means that the minimal minimizer can be returned at the same time as an SFM algorithm completes. Therefore, in this paper, the complexity of finding the minimum and minimal minimizer of the SFM problem $\min \Set{f(X) \colon X \subseteq V}$ is $O(\SFM(|V|))$.

\section{Divide-and-conquer (DC) Algorithm \cite{MiloDivConq2011} } \label{app:DV}

The authors in \cite{MiloDivConq2011} proposed a divide-and-conquer (DC) algorithm (as shown in Algorithm~\ref{algo:DC}) for solving the asymptotic minimum sum-rate problem based on the results as follows. For a subset $X \subseteq V$, let $\RACO(X)$ and $\Pat^*_X$ be the minimum sum-rate and the fundamental partition, respectively, for the asymptotic minimum sum-rate problem for attaining the omniscience in $X$. Then, for all optimal rate vectors $\rv_X \in \RRACO(X)$, we have $r(C) = \RACO(X)- H(X) - H(C), \forall C \in \Pat^*_X$\cite[Lemma 1]{MiloDivConq2011}.\footnote{Here, $\RACO(X)- H(X) + H(C)$ corresponds to $\FuHash{\RACO(V)}(C) = \RACO(V) - H(V) + H(C)$ in that $X$ is the ground set for the omniscience problem in $X$. }
Then, for all $C \in \Pat^*_X $ such that $|C| = 1$, the optimal rate of the only user in $C$ is determined; for all $C \in \Pat^*_X$ such that $|C| > 1$, the optimal rate can be determined by recursively solving the omniscience problem in $C$. Therefore, by knowing the fundamental partition $\Pat^*$ for the omniscience in $V$, an optimal rate vector in $\RRACO(V)$ can be solved by recursively breaking the non-singleton subset in the $\Pat^*$ until the rates of all individual users are determined. This idea is implemented by the DC algorithm, where $\RACO(X)$ and $\Pat^*_X$ are determined by the DA algorithm in \cite[Section 3]{MinAveCost}. In step 3 in the DC algorithm, $\Delta{r}$ is the excessive rate, the additional rate that should be transmitted by $X$ in addition to $\RACO(V)$. $\Delta{r}$ can be transmitted by any users in $X$ after the omniscience is attained in $X$ \cite[Lemma 3]{MiloDivConq2011}.

	\begin{algorithm} [t]
	\label{algo:DC}
	\small
	\SetAlgoLined
	\SetKwInOut{Input}{input}\SetKwInOut{Output}{output}
	\SetKwFor{For}{for}{do}{endfor}
    \SetKwRepeat{Repeat}{repeat}{until}
    \SetKwIF{If}{ElseIf}{Else}{if}{then}{else if}{else}{endif}
	\BlankLine
	\Input{a subset $X \subseteq V$,  and $r(X)$ if $X \neq V$}
	\Output{a rate vector $\rv_X \in \RRACO(X)$}
	\BlankLine
        $(\RACO(X),\Pat^*_X) \leftarrow \DA(\emptyset,X)$\;
        \lForAll{$C \in \Pat^*_X$}{$r(C) \leftarrow \RACO(X)- H(X) + H(C)$}
        \lIf{$X \neq V$}{$\Delta{r} \leftarrow r(X)-\RACO(X)$, choose any $C \in \Pat^*_X$ and let $r(C) \leftarrow r(C)+\Delta{r}$}
        \lForAll{$C \in \Pat^*_X \colon |C| >1$}{$\rv_C \leftarrow \DC(C,r(C))$}
	\caption{divide-and-conquer (DC) algorithm \cite{MiloDivConq2011}}
	\end{algorithm}

\begin{example} \label{ex:auxDC}
    By applying the DC algorithm to the system in Example~\ref{ex:aux}, we first determine that the minimum sum-rate $\RACO(V) = \frac{13}{2}$ and fundamental partition $\Pat^* = \Set{\Set{1,4,5},\Set{2},\Set{3}}$. We also know that we must have $r(\Set{1,4,5}) = \frac{11}{2}$, $r(\Set{2}) = \frac{1}{2}$ and $r(\Set{3}) = \frac{1}{2}$ for attaining the omniscience in $V$ with the minimum sum-rate $\RACO(V) = \frac{13}{2}$. By recursively calling the DC algorithm, the individual rates in $\Set{1,4,5}$ are determined as $r_1 = 1$, $r_4 = \frac{5}{2}$ and $r_5 = 0$ so that we have the optimal rate vector $\rv_V = (1,\frac{1}{2},\frac{1}{2},\frac{5}{2},0) \in \RRACO(V)$ at the output.
\end{example}

Since the DA algorithm completes in $O(|V|^2 \cdot \SFM(|V|))$ time \cite[Theorem 5]{MinAveCost}, the complexity of the DC algorithm is $O(|V|^3 \cdot \SFM(|V|))$. However, the recursive splitting of the non-singleton subset in the fundamental partition $\Pat^*$ in the DC algorithm is not necessary since an optimal rate vector $\rv_V \in \RRACO(V)$ is obtained when $\Pat^*$ is determined by the DA algorithm.\footnote{In fact, the DC algorithm utilizes $\RACO(V)$ and $\Pat^*$ determined by the DA algorithm in \cite[Section 3]{MinAveCost} while discards the optimal rate vector $\rv_V \in B(\FuHashHat{\RACO(V)},\leq)$. This is not surprising since the study in \cite{MinAveCost} aims to solve a clustering problem, where the optimal partition $\Pat^*$ is of the most interest. Therefore, although a rate vector $\rv_V \in B(\FuHashHat{\RACO(V)},\leq)$ is returned as an auxiliary result, it is not explicitly stated in \cite{MinAveCost} that $B(\FuHashHat{\RACO(V)},\leq)=\RRACO(V)$ so that this rate vector is the solution to the asymptotic minimum sum-rate problem.}

\section{Proof of Theorem~\ref{theo:WeightedSumRate}} \label{app:theo:WeightedSumRate}

    According to Theorem~\ref{theo:ExOutput}, the MDA algorithm returns an optimal rate vector $\rv_V \in \RRACO(V)$ that is an extreme point in $B(\FuHashHat{\RACO(V)},\leq)$, i.e., $\rv_V \in \EX(\FuHashHat{\RACO(V)})$. Based on \cite[Corollary 3.17]{Fujishige2005}, for linear ordering w.r.t. $\wv_V$, the output $\rv_V$ is also the minimizer of $\min \Set{ \wv_V^{\intercal} \rv_V \colon \rv_V \in \RRACO(V) }$, where $\RRACO(V) = B(\FuHashHat{\RACO(V)},\leq)$. It is shown in the proof of Corollary~\ref{coro:IntFracFinite} that, for a finite linear source model, when the input $\alpha = \RNCO(V)$, the CoordSatCapFus algorithm outputs a rate vector $\rv_V \in \EX(\FuHashHat{\RNCO(V)})$, which, according to \cite[Corollary 3.17]{Fujishige2005} is the minimizer of $\min \Set{ \wv_V^{\intercal} \rv_V \colon \rv_V \in \RRNCO(V)}$ where $\RRNCO(V) = B(\FuHashHat{\RNCO(V)},\leq)$. Since the outputs of the CoordSatCapFus and CoordSatCap algorithms are the same, it also applies to the CoordSatCap algorithm. \hfill\IEEEQED

\section{Proofs of Theorem~\ref{theo:FundPatMinSep} and Corollary~\ref{coro:MinSepMConvex}}\label{app:FundPatMinSep}

The proof of Theorem~\ref{theo:FundPatMinSep} relies on the method for determining the minimal separators of a submodular function that is proposed in \cite[Lemma 3.41]{Fujishige2005} \cite[Theorems 3.1 and 3.2]{Bilxby1985}.\footnote{This method was first proposed in \cite[Theorems 3.1 and 3.2]{Bilxby1985} for polymatroid rank functions and then generalized to \cite[Lemma 3.41]{Fujishige2005} for submodular functions. Here, the minimal separators also correspond to the principal structure of a submodular system in \cite{FujishigePincStruct}.}

\begin{lemma}[{\cite[Lemma 3.41]{Fujishige2005}}] \label{lemma:MinSepMethod}
    For any rate vector $\rv_V \in B(\FuHashHat{\RACO(V)},\leq)$, let $\hat{X}_i$ be the minimal minimizers of $\min \Set{ \FuHashHat{\RACO(V)} - r(X) \colon i \in X \subseteq V }$. Initiate the minimal minimizer set $\Pat^* = \Set{\hat{X}_i \colon i \in V}$ and repeatedly merge any two distinctive elements $\hat{X}_{i}, \hat{X}_{j} \in \Pat^*$ such that $\hat{X}_{i} \cap \hat{X}_{j} \neq \emptyset$, i.e., do $\Pat^* \leftarrow ( \Pat^* \setminus \Set{\hat{X}_{i},\hat{X}_{j}} ) \sqcup \Set{ \hat{X}_{i} \cup \hat{X}_{j} } $, until there are no such elements left. The resulting $\Pat^*$ is the set of minimal separators of $\FuHashHat{\RACO(V)}$.
\end{lemma}

\begin{lemma} \label{lemma:MinSep}
    Let $\Phi$ be any linear ordering of $V$. For all $\alpha \in \RealP$, $i \in \Set{1,\dotsc,|V|}$ and $\rv_V \in P(\FuHash{\alpha},\leq)$, if $\hat{\xi}_{\phi_i}$ and $\hat{X}_{\phi_i}$ are the minimum and minimal minimizer of $\min \Set{ \FuHash{\alpha}(X)-r(X) \colon \phi_i \in X \subseteq V }$, respectively, then $\hat{\xi}_{\phi_i}$ and $\hat{X}_{\phi_i}$ are also the minimum and minimal minimizer of $\min \Set{ \FuHashHat{\alpha}(X)-r(X) \colon \phi_i \in X \subseteq V }$, respectively.
\end{lemma}
\begin{IEEEproof}
    Let $\rv'_V = \rv_V + \hat{\xi}_{\phi_i} \chi_{\phi_i}$. Then, we have $\rv'_V \in P(\FuHash{\alpha},\leq)$ and $\hat{X}_{\phi_i}$ is $\rv'_V$-tight, i.e., $\FuHash{\alpha}(\hat{X}_{\phi_i}) = r'(\hat{X}_{\phi_i})$ \cite[proof of Theorem 3.19]{Fujishige2005}. Since $P(\FuHash{\alpha},\leq)=P(\FuHashHat{\alpha},\leq)$, $\rv'_V \in P(\FuHashHat{\alpha},\leq)$ and we have $r'(\hat{X}_{\phi_i}) \leq \FuHashHat{\alpha}(\hat{X}_{\phi_i})$ \cite[Theorems 2.5(i) and 2.6(i)]{Fujishige2005}. On the other hand, $\FuHashHat{\alpha}(X) \leq \FuHash{\alpha}(X)$ for all $X \subseteq V$ by the definition of the Dilworth truncation. Then, $\FuHashHat{\alpha}(\hat{X}_{\phi_i}) \leq \FuHash{\alpha}(\hat{X}_{\phi_i}) = r'(\hat{X}_{\phi_i})$. Therefore, $r'(\hat{X}_{\phi_i})=\FuHashHat{\alpha}(\hat{X}_{\phi_i})$, which means $\hat{\xi}_{\phi_i}$ is also the minimum of $\min \Set{ \FuHashHat{\alpha}(X)-r(X) \colon \phi_i \in X \subseteq V }$.

    We prove that $\hat{X}_{\phi_i}$ is the minimal minimizer of $\min \Set{ \FuHashHat{\alpha}(X)-r(X) \colon \phi_i \in X \subseteq V }$ by contradiction. Assume that $\hat{X}_{\phi_i}$ is not the minimal minimizer, i.e., there exists $X \subsetneq \hat{X}_{\phi_i}$ such that $\phi_i \in X$ and $\FuHashHat{\alpha}(X) - r(X) = \hat{\xi}_{\phi_i}$. Then, $r'(X)=\FuHashHat{\alpha}(X) < \FuHash{\alpha}(X)$, where the last inequality is strict due to the fact that $X$ is not the minimizer of $\min \Set{ \FuHash{\alpha}(X)-r(X) \colon \phi_i \in X \subseteq V }$. According to the definition of the Dilworth truncation, $\FuHashHat{\alpha}(X) < \FuHash{\alpha}(X)$ means that there exists a $\Pat^* \in \Pi'(X)$ such that
    $$ r'(X) = \FuHashHat{\alpha}(X) = \FuHash{\alpha}[\Pat^*] = r'[\Pat^*]. $$
    On one hand, we have $r'(C) \leq \FuHash{\alpha}(C), \forall C \in \Pat^*$ since $\rv'_V \in P(\FuHash{\alpha},\leq)$. On the other hand, $\FuHash{\alpha}(C) - r'(C) = \sum_{C' \in \Pat^* \setminus \Set{C}} (r'(C')-\FuHash{\alpha}(C')) \leq 0$, i.e., $r'(C) \geq \FuHash{\alpha}(C)$ for all $C \in \Pat^*$. Then, $r'(C) = \FuHash{\alpha}(C)$ for all $C \in \Pat^*$. Let $\hat{C} \in \Pat^*$ such that $\phi_i \in \hat{C}$. We have $\FuHash{\alpha}(\hat{C}) = r'(\hat{C})$, i.e., $\FuHash{\alpha}(\hat{C})-r(\hat{C}) = \hat{\xi}_{\phi_i}$. Here, $\hat{C} \subsetneq X$ necessarily, which means $\hat{X}_{\phi_i}$ is not the minimal minimizer of $\min \Set{ \FuHash{\alpha}(X)-r(X) \colon \phi_i \in X \subseteq V }$. This contradicts the given condition. Therefore, we must have $\hat{X}_{\phi_i}$ being the minimal minimizer of $\min \Set{ \FuHashHat{\alpha}(X)-r(X) \colon \phi_i \in X \subseteq V }$.
\end{IEEEproof}

Recall that we have $\rv_V \in B(\FuHashHat{\RACO(V)},\leq)$ and the fundamental partition $\Pat^*$ at the output of the CoordSatCap algorithm by inputting $\alpha=\RACO(V)$. For the base point $\rv_V \in B(\FuHashHat{\RACO(V)},\leq)$, $\hat{X}_{i}$ is the minimal minimizer of $\min\Set{\FuHashHat{\RACO(V)}(X)-r(X), i \in X \subseteq V}$ for all $i \in V$ according to Lemma~\ref{lemma:MinSep}. If we implement the method in Lemma~\ref{lemma:MinSepMethod} over all $\hat{X}_{i}$s, we have the set of minimal separators of $\FuHashHat{\RACO(V)}$ the same as the fundamental partition $\Pat^*$. This proves Theorem~\ref{theo:FundPatMinSep}. \hfill \IEEEQED

Corollary~\ref{coro:MinSepMConvex} is proved as follows. For all $C \in \Pat^*$, $g(\rv_C^*) \leq g(\rv_C^* + \epsilon (\chi_i - \chi_j)),\forall i, j \in C$ according to Theorem~\ref{theo:LocToGlobMConvex}. On the other hand, according to property (c) in Lemma~\ref{lemma:MinSepProp}, if $i \in C$ and $i \in C'$ for any $C,C'\in\Pat^*$ such that $C \neq C'$, $ \rv_V + \epsilon(\chi_i-\chi_j) \notin B(\FuHashHat{\RACO(V)},\leq) = \RRACO(V)$ for all $\epsilon > 0$. Then, we have $ g(\rv_V^*) \leq g(\rv_V^* + \epsilon (\chi_i - \chi_j))$ for all $i,j \in V$ and $\epsilon > 0$ such that $\rv_V + \epsilon (\chi_i - \chi_j) \in B(\FuHashHat{\RACO(V)},\leq) = \RRACO(V)$. Therefore, according to Theorem~\ref{theo:LocToGlobMConvex}, $\rv_V^* = \oplus_{C\in\Pat^*} \rv_C^*$ is the minimizer of $\min\Set{g(\rv_V) \colon \rv_V \in B(\FuHashHat{\RACO(V)},\leq) = \RRACO(V)}$. \hfill \IEEEQED

%

\bibliography{COFusionBIB}

\begin{IEEEbiographynophoto}
{Ni Ding} received the B.C.A. degree from the Shanghai Second Polytechnic University, China, the B.E. degree in Telecommunications (1st class honours) from the University of New South Wales, Australia, and the PhD degree from the Australian National University, Australia, in 2005, 2012 and 2017, respectively. From September 1998 to August 2006, she was an associate engineer in Shanghai Telecom, China. She is now a postdoctoral fellow at Data 61, The Commonwealth Scientific and Industrial Research Organisation (CSIRO), Australia. Her research interests generally include optimizations in information theory, wireless communications, signal processing and machine learning. She is currently interested in the discrete and combinatorial optimization problems raised in discrete event control in cross-layer adaptive modulation, source coding and network coding problems and game theory (in particular, the games with strong structures, e.g., supermodular and convex games).
\end{IEEEbiographynophoto}

\begin{IEEEbiographynophoto}
{Chung Chan} received his B.Sc., M.Eng. and Ph.D. from the Electrical Engineering and Computer Science Department at the Massachusetts Institute of Technology (MIT) in 2004, 2005 and 2010 respectively. He is currently a Research Assistant Professor at the Institute of Network Coding, the Chinese University of Hong Kong, and the secretary of the IEEE Information Theory Society Hong Kong Chapter. He will join the Department of Computer Science, City University of Hong Kong, as an Assistant Professor in 2018. His research interest is to develop general information measures and flow models from network information theory that are applicable to practical problems in machine learning. His research topics include the development of network link models using matroids, the derivation of theoretical limits and optimal strategies for the problems of multiterminal source coding, data exchange, and secret generation. His most significant contribution is the extension of Shannon’s mutual information to the multivariate case, and the discovery of its connections to various problems in information theory and machine learning.
\end{IEEEbiographynophoto}

\begin{IEEEbiographynophoto}
{Qiaoqiao Zhou} received his B.B.A. in business administration and M.S. in electrical engineering from Beijing University of Post and Telecommunication, China, in 2011 and 2014, respectively. In 2014-2015 he was a research assistant at Institute of Network Coding, the Chinese University of Hong Kong. He is currently a Ph.D. student at Department of Information Engineering, the Chinese University of Hong Kong. His research interests include information-theoretic security and machine learning.
\end{IEEEbiographynophoto}

\begin{IEEEbiographynophoto}
{Rodney A. Kennedy} received the B.E. degree (1st class honours and university medal) from the University of New South Wales, Sydney, Australia, the M.E. degree from the University of Newcastle, and the Ph.D. degree from the Australian National University, Canberra. Since 2000 he has been a Professor in engineering at the Australian National University, Canberra, Australia.
He has co-authored over 350 refereed journal or conference papers and a book “Hilbert Space Methods in Signal Processing” (Cambridge Univ. Press, 2013). He has been a Chief Investigator in a number of Australian Research Council Discovery and Linkage Projects. His research interests include digital signal processing, digital and wireless communications, and acoustical signal processing.
\end{IEEEbiographynophoto}

\begin{IEEEbiographynophoto}
{Parastoo Sadeghi}is an Associate Professor at the Research School of Engineering, Australian National University, Canberra, Australia. She received the B.E. and M.E. degrees in electrical engineering from Sharif University of Technology, Tehran, Iran, in 1995 and 1997, respectively, and the Ph.D. degree in electrical engineering from the University of New South Wales, Sydney, Australia, in 2006. From 1997 to 2002, she was a Research Engineer and then a Senior Research Engineer at Iran Communication Industries, Tehran, and at Deqx (formerly known as Clarity Eq), Sydney. She has visited various research institutes, including the Institute for Communications Engineering, Technical University of Munich, in 2008 and MIT in 2009 and 2013.

Dr. Sadeghi has co-authored around 140 refereed journal or conference papers and a book on Hilbert Space Methods in Signal Processing (Cambridge Univ. Press, 2013). She has been a Chief Investigator in a number of Australian Research Council Discovery and Linkage Projects. Her research interests are mainly in the areas of network coding, wireless communications systems, and signal processing. Dr. Sadeghi was the recipient of two IEEE Region 10 Student Paper Awards for her research on the information theory of time-varying fading channels in 2003 and 2005.
\end{IEEEbiographynophoto}

\vfill

\end{document}